\documentclass{elsart}

\usepackage{natbib}

\usepackage{natbib}
\usepackage{amsmath}
\usepackage{graphicx}
\usepackage{yfonts}
\usepackage{color}
\usepackage{epstopdf}
\usepackage{amssymb}

\begin{document}

\begin{frontmatter}

\title{(Never) Mind your $p$'s and $q$'s: \\ Von Neumann versus Jordan on the Foundations of Quantum Theory.}

\thanks[mpiwg]{This paper was written as part of a joint project in the history of quantum physics of the {\it Max Planck Institut f\"{u}r Wissenschaftsgeschichte} and the {\it Fritz-Haber-Institut} in Berlin.}

\author[duncan]{Anthony Duncan},
\author[janssen]{Michel Janssen\corauthref{cor}}
\corauth[cor]{Corresponding author. Address: Tate Laboratory of Physics, 116 Church St.\ SE, Minneapolis, MN 55455, USA, Email: janss011@umn.edu}
\address[duncan]{Department of Physics and Astronomy, University of Pittsburgh}
\address[janssen]{Program in the History of Science, Technology, and Medicine, \\ University of Minnesota}

\begin{abstract}
In two papers entitled ``On a new foundation [{\it Neue Begr\"undung}] of quantum mechanics," Pascual \citet{Jordan 1927b, Jordan 1927f} presented his version of what came to be known as the Dirac-Jordan statistical transformation theory.  As an alternative that avoids the mathematical difficulties facing the approach of Jordan and  Paul A.\ M.\ \citet{Dirac 1927}, John \citet{von Neumann 1927a} developed the modern Hilbert space formalism of quantum mechanics. In this paper, we focus on Jordan and von Neumann. Central to the formalisms of both are expressions for conditional probabilities of finding some value for one quantity given the value of another. Beyond that Jordan and von Neumann had very different views about the appropriate formulation of problems in quantum mechanics. For Jordan, unable to let go of the analogy to classical mechanics, the solution of such problems required the identification of sets of canonically conjugate variables, i.e., $p$'s and $q$'s. For von Neumann, not constrained by the analogy to classical mechanics,  it required only the identification of a maximal set of commuting operators with simultaneous eigenstates. He had no need for $p$'s and $q$'s. Jordan and von Neumann also stated the characteristic new rules for probabilities in quantum mechanics somewhat differently. \citet{Jordan 1927b} was the first to state those rules in full generality. Von Neumann (1927a) rephrased them and, in a subsequent paper \citep{von Neumann 1927b}, sought to derive them from more basic considerations. In this paper we reconstruct the central arguments of these 1927 papers by Jordan and von Neumann and of a paper on Jordan's approach by \citet{Hilbert-von Neumann-Nordheim}. We highlight those elements in these papers that bring out the gradual loosening of the ties between the new quantum formalism and classical mechanics.
\end{abstract}

\begin{keyword} 
Pascual Jordan \sep John von Neumann \sep transformation theory  \sep probability amplitudes \sep canonical transformations \sep Hilbert space \sep spectral theorem 
\end{keyword}

\end{frontmatter}

\section{Introduction}

\subsection{The Dirac-Jordan statistical transformation theory}

On Christmas Eve 1926, Paul A.\ M.\ Dirac, on an extended visit to Niels Bohr's institute in Copenhagen, wrote to Pascual Jordan, assistant to Max Born in G\"ottingen:\footnote{Reminiscing about the early days of quantum mechanics, \citet{Jordan 1955} painted an idyllic picture of G\"ottingen during this period, both of the town itself and of its academic life. J.\ Robert Oppenheimer, who spent a year there in 1926--1927, also remembered a darker side:
``[A]lthough this society was extremely rich and warm and helpful to me, it was parked there in a very very miserable German mood which probably in Thuringia was not as horrible as a little bit further east; G\"ottingen is not in Thuringia ---not as bad as in Thuringia---but it was close enough to Thuringia to be bitter, sullen, and, I would say, discontent and angry and loaded with all those ingredients which were later to produce a major disaster" (Interview with Oppenheimer for the {\it Archive for History of Quantum Physics} [AHQP], session 2, p.\ 5; quoted in part by Bird and Sherwin, 2005, pp.\ 57--58). For detailed references to material in the AHQP, see \citet{Kuhn et al. 1967}.}
\begin{quotation}
Dr.\ Heisenberg has shown me the work you sent him, and as far as I can see it is equivalent to my own work in all essential points. The way of obtaining the results may be rather different though \ldots\ I hope you do not mind the fact that I have obtained the same results as you, at (I believe) the same time as you. Also, the Royal Society publishes papers more quickly than the Zeits.\ f.\ Phys., and I think my paper will appear in their January issue. I am expecting to go to G\"ottingen at the beginning of February, and I am looking forward very much to meeting you and Prof.\ Born there (Dirac to Jordan, December 24, 1916, AHQP).\footnote{This letter is quoted almost in its entirety (cf.\ note \ref{MR2}) and with extensive commentary by \citet[p.\ 72, pp.\ 83--87]{Mehra Rechenberg}.\label{MR1}}
\end{quotation}
Dirac's paper, ``The physical interpretation of the quantum dynamics" \citep{Dirac 1927}, had been received by the Royal Society on December 2. {\it Zeitschrift f\"ur Physik} had received Jordan's paper, ``On a new foundation [{\it Neue Begr\"un\-dung}] of quantum mechanics" \citep{Jordan 1927b}, on December 18.\footnote{We will refer to this paper as {\it Neue Begr\"undung} I to distinguish it from {\it Neue Begr\"undung} II, submitted June 3, 1927 to the same journal, a sequel in which Jordan tried both to  simplify and  generalize his theory \citep{Jordan 1927f}. A short version of {\it Neue Begr\"undung} I was presented to the G\"ottingen Academy by Born on behalf of Jordan in the session of January 14, 1927 \citep{Jordan 1927c}.} In both cases it took a month for the paper to get published: Dirac's appeared January 1, Jordan's January 18.\footnote{The extensive bibliography of \citet{Mehra Rechenberg} gives the dates papers were received and published for all the primary literature it lists.} What Dirac and Jordan, independently of one another, had worked out and presented in these papers has come to be known as the Dirac-Jordan (statistical) transformation theory.\footnote{The term `transformation theory'---and even the term `statistical transformation theory'---is sometimes used more broadly (see cf.\ note \ref{nomenclature}).}
As Jordan wrote in a volume in honor of Dirac's 70th birthday:
\begin{quotation}
After Schr\"odinger's beautiful papers [Schr\"odinger, 1926a], I formulated what I like to call the statistical transformation theory of quantum mechanical systems, answering generally the question concerning the probability of finding by measurement of the observable $b$ the eigenvalue $b'$, if a former measurement of another observable $a$ had given the eigenvalue $a'$. The same answer in the same generality was developed in a wonderful manner by {\it Dirac} \citep[p.\ 296; emphasis in original]{Jordan 1973}.
\end{quotation}
 In our paper, we focus on Jordan's version of the theory and discuss Dirac's version only to indicate how it differs from Jordan's.\footnote{For discussions of Dirac's version, see, e.g., \citet[pp.\ 302--305]{Jammer 1966}, \citet[pp.\ 39--43]{Kragh 1990}, \citet[pp.\ 337--345]{Darrigol 1992}, \citet[72--89]{Mehra Rechenberg}, and \citet[543--548]{Rechenberg 2010}.\label{dirac}}

Exactly one month before Dirac's letter to Jordan, Werner Heisenberg, Bohr's assistant in the years 1926--1927, had already warned Jordan that he was about to be scooped. In reply to a letter, apparently no longer extant, in which Jordan must have given a preview of {\it Neue Begr\"undung} I, he wrote:
\begin{quotation}
I hope that what's in your paper isn't exactly the same as what's in a paper Dirac did here. Dirac's basic idea is that the physical meaning of $S_{\alpha \beta}$, given in my note on fluctuations, can greatly be generalized, so much so that it covers all physical applications of quantum mechanics there have been so far, and, according to Dirac, all there ever will be (Heisenberg to Jordan, November 24, 1926, AHQP).\footnote{\citet[p.\ 72]{Mehra Rechenberg} quote more extensively from this letter. They also quote and discuss (ibid., p.\ 78) a similar remark in a letter from Heisenberg to Wolfgang Pauli, November 23, 1926 \citep[Doc.\ 148, p.\ 357]{Pauli 1979}. In this letter, Heisenberg talks about Dirac's ``extraordinary grandiose [{\it gro\ss z\"ugige}] generalization" (quoted by Kragh, 1990, p.\ 39).}
\end{quotation} 
The ``note on fluctuations" is a short paper received by {\it Zeit\-schrift f\"ur Physik} on November 6 but not published until December 20. In this note, \citet{Heisenberg 1926c}, drawing on the technical apparatus of an earlier paper on resonance phenomena \citep{Heisenberg 1926a}, analyzed a simple system of a pair of identical two-state atoms perturbed by a small interaction to allow for the flowing back and forth of the amount of energy separating the two states of the atoms.\footnote{As \citet[p.\ 501]{Heisenberg 1926c} explained in the introduction of his paper, fluctuations were of interest because, as Albert \citet{Einstein 1905, Einstein 1909a, Einstein 1909b} had shown, they provide a tell-tale sign of quantum discontinuities. For discussion of Heisenberg's (1926c) note on fluctuations in the context of the debate over Jordan's derivation of Einstein's (1909a,b) formula for fluctuations in black-body radiation in the {\it Dreim\"annerarbeit} of \citet{dreimaenner}, see \citet[pp.\ 643--645]{Duncan and Janssen 2008}. For further discussion of Heisenberg's (1926c) note, see \citet[pp.\ 541--543]{Rechenberg 2010}.\label{fluctuations}}  The $S_{\alpha \beta}$'s are the elements of a matrix $S$ implementing a canonical transformation, $f' = S^{-1} f S$, from a quantum-mechanical quantity $f$ for the unperturbed atom to the corresponding quantity $f'$ for the perturbed one. The discrete two-valued indices $\alpha$ and $\beta$ label the perturbed and unperturbed states, respectively. Heisenberg proposed the following interpretation of these matrix elements:
\begin{quotation}
{\it If the perturbed system is in a state $\alpha$, then $| S_{\alpha \beta} |^2$ gives the probability that (because of collision processes, because the perturbation suddenly stops, etc.) the system is found to be in state $\beta$} 
\citep[p.\ 505; emphasis in the original]{Heisenberg 1926c}.
\end{quotation}
\citet[pp.\ 504--505]{Heisenberg 1926c} emphasized that the same analysis  applies to fluctuations of other quantities (e.g., angular momentum) and to other quantum systems with two or more not necessarily identical components as long as their spectra all share the same energy gap.

In the introduction of his paper on transformation theory,  \citet[p.\ 622]{Dirac 1927} briefly described Heisenberg's proposal, thanked him for sharing it before publication, and announced that it is ``capable of wide extensions." Dirac, in fact, extended Heisenberg's interpretation of $S_{\alpha \beta}$ to {\it any} matrix implementing a transformation of some quantum-mechanical quantity $g$ from one matrix representation, written as $g(\alpha' \alpha'')$, to another, written as $g(\xi'\xi'')$. The primes on the indices labeling rows and columns distinguish the numerical values of the quantities $\alpha$ and $\xi$ from those quantities themselves (ibid., p.\ 625). Depending on the spectrum of the associated quantities, these indices thus take on purely discrete values, purely continuous ones, or a mix of both. Dirac  wrote all equations in his paper as if the indices only take on continuous values (ibid.). He introduced the compact notation $(\xi'/\alpha')$ for the transformation matrix from the $\alpha$-basis to the $\xi$-basis (ibid., p.\ 630).  In the spirit of Heisenberg's proposal, $|(\xi'/\alpha')|^2 d\xi'$ is interpreted as the probability that $\xi$ has a value between $\xi'$ and $\xi'+ d\xi'$ given that $\alpha$ has the value $\alpha'$. Although the notation is Dirac's, this formulation of the interpretation is  Jordan's (1927b, p.\ 813).\footnote{For careful discussion of Dirac's (1927, secs.\ 6--7, pp.\ 637--641) own formulation, see \citet[pp.\ 342--343]{Darrigol 1992}.}

It is easy to understand why Heisenberg would have been excited about what he saw as Dirac's generalization of his own work.  He felt strongly that the interpretation of the quantum formalism should naturally emerge from matrix mechanics without any appeal to wave mechanics. For this reason, he initially disliked Born's statistical interpretation of the wave function as well as Bohr's concept of complementarity, with its emphasis on wave-particle duality.\footnote{Heisenberg  explicitly said so when interviewed by Kuhn for the AHQP project. Talking about Born's statistical interpretation of the wave function and his own proposal in his note on fluctuations, he said: ``I definitely wanted to keep always on the quantum mechanical side and not make any concession to the Schr\"odinger side" (session 10, p.\ 6). Talking about complementarity, he likewise said: ``I didn't want to go too much into the Schr\"odinger line of thought. I just wanted to stick to the matrix line" (session 11, p.\ 11). He made a similar statement in his autobiography: ``Bohr was trying to allow for the simultaneous existence of both particle and wave concepts \ldots\ I disliked this approach. I wanted to start from the fact that quantum mechanics as we then knew it [i.e., matrix mechanics] already imposed a unique physical interpretation \ldots\ For that reason I was---certainly quite wrongly---rather unhappy about a brilliant piece of work Max Born had done \ldots\ I fully agreed with Born's thesis as such, but disliked the fact that it looked as if we still had some freedom of interpretation"  \citep[p.\ 76]{Heisenberg 1971}. The following remark in the conclusion of one of the two papers in which Born first introduced his statistical interpretation makes it easy to understand Heisenberg's initial animosity. \citet[p.\ 826]{Born 1926b} wrote: ``Schr\"odinger's form [of quantum mechanics] appears to account for the facts in by far the easiest way; moreover, it makes it possible to retain the usual conception of space and time in which events take place in perfectly familiar fashion."\label{Teil-Ganze}} Transformation theory reconciled him with both ideas. It showed that wave-particle duality was just one example of a much broader plurality of equivalent forms in which quantum mechanics can be expressed.\footnote{Heisenberg made this contrast between the dualism of complementarity and the pluralism of transformation theory in one of the passages from the AHQP interview referred to in the preceding note (session 11, p.\ 11).} And it showed that the Schr\"odinger energy eigenfunctions can be seen as elements of the transformation matrix diagonalizing the Hamiltonian. As Dirac put it: 
\begin{quotation}
{\it The eigenfunctions of Schr\"odinger's wave equation are just the transformation functions \ldots\ that enable one to transform from the ($q$) scheme of matrix representation to a scheme in which the Hamiltonian is a diagonal matrix} \citep[p.\ 635; emphasis in the original]{Dirac 1927}.\footnote{This passage is quoted in most discussions of Dirac's paper. See, e.g., \citet[p.\ 403]{Jammer 1966}, \citet[p.\ 40]{Kragh 1990}, and \citet[p.\ 547]{Rechenberg 2010}.}
\end{quotation}
\citet[p.\ 810, p.\ 822]{Jordan 1927b} clearly recognized this too. The probability interpretation could thus either be given in terms of Schr\"odinger wave functions or in terms of transformation matrices.\footnote{The passage from Heisenberg's autobiography quoted in note \ref{Teil-Ganze} continues: ``I was firmly convinced that Born's thesis itself was the necessary consequence of the fixed interpretation of special magnitudes in quantum mechanics. This conviction was strengthened further by two highly informative mathematical studies by Dirac and Jordan" \citep[pp.\ 76--77]{Heisenberg 1971}.}

Heisenberg explicitly made the connection between wave functions and transformation matrices in the letter to Jordan from which we quoted above:
\begin{quotation}
$S$ is the solution of a transformation to principal axes and also of a differential equation \`a la Schr\"odinger, though by no means always in $q$-space. One can introduce matrices of a very general kind, e.g., $S$ with indices $S(q,E)$. The $S(q,E)$ that solves Born's transformation to principal axes in Qu.M.\ II (Ch.\ 3, Eq.\ (13)) is Schr\"odinger's $S(q,E) = \psi_E(q)$  (Heisenberg to Jordan, November 24, 1926, AHQP).\footnote{Heisenberg added: ``The more general cases can be seen as a simple application of Lanczos's ``field equations" (the same might be true of your $\varphi(y) = \int K(x,y) \, \psi(x) \, dx$??). I have said all along that this Lanczos is not bad" (ibid.). Cornelius \citet{Lanczos 1926} wrote a paper on the connection between matrices and integral kernels (such as $K(x,y)$ in the equation above). This paper is cited by both \citet[p.\ 624]{Dirac 1927} and \citet[p.\ 832]{Jordan 1927b}. For discussion of Lanzcos's (1926) contribution, see, e.g., \citet[pp.\ 665--668]{Mehra Rechenberg 1987}.\label{kernel}}
\end{quotation}
Eq.\ (13) in Ch.\ 3 of ``Qu.M.\ II," the famous {\it Dreim\"annerarbeit}, is $H(pq) = SH(p^0q^0)S^{-1} = W$ \citep[p.\ 351]{dreimaenner}. This equation summarizes how one solves problems in matrix mechanics: one has to find the matrix $S$ for a ``transformation to principal axes," i.e., a canonical transformation from initial coordinates $(p^0q^0)$ to new coordinates $(pq)$ in which the Hamiltonian $H$ becomes the diagonal matrix $W$. In Ch.\ 3 of the {\it Dreim\"annerarbeit}, i.e., prior to Schr\"odinger's work, Born, who wrote the chapter, had already come close to making the connection Heisenberg makes here between transformation matrices $S$ and solutions $\psi_E(q)$ of the time-independent Schr\"odinger equation.\footnote{For a reconstruction in modern language of the  steps Born was still missing, see \citet[p.\ 355]{Duncan and Janssen 2009}.} 

Expanding on his comments in the letter to Jordan, Heisenberg began his next paper---the sequel to the paper on resonance phenomena \citep{Heisenberg 1926a} that he had used for his note on fluctuations \citep{Heisenberg 1926c}---with a three-page synopsis of the remarkable new formalism that subsumed both wave and matrix mechanics \citep[sec.\ 1, pp.\ 240--242]{Heisenberg 1927a}. This paper was  received by {\it Zeitschrift f\"ur Physik} on December 22, 1926. The new formalism was of such recent vintage at that point that  \citet[ p.\ 240, note]{Heisenberg 1927a} had to list the three main sources he cited for it \citep{Dirac 1927, Jordan 1927b, Heisenberg 1926c} as   ``forthcoming" [{\it im Erscheinen}\,]. He made much of the third item, his own note on fluctuations. 
Jordan,  he wrote, ``has independently found results that are equivalent to those of the Dirac paper {\it and those of a preceding paper by the author}\," \citep[p.\ 240, note; our emphasis]{Heisenberg 1927a}. He referred to the latter again when he introduced the probability interpretation of {\it arbitrary} transformation matrices (ibid., p.\ 242), Dirac's far-reaching generalization of  his own proposal in his note on fluctuations. Heisenberg cited Jordan's paper in addition to his own note but not Dirac's paper. Moreover, he did not mention Born's (1926a,b) statistical interpretation of the wave function anywhere in this brief exposition of the new general formulation of quantum mechanics.
This undoubtedly reflects Heisenberg's initial hostility toward Born's seminal contribution (see note \ref{Teil-Ganze}).\footnote{In his paper on the uncertainty principle a few months later, \citet[p.\ 176, note]{Heisenberg 1927b} did cite \citet{Born 1926b}, but only {\it after} \citet{Einstein 1925}, \citet{dreimaenner}, and \citet{Jordan 1926a}, three papers that have nothing to do with the statistical interpretation of wave functions or transformation matrices. In fairness to Heisenberg, however, we should note that he now also cited an article by \citet{Born 1927c} on quantum mechanics and statistics in {\it Die Naturwissenschaften}.}

In his later reminiscences, Heisenberg did give Born and Dirac their due (cf.\ note \ref{Teil-Ganze}).
Discussing the statistical interpretation of the quantum formalism in his contribution to the Pauli memorial volume, for instance, he mentioned Born, Dirac, and Pauli along with his own modest contribution (see the italicized sentence below). On this occasion, however, he omitted Jordan, perhaps because, for reasons we will discuss below, he much preferred Dirac's version of transformation theory over Jordan's. Heisenberg wrote:
\begin{quotation}
In the summer of 1926, Born developed his theory of collisions and, building on an earlier idea of Bohr, Kramers, and Slater, correctly interpreted the wave [function] in multi-dimensional configuration space as a probability wave.\footnote{Heisenberg repeatedly claimed that Born's statistical interpretation of the wave function owed a debt to the theory of \citet{BKS}, according to which so-called `virtual radiation' determines the probability of atomic transitions. Born himself took exception to this claim (AHQP interview with Heisenberg, session 4, p.\ 2). Yet, in his insightful discussion of Born's contribution, \citet[p.\ 286]{Jammer 1966} follows Heisenberg in identifying BKS as one of its roots. In addition to the quantum collision papers \citep{Born 1926a, Born 1926b, Born 1927a}, \citet[pp.\ 283--290]{Jammer 1966} discusses two papers elaborating on the statistical interpretation of the wave function \citep{Born 1926c, Born 1927b}. In the second of these, \citet[p.\ 356, note]{Born 1927b} emphasized that his theory ``is {\it not} equivalent to that of Bohr, Kramers, Slater."  \citet[p.\ 804]{Born 1926b} did explicitly acknowledge one of the other roots identified by \citet[p.\ 285]{Jammer 1966}: Einstein's proposal to interpret the electromagnetic field as a guiding field
for light quanta. For another discussion of Born's statistical interpretation of the wave function, see \citet[pp.\ 36--55]{Mehra Rechenberg}.}
Pauli thereupon explained to me in a letter that Born's interpretation is only a special case of a much more general interpretation prescription. He pointed out that one could, for instance, interpret $|\psi(p)|^2 dp$ as the probability that the particle has a momentum between $p$ and $p+dp$.\footnote{See Pauli to Heisenberg, October 19, 1926 \citep[Doc.\ 143; the relevant passage of this 12-page letter can be found on pp.\ 347--348]{Pauli 1979}. When listing his sources for the presentation of transformation theory mentioned above, \citet[p.\ 240, note]{Heisenberg 1927a} referred to that same letter: ``Several of these results [of \citet{Dirac 1927}, \citet{Jordan 1927b}, and \citet{Heisenberg 1926c}] had already been communicated earlier and independently to me by Mr. W.\ Pauli." \label{pauli-letter}} {\it This fit well with my own considerations about fluctuation phenomena}. In the fall of 1926, Dirac developed his transformation theory, in which then in all generality the absolute squares of matrix elements of unitary transformation matrices were interpreted as  probabilities \citep[p.\ 44; our emphasis]{Heisenberg 1960}.
\end{quotation}
Born's work was undoubtedly more important for the development of the Dirac-Jordan statistical transformation theory than Heisenberg's. Before acknowledging Heisenberg's note on fluctuations, \citet[p.\ 621]{Dirac 1927}, in fact, had already acknowledged both the preliminary announcement and the full exposition of Born's (1926a,b) theory of quantum collisions, published in July and September of 1926, respectively. These papers contained the statistical interpretation of the wave function for which Born was awarded part of the 1954 Nobel Prize in physics.

Concretely, \citet[p.\ 805]{Born 1926b} suggested that, given a large number of systems in a superposition $\psi(q) =  \sum_n c_n \, \psi_n(q)$ of energy eigenfunctions $\psi_n(q)$, the fraction of systems with eigenfunctions $\psi_n(q)$ is given by the absolute square of the complex expansion coefficients $c_n$.\footnote{We will interpret `fraction of  systems with a particular energy eigenfunction' as `fraction of systems found  to be in a particular energy eigenstate upon measurement of the energy,' even though Born did not explicitly say this. A careful distinction between pure and mixed states  also had yet to be made (see von Neumann, 1927b, which we will discuss in Section 6).} 
In the preliminary version of the paper, \citet[p.\ 865]{Born 1926a} introduced his probability interpretation examining the case of inelastic scattering of an electron by an atom. He wrote the wave function of the electron long after and far away from the point of interaction as a superposition of wave functions for free electrons  flying off in different directions. As in the case of the expansion in terms of energy eigenstates mentioned above, Born interpreted the absolute square of the coefficients in this expansion in terms of free electron wave functions as the probability that the electron flies off in a particular direction. He famously only added in a footnote that this probability is not given by these coefficients themselves but by their absolute square \citep[p.\ 865]{Born 1926a}.

In {\it Neue Begr\"undung} I,  \citet[p.\ 811]{Jordan 1927b} cited Born's second longer paper on quantum collisions and a more recent one in which he elaborated on his statistical interpretation of the wave function \citep{Born 1926b, Born 1926c}. In the latter, \citet[p.\ 174]{Born 1926c} noted, although he did not use that term at this point, that his probability interpretation typically leads to the occurrence of {\it interference terms} (see note \ref{2slits} for a simple example).\footnote{According to \citet[p.\ 174, note]{Born 1926c}, the same result was obtained by \citet{Dirac 1926}. Born was probably referring to the phase averaging occurring in this paper \citep[pp.\ 675--677]{Dirac 1926}. The connection strikes us as rather tenuous. However, \citet[p.\ 290]{Jammer 1966}, citing Born's acknowledgment of Dirac's paper, claims that the result was ``incorporated into the construction of the transformation theory of quantum mechanics." Unfortunately, Jammer does not identify which passage in Dirac's paper he takes Born to be referring to.}$\!^,\!$\footnote{Had Heisenberg not resented Born's use of wave mechanics so strongly, he could not have helped but notice that \citet{Born 1926c} had already proposed to interpret elements of transformation matrices as probabilities before he himself did in his note on fluctuations (Born's paper appeared December 6, 1926, i.e., after  submission but before publication of Heisenberg's note). Born wrote: ``{\it If because of a perturbation, persisting for a finite time $T$, the system is taken from one state to another, then the distribution constants, the squares of which give the states' probabilities} [i.e., the  coefficients $c_n$ in $\psi(q) =  \sum_n c_n \, \psi_n(q)$], {\it undergo an orthogonal transformation}[:] $C_n = \sum_m b_{mn} c_m$, $\sum_k b_{mk} b^*_{nk} = \delta_{nm}$. 
{\it The squares of the coefficients are the transition probabilities}[:] $\Phi_{nm} = |b_{nm}|^2$, $\sum_m \Phi_{nm} = 1$ \citep[p.\ 176; emphasis in the original; quoted by Mehra and Rechenberg, 2000--2001, p. 52]{Born 1926c}.} 
These two papers by \citet{Born 1926b, Born 1926c} are mentioned much more prominently in {\it Neue Begr\"undung} I than Heisenberg's (1926b) note on fluctuations, which is cited only as providing an example of a special case in which there are no interference terms  \citep[p.\ 812, note]{Jordan 1927b}.

In a two-part overview of recent developments in quantum theory that appeared in {\it Die Naturwissenschaften}  in July and August 1927, \citet[Pt.\ 2, p.\ 645]{Jordan 1927i} accorded Heisenberg's note a more prominent role, recognizing it, along with one of his own papers, as an important step toward statistical transformation theory. Jordan's (1927a) paper was submitted about three weeks after Heisenberg's (1926c) but was independent of it.\footnote{Jordan's paper was received by {\it Zeitschrift f\"ur Physik} on November 25, 1926, and published January 2, 1927; Heisenberg's, as we already noted, was received November 6 and published December 20, 1926. In the letter from which we already quoted several passages above, Heisenberg praised Jordan's paper: ``Your resonance example is very pretty" (Heisenberg to Jordan, November 24, 1926, AHQP).} The two papers are remarkably similar. Both are concerned with the reconciliation of two descriptions of the energy exchange between two quantum systems, a continuous description in terms of a mechanism similar to beats between two waves and a description in terms of quantum jumps (cf.\ note \ref{fluctuations}). Both argued, against Schr\"odinger,\footnote{Like Heisenberg, Jordan profoundly disliked Schr\"odinger's interpretation of the wave function while appreciating wave mechanics as providing a reservoir of mathematical tools in the service of matrix mechanics. \citet{Jordan 1927g} bluntly stated this opinion in his review of  a volume collecting Schr\"odinger's papers on wave mechanics (Schr\"odinger, 1927; for a long quotation from Jordan's review, see Mehra and Rechenberg, 2000--2001, pp.\ 59--60). Although Jordan did not mention Born's interpretation of the wave function in his review (he only criticized Schr\"odinger's), he did not share Heisenberg's initial resistance to Born's work. In his habilitation lecture, for instance, \citet[p.\ 107]{Jordan 1927d} praised the ``very clear and impressive way" in which Born had introduced this interpretation.} 
that despite appearances to the contrary quantum jumps are unavoidable. As \citet[Pt.\ 2, pp.\ 645--646]{Jordan 1927i} concluded in his {\it Naturwissenschaften} article, the probabilistic nature of the laws of quantum mechanics is key to  reconciling the continuous and discontinuous descriptions. In the letter Heisenberg mentioned above (see note \ref{pauli-letter}), Pauli had drawn the same conclusion from his analysis of an ingenious example of his own device that he told Heisenberg was ``a pure culture of your favorite resonance phenomenon" 
\citep[p.\ 344]{Pauli 1979}. Instead of two quantum systems exchanging energy, Pauli considered one quantum system, a particle constrained to move on a closed ring, which periodically encounters a small obstacle. He considered the case in which the particle is in a state in which it should alternate between rotating clockwise and rotating counterclockwise. That was only possible, Pauli explained, if we accept the conclusion of Born's analysis of quantum collisions that there is a definite probability that the system reverses course upon hitting the obstacle, which  classically would not be large enough to change the direction of the system's rotation. Unlike Pauli in this letter  or \citet{Heisenberg 1926c} in his note on fluctuations, \citet{Jordan 1927a} did not elaborate on exactly {\it how} the statistical element in his example of a resonance phenomenon should formally be introduced, via wave functions {\it \`a la} Born or via transformation matrices {\it \`a la} Heisenberg. In the {\it Naturwissenschaften} article,  he did not resolve this issue either. Instead, \citet[Pt.\ 2, p.\ 646]{Jordan 1927i} wrote in the paragraph immediately following his discussion of resonance phenomena that the statistical nature of the quantum laws ``manifests itself in many ways even more impressively and intuitively" in Born's (1926b) analysis of quantum collisions. This strongly suggests that Jordan's statistical interpretation of quantum mechanics in {\it Neue Begr\"undung} I owed more to Born's statistical interpretation of wave functions than to Heisenberg's statistical  interpretation of transformation matrices.

Jordan was particularly taken with Pauli's further development of Born's ideas even though he probably did not see the letter with Pauli's probability interpretation of $|\psi(p)|^2 dp$ mentioned above (see note \ref{pauli-letter}).
As \citet[p.\ 544]{Rechenberg 2010} points out, citing a letter from Heisenberg to Pauli of October 28, 1926 \citep[p.\ 349]{Pauli 1979}, Pauli's letter made the rounds in Copenhagen and Dirac, for instance, certainly read it, but Jordan was in G\"ottingen at the time and would not have had access to it. Instead, \citet[p.\ 811, note]{Jordan 1927b} referred to a forthcoming paper by Pauli on gas degeneracy. In that paper, \citet[p.\ 83,  note]{Pauli 1927a} only suggested that $|\psi(q)|^2 dq$ (with $q \equiv (q_1, \ldots, q_f)$, where $f$ is the number of degrees of freedom of the system) be interpreted as the probability of finding the system at a position somewhere in the region $(q, q + dq)$. As Jordan emphasized in the {\it Naturwissenschaften} article discussed above, even this suggestion is subtly different from what Born had proposed.\footnote{Discussing ``the statistical interpretations of the Schr\"odinger function given by Born and Pauli,"  \citet[Pt.\ 2, p.\ 647]{Jordan 1927i} added parenthetically: ``(it should be emphasized that Born's hypothesis and Pauli's hypothesis, though related, are at first independent of one another)." Unfortunately, Jordan did not explain what he meant by the qualification ``at first [{\it zun\"achst}\,]" in this clause.}  \citet{Born 1926a, Born 1926b} considered probabilities such as the probability that a system has a particular energy or the probability that an electron scatters in a particular direction (or, equivalently, has a certain momentum). Those probabilities are given by the absolute squares of  coefficients in some expansion of the wave function (in configuration space), 
either as a superposition of energy eigenstates or, in the case of collisions, as a superposition (asymptotically) of wave functions for free particles with specified outgoing momenta.
Pauli seems to have been the first to ask about the probability that a system has a particular position. That probability is given by the absolute square of the wave function itself rather than by the absolute square of the coefficients in some expansion of it.\footnote{Although Born did not give the probability interpretation of $|\psi(q)|^2 dq$ in his papers on collision theory, he did state a completeness relation highly suggestive of it: $\int |\psi(q)|^2 dq = \sum |c_n|^2$ \citep[p.\ 805]{Born 1926b}.}  The generalization in Pauli's letter to Heisenberg from wave functions in configuration space to wave functions in momentum space is not mentioned in  {\it Neue Begr\"undung} I. It is certainly possible, however, that Pauli had shared this generalization with Jordan, as the two of them regularly saw each other during this period and even went on vacation together.\footnote{AHQP interview with Jordan, session 3, p.\ 15, cited by \citet[p.\ 66]{Mehra  Rechenberg} and by \citet[p.\ 359]{Duncan and Janssen 2009}.} As a result, it is impossible to reconstruct what exactly Jordan got from Pauli, whose name occurs no less than seven times in the first three pages of {\it Neue Begr\"undung} I.

Jordan wanted to use what he had learned from Pauli to provide a new unified foundation for the laws of quantum mechanics by showing that they can be derived as ``consequences of a few simple statistical assumptions" \citep[pp.\ 810--811]{Jordan 1927b}. The central quantities in Jordan's formalism are what he called ``probability amplitudes.'' This echoes Born's (1926b, p.\ 804) term ``probability waves" for Schr\"odinger wave functions but is an extremely broad generalization of Born's concept. Moreover, \citet[p.\ 811]{Jordan 1927b} credited Pauli rather than Born with suggesting the term. Jordan defined a complex probability amplitude, $\varphi(a,b)$, for two arbitrary quantum-mechanical quantities $\hat{a}$ and $\hat{b}$ with fully continuous spectra.\footnote{This is the first of several instances where we will enhance Jordan's own notation. In {\it Neue Begr\"undung} I, Jordan used different letters for quantities and their values. We will almost always use the same letter for a quantity and its values and use a hat to distinguish the former from the latter. The main exception will be the Hamiltonian $\hat{H}$ and the energy eigenvalues $E$. As we noted above, \citet{Dirac 1927} used primes to distinguish the value of a quantity from the quantity itself.\label{notation1}} 
At this point, he clearly labored under the illusion that it would be relatively straightforward to generalize his formalism to cover quantities with wholly or partly discrete spectra as well. In {\it Neue Begr\"undung} II, \citet{Jordan 1927f} would  discover that such a generalization is highly problematic. Unaware of these complications and following Pauli's lead, Jordan interpreted $|\varphi(a,b)|^2 da$
as the conditional probability ${\rm Pr}(a|b)$ for finding a value between $a$ and $a + da$ for $\hat{a}$ given that the system under consideration has been found to have the value $b$ for the quantity $\hat{b}$.\footnote{The notation ${\rm Pr}(a|b)$ is ours and is not used in the sources we discuss. At this point, we suppress another wrinkle that will loom large in our paper, the so-called `supplementary amplitude' ({\it Erg\"anzungsamplitude}; see, in particular, Section 2.4).\label{notation2}} 

Eigenfunctions $\psi_E(q)$ with eigenvalues $E$ for some Hamiltonian of a one-dimensional system in configuration space are examples of Jordan's probability amplitudes $\varphi(a,b)$. The quantities $\hat{a}$ and $\hat{b}$ in this case are the position $\hat{q}$ and the Hamiltonian $\hat{H}$, respectively. Hence $|\psi_E(q)|^2 dq = |\varphi(q, E)|^2 dq$ gives the conditional probability ${\rm Pr}(q|E)$ that $\hat{q}$ has a value between $q$ and $q +dq$ given that $\hat{H}$ has the value $E$. This is the special case of Jordan's interpretation given in Pauli's paper on gas degeneracy (in $f$ dimensions). 

As indicated above, Jordan took an axiomatic approach in his {\it Neue Begr\"un\-dung} papers. Here we clearly see the influence of the mathematical tradition in G\"ottingen \citep{Lacki 2000}. In fact, Jordan had been Richard Courant's assistant before becoming Born's. In his {\it Neue Begr\"un\-dung} papers, Jordan began with a series of postulates about his probability amplitudes and the rules they ought to obey (the formulation and even the number of these postulates varied) and then developed a formalism realizing these postulates.  

A clear description of the task at hand can be found in a paper by David Hilbert, Lothar Nordheim, and the other main protagonist of our story, John von Neumann, who had come to G\"ottingen in the fall of 1926 on a fellowship of the International Education Board \citep[pp.\ 401--402]{Mehra Rechenberg}. Born in 1903, von Neumann was a year younger than Dirac and Jordan, who, in turn, were a year younger than Heisenberg and two years younger than Pauli. The paper by Hilbert, von Neumann, and Nordheim grew out of Hilbert's course on quantum mechanics in 1926/1927 for which Nordheim prepared most of the notes.\footnote{See p.\ 13 of the transcript of the interview with Nordheim for the Archive of the History of Quantum Physics (AHQP) \citep[p.\ 361]{Duncan and Janssen 2009}.} The course concluded with an exposition of {\it Neue Begr\"undung} I \citep[pp.\ 698--706]{Sauer and Majer 2009}. The notes for this part of the course formed the basis for a paper, which was submitted in April 1927 but not published, for whatever reason, until the beginning of 1928.\footnote{\citet[pp.\ 310--312]{Jammer 1966} talks about the ``Hilbert-Neumann-Nordheim transformation theory" as if it were a new version of the theory, going beyond ``its predecessors, the theories of Dirac and Jordan" (ibid., p.\ 312). However, what Jammer sees as the new element, the identification of probability amplitudes with the kernels of certain integral operators (ibid., p.\ 311), is part and parcel of Jordan's papers (see note \ref{kernel} as well as Sections 2 and 4).\label{jammer blooper}} In the introduction, the authors described the strategy for formulating the theory:
\begin{quotation}
One imposes certain physical requirements on these probabilities, which are suggested by earlier experience and  developments, and the satisfaction of which calls for certain relations between the probabilities. Then, secondly, one searches for a simple analytical apparatus in which quantities occur that satisfy these relations exactly \citep[p.\ 2--3; cf. Lacki, 2000, p.\ 296]{Hilbert-von Neumann-Nordheim}.
\end{quotation}
After everything that has been said so far, it will not come as a surprise that the quantities satisfying the relations that Jordan postulated for his probability amplitudes are {\it essentially}\footnote{The qualification is related to the complication mentioned in note \ref{notation2} and will be explained at the end of this subsection.} the transformation matrices central to Dirac's (1927) presentation of the statistical transformation theory. An example, based on lecture notes by Dirac,\footnote{AHQP, ``Notes for Dirac's first lecture course on quantum mechanics, 115 pp. [Oct 1927\,?]" \citep[p.\ 32]{Kuhn et al. 1967}. As \citet[p.\ 344]{Darrigol 1992} notes, ``Dirac did try [on pp.\ 6--9  of these lecture notes] to lay out his competitor's theory, but he quickly returned to his own, which he found simpler and more elegant."\label{Dirac notes}} will give a rough illustration of how this works.

One of the features that Jordan saw as characteristic of quantum mechanics and that he therefore included among his postulates was that in quantum mechanics the simple addition and multiplication rules of ordinary probability theory for mutually exclusive and independent outcomes, respectively, apply to probability {\it amplitudes} rather than to the probabilities themselves. \citet[p.\ 812]{Jordan 1927b} used the phrase  ``interference of probabilities" for this feature. He once again credited Pauli with the name for this phenomenon, even though \citet[p.\ 804]{Born 1926b} had already talked about the ``interference of \ldots\ ``probability waves"\," in his paper on quantum collisions. As we saw above, \citet[p.\ 174]{Born 1926c}  had discussed the feature itself in a subsequent paper, albeit only in  the special case of Schr\"odinger wave functions. \citet{Jordan 1927b} was the first, at least in print, who explicitly recognized this feature in full generality. It is implicit in Dirac's (1927) version of transformation theory,  but Dirac may well have shared the skepticism about the interference of probabilities that Heisenberg expressed in a letter to Jordan (we will quote and discuss the relevant passage below).

As we will see in Section 2.1, Jordan's postulate about the addition and multiplication of probability amplitudes basically boils down to the requirement that  the amplitudes $\varphi(a,c)$, $\psi(a,b)$ and $\chi(b,c)$ for the quantities $\hat{a}$, $\hat{b}$, and $\hat{c}$ with purely continuous spectra satisfy the relation $\varphi(a,c) = \int  db \, \psi(a,b) \, \chi(b,c)$. It is easy to see intuitively, though much harder to prove rigorously (see Section 1.2), that this relation is indeed satisfied if these three amplitudes are equated with the transformation matrices---in the notation of \citet{Dirac 1927} explained above---$(a/c)$, $(a/b)$, and $(b/c)$, respectively. These matrices relate wave functions in $a$-, $b$-, and $c$-space to one another. From $\psi(a) = \int db \, (a/b) \, \psi(b)$ and $\psi(b) = \int dc \, (b/c) \, \psi(c)$, it follows that
$$
\psi(a) = \int \!\!\! \int db \, dc \, (a/b) \, (b/c) \, \psi(c).
$$ 
Comparing this expression to $\psi(a) = \int dc \, (a/c) \, \psi(c)$, we see that $(a/c) = \int db \, (a/b) \, (b/c)$, in accordance with Jordan's postulates.\footnote{Dirac lecture notes, p.\ 7 (cf.\ note \ref{Dirac notes}). Dirac used $\alpha$, $\beta$, and $\gamma$ instead of $a$, $b$, and $c$, and considered a purely discrete rather than a purely continuous case, thus using sums instead of integrals. Dirac actually used the notation $(.|.)$ in his handwritten letters and manuscripts during this period, but apparently did not mind that this was rendered as $(./.)$ in print.}

This brings us to an important difference between Jordan's and Dirac's versions of their statistical transformation theory. For Dirac, the transformation element was primary, for Jordan the statistical element was. Most of Dirac's (1927) paper is devoted to the development of the formalism that allowed him to represent the laws of quantum mechanics in different yet equivalent ways  (secs.\ 2--5, pp.\ 624--637). The probability interpretation of the transformation matrices is then grafted onto this formalism in the last two sections (ibid., secs.\ 6--7, 637--641). Jordan's (1927b) paper begins with the axioms about probability (Pt.\ I, secs.\ 1--2, pp.\ 809--816). It is then shown that those can be implemented by equating probability amplitudes with transformation matrices, or, to be more precise, with integral kernels of canonical transformations (Pt.\ 2, secs.\ 1--6, pp.\ 816--835). 

Heisenberg strongly preferred Dirac's version of statistical transformation theory over Jordan's. For one thing, he disliked Jordan's axiomatic approach. As he told Kuhn in his interview for the AHQP project:
\begin{quotation}
Jordan used this transformation theory for deriving what he called the axiomatics of quantum theory \ldots\ This I disliked intensely \ldots\
Dirac kept within the spirit of quantum theory while Jordan, together with Born, went into the spirit of the mathematicians (AHQP interview with Heisenberg, session 11, pp.\ 7--8; quoted by Duncan and Janssen, 2009, pp.\ 360--361).
\end{quotation}
Presumably, the axiomatic approach in and of itself would not have presented too much of an obstacle for Heisenberg.  As Paul Ehrenfest told Jordan, who still remembered it with amusement when interviewed decades later: ``Since you wrote the paper axiomatically, that only means that one has to read it back to front."\footnote{AHQP interview with Jordan, session 3, p. 17; quoted by \citet[p.\ 69]{Mehra Rechenberg} and \citet[p.\ 360]{Duncan and Janssen 2009}.} In that case, one would encounter probability amplitudes in the guise of transformation matrices first, as in Dirac's version of the theory. 

Heisenberg had more serious reservations about Jordan's version of theory, which initially made it difficult for him even to understand {\it Neue Begr\"undung} I. He expressed his frustration in a letter to Pauli a few weeks after the paper was published. After praising Jordan's (1927d) habilitation lecture which had just appeared in {\it Die Naturwissenschaften}, Heisenberg wrote: 
\begin{quotation}
I could not understand Jordan's big paper in [{\it Zeitschrift f\"ur Physik}]. The ``postulates" are so intangible and undefined, I cannot make heads or tails of them (Heisenberg to Pauli, February 5, 1927; Pauli, 1979, p.\ 374).\footnote{Even Dirac's (1927) paper, which for modern readers is much more accessible than Jordan's {\it Neue Begr\"undung} I, was difficult to understand for many of his contemporaries. In a letter of June 16, 1927, Ehrenfest told Dirac that he and his colleagues in Leyden had had a hard time with several of his papers, including the one on transformation theory \citep[p.\ 46]{Kragh 1990}.} 
\end{quotation}

About a month later, Heisenberg wrote to Jordan himself, telling him that he was working
\begin{quotation}
on a fat paper [Heisenberg, 1927b, on the uncertainty princple]
that one might characterize as physical commentary on your paper and Dirac's. You should not hold it against me that I consider this necessary. The essence from a mathematical point of view is roughly that it is possible with your mathematics to give an exact formulation of the case in which $p$ and $q$ are {\it both} given with a certain {\it accuracy}  (Heisenberg to Jordan, March 7, 1927, AHQP; emphasis in the original).\footnote{Heisenberg's (1927b) paper was received by {\it Zeitschrift f\"ur Physik} on March 23, 1927. 
Shortly before he submitted the paper, he wrote to Jordan for help with a sign error in a derivation in {\it Neue Begr\"undung} I that he needed for his paper (see note \ref{F-verschwommen}). 
For discussion of Heisenberg's reliance on transformation theory for his uncertainty paper, see \citet[pp.\ 326--328]{Jammer 1966}, \citet[pp.\ 148--151, and pp.\ 159--161]{Mehra Rechenberg}, and Beller (1985, 1999, pp.\ 91--95).\label{beller}} 
\end{quotation}
Heisenberg now had a clearer picture of Jordan's approach and of how it differed from the approach of Dirac, who had meanwhile left Copenhagen and had joined Born and Jordan in G\"ottingen. After registering some disagreements with Dirac, Heisenberg turned to his disagreements with Jordan:
\begin{quotation}
With you I don't quite agree in that, in my opinion, the relation $\int \varphi(x,y)$\, $\psi(y,z) \, dz$
has nothing to do with the laws of probability. In all cases in which one can talk about probabilities the usual addition and multiplication of probabilities is valid, {\it without} ``interference." {\it With} Dirac I believe that it is more accurate to say: all statistics is brought in only through our experiments (Heisenberg to Jordan, March 7, 1927, AHQP; emphasis in the original).
\end{quotation}
The reservation expressed in the first two sentences
is largely a matter of semantics. Heisenberg did not dispute that the relation given by Jordan, which we wrote as $\varphi(a,c) = \int  db \, \psi(a,b) \, \chi(b,c)$ above, holds in quantum mechanics (as we will see in Section 1.2, it corresponds to the familiar completeness and orthogonality relations). Heisenberg also did not dispute that this relation describes interference phenomena.\footnote{A simple example, the familiar two-slit experiment, immediately makes that clear. Let the quantity $\hat{c}$ (value $c$) be the position where photons are emitted; let $\hat{b}$ (values $b_1$ and $b_2$) be the positions of the two slits; and let $\hat{a}$ (value $a$) be the position where photons are detected. On the assumption (vindicated by modern theory) that the relation also applies to quantities with discrete values (such as $\hat{b}$ in this example), the integral on its right-hand side reduces to a sum of two terms in this case:  $\varphi(a,c) = \psi(a,b_1) \, \chi(b_1,c) \, + \, \psi(a,b_2) \, \chi(b_2,c)$. Multiplying the left- and the right-hand side by their complex conjugates, we find that the probability ${\rm Pr}(a|c)$ is equal to the sum of the probabilities ${\rm Pr}(a \, \& \, b_1 |c)$ and ${\rm Pr}(a \, \& \, b_2 |c)$  {\it plus} the interference terms $\psi(a,b_1) \, \psi^*(a,b_2) \, \chi(b_1,c) \, \chi^*(b_2,c)$ and $\psi(a,b_2) \, \psi^*(a,b_1) \, \chi(b_2,c)  \, \chi^*(b_1,c)$.\label{2slits}}
What Heisenberg objected to was Jordan's way of looking upon this relation as a consequence of his addition and multiplication rules for probability amplitudes. This did not materially affect Jordan's theory as \citet[p.\ 813]{Jordan 1927b} only ever used those rules to derive this particular relation.\footnote{Heisenberg elaborated on his criticism of Jordan on this score in his uncertainty paper \citep[pp.\ 183--184, p.\ 196]{Heisenberg 1927b}. Von Neumann (1927a, p.\ 46; cf.\ note \ref{HvJ2})
initially followed Jordan (see also Hilbert, von Neumann, and Nordheim, 1928, p.\  5; cf.\ note \ref{HvJ1}), but he changed his mind after reading Heisenberg's uncertainty paper \citep[p.\ 246; cf.\ note \ref{HvJ3}]{von Neumann 1927b}.\label{HvJ0}}

The reference to Dirac for Heisenberg's second reservation is to the concluding sentence of Dirac's paper on transformation theory:
\begin{quotation}
The notion of probabilities does not enter into the ultimate description of mechanical processes: only when one is given some information that involves a probability \dots\ can one deduce results that involve probabilities \citep[p.\ 641]{Dirac 1927}.\footnote{Quoted and discussed by \citet[p.\ 42]{Kragh 1990}. Dirac's position, in turn, was  undoubtedly influenced by the exchange between Pauli and Heisenberg that led to the latter's uncertainty paper. Recall, for instance, that Dirac read the long letter from Pauli to Heisenberg of October 19, 1926 \citep[quoted and discussed by Mehra and Rechenberg, 2000--2001, pp.\ 145--147]{Pauli 1979}.}
\end{quotation}
Instead of elaborating on this second reservation, Heisenberg told Jordan that he had just written a 14-page letter about these matters to Pauli and suggested that Jordan have Pauli send him this letter for further details. In this letter, the blueprint for his uncertainty paper, Heisenberg told Pauli:
\begin{quotation}
One can, like Jordan, say that the laws of nature are statistical. But one can also, and that to me seems considerably more profound, say with Dirac that all statistics are brought in only through our experiments. That we do not know at which position the electron will be the moment of our experiment, is, in a manner of speaking, only because we do not know the phases, if we do know the energy \ldots\ and in {\it this} respect the classical theory would be no different. That we {\it cannot} come to know the phases without \ldots\ destroying the atom is characteristic of quantum mechanics (Heisenberg to Pauli, February 23, 1927; Pauli, 1979, Doc.\ 154, p.\ 377; emphasis in the original; see also Heisenberg, 1927b, p.\ 177).
\end{quotation}
So Heisenberg---and with him Dirac---held on to the idea that nature itself is deterministic and that all the indeterminism that quantum mechanics tells us we will encounter is the result of unavoidable disturbances of nature in our experiments. Jordan, by contrast, wanted at least to keep open the possibility that nature is intrinsically indeterministic (as did Born [1926a, p.\ 866; 1926b, pp.\ 826--827]). \citet{Jordan 1927d} made this clear in his habilitation lecture.\footnote{This lecture, published in {\it Die Naturwissenchaften} and, in a translation prepared by Oppenheimer, in {\it Nature}, shows that Jordan's position was more nuanced than Heisenberg made it out to be. First, Jordan clearly recognized that the laws governing his probability amplitudes are fully deterministic \citep[ pp.\ 107--108 in the German original; pp.\ 567--568 in the English translation]{Jordan 1927d}. Second, whether nature is indeterministic remained an open question for him. As an example he considered the time of a single quantum jump, which, he wrote, ``under certain circumstances is a measurable quantity. What predictions can our theory make on this point? The most obvious answer is that the theory only gives averages, and can tell us, on the average, how many quantum jumps will occur in any interval of time. Thus, we must conclude, the theory gives the probability that a jump will occur at a given moment; and thus, so we might be led to conclude, the exact moment is indeterminate, and all we have is a probability for the jump. But this last conclusion does not necessarily follow from the preceding one: it is an additional hypothesis" (ibid., p.\ 109 [German]; p.\ 569 [English]). In general, he reiterated in the conclusion of his habilitation lecture, ``[t]he circumstance that the quantum laws are laws of averages, and can only be applied statistically to specific elementary processes, is not a conclusive proof that the elementary laws themselves can only be put in terms of probability \ldots\ Probably we shall find that an incomplete determinism, a certain element of pure chance, is intrinsic in these elementary physical laws. But, as I have said, a trustworthy decision will only be possible after a further analysis of quantum mechanics" (ibid., p.\ 110 [German]; p.\ 569 [English]).\label{jordan indeterminism}} In his {\it Neue Begr\"undung} papers, Jordan did not discuss the nature of the probabilities he introduced. He did not even properly define these probabilities. This was done only by \citet{von Neumann 1927b} with the help of the notion of ensembles of systems from which one randomly selects members, a notion developed by Richard von Mises and published in book form the following year \citep[see note \ref{von mises}]{von Mises 1928}.

As far as the issue of determinism versus indeterminism is concerned, Dirac thus stayed closer to classical theory than Jordan. In other respects, however, Jordan stayed closer. Most importantly, Jordan's use of canonical transformations is much closer to their use in classical mechanics than Dirac's.\footnote{\citet[p.\ 343]{Darrigol 1992} also recognizes this but, pointing to Jordan's unusual definition of canonically conjugate quantities (see below), concludes that ``Jordan departed much more from the classical model than did Dirac." This may be true, as we will see, for where Jordan ended up with {\it Neue Begr\"undung} II, but he started out following the classical model much more closely than Dirac.} In the letter to Jordan from which we quoted right at the beginning of our paper, Dirac clearly identified part of the difference in their use of of canonical transformations:
\begin{quotation}
In your work I believe you considered transformations from one set of dynamical variables to another, instead of a transformation from one scheme of matrices representing the dynamical variables to another scheme representing the same dynamical variables, which is the point of view adopted throughout my paper. The mathematics appears to be the same in the two cases, however (Dirac to Jordan, December 24, 1916, AHQP).\footnote{This is the one paragraph of this letter not quoted by \citet[cf.\ note \ref{MR1}]{Mehra Rechenberg}.\label{MR2}}
\end{quotation}
Traditionally, canonical transformations had been used the way Jordan used them, as transformations to new variables, and not the way Dirac used them, as transformations to new representations of the same variables. Canonical transformations had been central to the development of matrix mechanics (Born and Jordan, 1925; Born, Heisenberg, and Jordan, 1926; see Section 2.2 below).
Prior to {\it Neue Begr\"undung} I, \citet{Jordan 1926b, Jordan 1926c} had actually published two important papers on the implementation of canonical transformations in matrix mechanics \citep{Lacki 2004, Duncan and Janssen 2009}. 
Jordan's use of canonical transformations in his {\it Neue Begr\"undung} papers was twofold. First, as already mentioned above, Jordan tried to show that integral kernels in canonical transformations have all the properties that probability amplitudes must satisfy according to his postulates. Second, he tried to use canonical transformations to derive differential equations for probability amplitudes for arbitrary quantities (such as the time-independent Schr\"odinger equation for $\varphi(q, E) \equiv \psi_E(q)$) from the trivial differential equations satisfied by the probability amplitude for $\rho(p,q)$ for some generalized coordinate $\hat{q}$  and its conjugate momentum $\hat{p}$ that he started from. The `mind your $p$'s and $q$'s' part of the title of our paper refers to the crucial role of canonical transformations and conjugate variables in Jordan's formalism. 

Both ways in which  Jordan relied on canonical transformations  in his {\it Neue Begr\"undung} papers turned out to be problematic and resulted in serious mathematical problems for his version of statistical transformation theory. These problems do not affect Dirac's version since \citet{Dirac 1927} only relied on canonical transformations in very loose sense. However, as we will see once we have introduced some modern tools to analyze the Dirac-Jordan theory in Section 1.2, the two versions of the theory do share a number of other mathematical problems. The `never mind your $p$'s and $q$'s' part of our title thus does not refer to Dirac's version of statistical transformation theory, but to the fundamentally different Hilbert space formalism that \citet{von Neumann 1927a} introduced as an alternative to the Dirac-Jordan theory (Section 1.3). 

To conclude this subsection, we briefly indicate how Jordan ran into problems with his twofold use of canonical transformations in {\it Neue Begr\"undung} and explain why those problems do not plague Dirac's version of the theory. We return to these problems in Section 1.2, relying more heavily on modern concepts and notation, and analyze them in greater detail in Sections 2 and 4, again helping ourselves to  modern tools. As \citet[p.\ 44]{Heisenberg 1960} wrote in the passage from his contribution to the Pauli memorial volume quoted above, Dirac only considered {\it unitary} transformations in his transformation theory. Unitary transformations of Hermitian operators preserve their Hermiticity. Unfortunately, there are many canonical transformations that are not unitary \citep[p.\ 356]{Duncan and Janssen 2009}. In {\it Neue Begr\"undung} I, Jordan therefore tried to develop a formalism that would allow quantities that, from a modern point of view, correspond to {\it non}-Hermitian operators. As long as the quantities $\hat{a}$ and $\hat{b}$ correspond to Hermitian operators, the probability amplitude $\varphi(a,b)$ is the integral kernel of a unitary transformation and the conditional probability ${\rm Pr}(a|b)$ is equal to $|\varphi(a,b)|^2 da$. In such cases, the integral kernel $\varphi(a,b)$ is simply identical to Dirac's unitary transformation matrix $(a/b)$. As soon as $\hat{a}$ and/or $\hat{b}$ are non-Hermitian, however, $\varphi(a,b)$ is the integral kernel of a non-unitary transformation and the expression for ${\rm Pr}(a|b)$ becomes more complicated, involving both the probability amplitude itself and what Jordan called a ``supplementary amplitude" ({\it Erg\"anzungsamplitude}). Following the lead of \citet{Hilbert-von Neumann-Nordheim}, Jordan dropped the {\it Erg\"anzungsamplitude} in {\it Neue Begr\"undung} II. The price he paid for this was what, from the point of view of classical mechanics, amounted to the rather arbitrary restriction of the canonical transformations allowed in his theory to unitary ones.

The other main problem that Jordan ran into with his use of canonical transformations is directly related to his axiomatic approach. Unlike Dirac, Jordan wanted to derive the entire theory from his statistical postulates. This meant, for instance, that Jordan did not include the canonical commutation relation, $[\hat{p},\hat{q}] \equiv \hat{p} \hat{q} - \hat{q} \hat{p} = h/2\pi i$ (where $h$ is Planck's constant), among his postulates. Nor did he assume the usual association of the momentum $\hat{p}$ conjugate to $\hat{q}$ with the differential operator $(\hbar/i) \partial/\partial q$ (with $\hbar \equiv h/2 \pi$) acting on wave functions in $q$-space. Dirac assumed both these elements. Instead of the commutation relation for $\hat{p}$ and $\hat{q}$, \citet[p.\ 814]{Jordan 1927b} assumed that the probability amplitude $\rho(p,q)$ for these two quantities had a particularly simple form---in our notation: $\rho(p,q) = e^{-ipq/\hbar}$. This then is how Planck's constant enters Jordan's formalism. Normally, the canonical commutation relation $[\hat{p},\hat{q}] = \hbar/i$ defines what it means for $\hat{p}$ and $\hat{q}$ to be canonically conjugate variables. Jordan, however, defined $\hat{p}$ and $\hat{q}$ to be canonically conjugate if and only if $\rho(p,q) = e^{-ipq/\hbar}$.\footnote{When Kuhn asked Jordan in his interview for the AHQP project why he had chosen this definition, Jordan said: ``I wanted to put the probability considerations first and thought that the multiplication, the operators, would follow from these. If one posits that $p$ and $q$ are two quantities for which there is this particular probability [amplitude] then one can draw conclusions from that and if one defines multiplication afterwards, then one gets the commutation rules. The goal was to not make the commutation rules the starting point but to obtain them as the result of something else. The basic rules of the formulation should be conceived of as statements about probability amplitudes" (session 3, pp.\ 16--17).\label{jordan interview}} This probability amplitude tells us that ``[f]or a given value of $\hat{q}$ all possible values of $\hat{p}$ are {\it equally probable}" \citep[ibid, p.\ 814; emphasis in the original; hats added]{Jordan 1927b}.\footnote{Jordan's habilitation lecture published about two months before Heisenberg's (1927b) paper (see note \ref{beller}), contains a statement that is even more suggestive of the uncertainty principle: ``With different experimental setups one can observe different coordinates. However, with a particular setup one can, at best, observe particular coordinates of an atom exactly, while it will then be impossible in that setup to observe the corresponding momenta exactly" \citep[p.\ 108, note 1, in the German version; this note was omitted in the English translation]{Jordan 1927d}.\label{jordan uncertainty}} The amplitude $\rho(p,q)$ trivially satisfies the equations $(p + (\hbar/i)\, \partial/\partial q)\,\rho = 0$ and $((\hbar/i) \, \partial/\partial p + q)\,\rho = 0$ (ibid.). Subjecting these basic equations to canonical transformations, Jordan argued for the usual association of quantum-mechanical quantities with differential operators acting on wave functions (e.g., $\hat{p}$ with $(\hbar/i) \partial/\partial q$). Once those associations have been made, his assumption about the form of $\rho(p,q)$ entails the usual commutation relation for $\hat{p}$ and $\hat{q}$. Performing more canonical transformation on the basic equations for $\rho(p,q)$, Jordan also tried to derive differential equations for probability amplitudes involving quantities related to the initial $\hat{p}$ and $\hat{q}$ via canonical transformations. In this manner, for instance, he tried to recover the time-independent Schr\"odinger equation. 

It was not until {\it Neue Begr\"undung} II, written in Copenhagen on an International Education Board fellowship and received by {\it Zeitschrift f\"ur Physik} on June 3, 1927, that Jordan realized that this strategy for deriving Schr\"odinger-type differential equations for his probability amplitudes was severely limited. He came to this realization as he was trying to extend his formalism, which in {\it Neue Begr\"undung} I was restricted to quantities with continuous spectra, to cover quantities with partly or wholly discrete spectra as well. The key problem (as we will show using modern tools at the end of Section 2.3), is that quantities related by a canonical transformation have the same spectrum. It follows that Jordan's procedure to get from the differential equations for probability amplitudes for one pair of quantities to those for another pair fails as soon as one pair has purely continuous spectra (such as  $\hat{p}$ and $\hat{q}$) and the other pair has partly or wholly discrete spectra (such as, typically, the Hamiltonian). In response to this problem, Jordan changed the way he used canonical transformations to the way Dirac had been using them all along, where the transformations are no longer to new quantities but rather to new representations of the same quantities \citep[pp.\ 16--17; we will quote the relevant passage in Section 4]{Jordan 1927f}. 

The treatment of quantities with partly or wholly discrete spectra in {\it Neue Begr\"undung} II necessitated further departures from the classical formalism of canonical transformations and conjugate variables. For quantities with fully continuous spectra, Jordan could show that his definition of conjugate variables in terms of a probability amplitude reduces to the standard definition in terms of a commutation relation. This is not true for quantities with partly or wholly discrete spectra. In {\it Neue Begr\"undung} II, Jordan gave a simple proof that such quantities can never satisfy the standard commutation relation $[\hat{p},\hat{q}]  = \hbar/i$ (see Section 4, Eqs.\ \eqref{NB2-2}--\eqref{reductio}). Jordan presented it as a point in favor of his formalism that his alternative definition of canonically conjugate variables  works for quantities with partly or wholly discrete spectra as well. Jordan's definition, however, led to such counter-intuitive propositions as different components of spin qualifying as canonically conjugate to one another.\footnote{In fact, Jordan's initial answer to Kuhn's question why he had adopted a new definition of canonically conjugate variables (see note \ref{jordan interview}) was that it would be applicable to spin as well. Only when Kuhn reminded him that this application must have come later, did Jordan give the answer we quoted in note \ref{jordan interview}. Later in the interview, Kuhn explicitly asked him how he felt about the idea of taking two components of spin to be conjugate to one another. ``Yes, wasn't that neat," Jordan answered, ``that was a success, wasn't it, that was pretty that that worked, that one could do it this way, a sensible generalization of $p$ and $q$" (session 3, p.\ 22). As Kuhn notes in defense of Jordan, the breakdown of the ordinary commutation relations did make room for Jordan's introduction of anti-commutation relations (ibid., p.\ 23; cf.\ Jordan, 1927h; Jordan and Wigner, 1928; and the coda to our paper)} 

All in all, the state of affairs by the end of {\it Neue Begr\"undung} II  can fairly be characterized as follows. Jordan was still trying to rely on the classical formalism of canonical transformations and conjugate variables to build up the formalism realizing the postulates for his quantum-mechanical probability amplitudes. However, these classical concepts had to be stretched almost beyond 
the breaking point to arrive at a satisfactory formulation of his quantum theory. 

\subsection{Mathematical challenges facing the Dirac-Jordan theory}

From a modern point of view, the realization of the axioms of {\it Neue Begr\"undung} is supplied by the Hilbert space formalism.\footnote{As \citet[p.\ 344]{Darrigol 1992} notes about Dirac's version of transformation theory: ``There is one feature of Dirac's original transformation theory that is likely to surprise the modern quantum physicist: the notion of state vector is completely absent. It was in fact introduced later by Weyl [1927] and von Neumann [1927a, see Section 6], and subsequently adopted by Dirac [1930] himself. In 1939 Dirac even split his original transformation symbol $(\xi'/\alpha')$ into two pieces $\langle \xi'|$ and $|\alpha' \rangle$, the ``bra" and the ``ket" vectors. The mathematical superiority of the introduction of state vectors is obvious, since it allows---albeit not without difficulty---an explicit construction of mathematical entities (rigged Hilbert spaces) that justify Dirac's manipulations." In a note added in a proof to his next paper (see Section 7), \citet[p.\ 256]{von Neumann 1927b} acknowledged that \citet{Weyl 1927} had independently introduced the notion of a pure state.\label{darrigol}} Jordan's probability amplitudes $\varphi(a,b)$ are then identified with `inner products' $\langle a | b \rangle$ of `eigenvectors' $| a \rangle$ and $| b \rangle$ of operators $\hat{a}$ and $\hat{b}$.\footnote{We will not introduce a special notation to distinguish between a physical  quantity and the operator acting in Hilbert space representing that quantity. In most cases it will be clear from context whether $\hat{a}$ stands for a quantity or an operator.}
The reason we used scare quotes in the preceding sentence is that for quantities with completely continuous spectra, to which Jordan restricted himself in {\it Neue Begr\"undung} I, the `eigenvectors' of the corresponding operators are {\it not} elements of Hilbert space. That in modern quantum mechanics they are nonetheless routinely treated {\it as if} they are vectors in Hilbert space with inner products such as $\langle a | b \rangle$ is justified by the spectral theorem for the relevant operators. 

Of course, neither the spectral theorem nor the notions of an abstract Hilbert space and of operators acting in it were available when Jordan and Dirac published their respective versions of transformation theory in 1927. The Hilbert space formalism and the spectral theorem were only introduced later that year, by  \citet{von Neumann 1927a}, and two more years passed before \citet{von Neumann 1929} published a rigorous proof of the spectral theorem.  So even though \citet{Dirac 1927} introduced the notation $(a/b)$ for what Jordan wrote as $\varphi(a,b)$,  Dirac, like Jordan, did {\it not} at that time conceive of these quantities as `inner products' of two more elementary quantities (see note \ref{darrigol}). Although Dirac later did accept the split (once again see note \ref{darrigol}), probability amplitudes remained the fundamental quantities  for Jordan \citep[p.\ 361]{Duncan and Janssen 2009}. 

Once the `inner-product' structure of probability amplitudes is recognized and justified with the help of the spectral theorem, Jordan's basic axioms about the addition and multiplication of probability amplitudes are seen to reduce to statements about orthogonality and completeness familiar from elementary quantum mechanics. For instance, as we mentioned in Section 1.1, Jordan's postulates demand that the probability amplitudes $\varphi(a,c)$, $\psi(a,b)$ and $\chi(b,c)$ for quantities $\hat{a}$, $\hat{b}$, and $\hat{c}$ with purely continuous spectra satisfy the relation $\varphi(a,c) = \int  db \, \psi(a,b) \, \chi(b,c) $. Once probability amplitudes are identified with `inner products' of `eigenvectors' (appropriately normalized, such that, e.g., $\langle a | a' \rangle$ $= \delta(a - a')$, where $\delta(x)$ is the Dirac delta function), the familiar completeness relation, $\langle a | c\rangle = \int db \, \langle a|b \rangle \langle b|c \rangle$ (which holds on account of the resolution of unity, $\hat{1} = \int db \, |b \rangle \langle b|$, corresponding to the spectral decomposition of  the operator $\hat{b}$), guarantees that $\varphi(a,c) = \int  db \, \psi(a,b) \, \chi(b,c) $. In this sense, the Hilbert space formalism thus provides a realization of Jordan's postulates.

In the absence of the Hilbert space formalism and the spectral theorem, Jordan relied on the formalism of canonical transformations to develop the analytical apparatus realizing his axiomatic scheme. As we saw in Section 1.1, his starting point was the probability amplitude $\rho(p,q) = e^{-ipq/\hbar}$ for some generalized coordinate $\hat{q}$  and its conjugate momentum $\hat{p}$.\footnote{In the parlance of modern quantum information theory, this is the statement that $\{ |p \rangle\}$ and $\{ |q \rangle \}$ are {\it mutually unbiased bases}.} This special probability amplitude, to reiterate, trivially satisfies two simple differential equations. Jordan then considered canonical transformations to other canonically conjugate variables $\hat{P}$ and $\hat{Q}$ and derived  differential equations for arbitrary probability amplitudes starting from the ones for $\rho(p,q)$. In this way, he claimed, one could recover both the time-independent and the time-dependent Schr\"odinger equations as examples of such equations. 

Both claims are  problematic. The recovery of the time-dependent Schr\"o\-din\-ger equation requires that we look upon the time $t$ not as a parameter as we would nowadays but as an operator to be expressed in terms of the operators $\hat{p}$ and $\hat{q}$. More importantly, Jordan's construction only gets us to the time-independent Schr\"odinger equation for Hamiltonians with fully continuous spectra. In {\it Neue Begr\"undung} I, Jordan deliberately restricted himself to quantities with completely continuous spectra, confident at that point that his approach could easily be generalized to quantities with wholly or partly discrete spectra. He eventually had to accept that this generalization fails. The problem, as he himself  recognized in {\it Neue Begr\"undung} II, is that two quantities $\hat{\alpha}$ and $\hat{p}$ related to each other via a canonical transformation (implemented by the similarity transformation $\hat{\alpha} = T \hat{p} T^{-1}$) always have the same spectrum.\footnote{See Eq.\ (\ref{trigger}) at the end of Section 2.3 for a simple proof of this claim in modern language.} Hence, no canonical transformation that can be implemented in this way can take us from quantities such as $\hat{p}$ and $\hat{q}$ with a completely continuous spectrum to a Hamiltonian with a wholly or partly discrete spectrum.

The quantities $\varphi(a, b)$ in Jordan's formalism do double duty as probability amplitudes  and as integral kernels of canonical transformations.  Even if we accept the restriction to quantities with fully continuous spectra for the moment, Jordan could not quite get his formalism to work, at least not at the level of generality  he had hoped for. In hindsight, we can see that another major hurdle was that canonical transformations from one set of conjugate variables to another, although they do preserve the spectra,  do {\it not} always preserve the Hermitian character of the operators associated with these variables in quantum mechanics \citep[secs.\ 5--6]{Duncan and Janssen 2009}. Initially Jordan tried to get around this problem through the introduction of the {\it Erg\"anzungsamplitude}. Once he dropped that notion, he had to restrict the allowed canonical transformations to those associated with unitary operators. In the modern Hilbert space formalism, the integral kernels of canonical transformations in Jordan's formalism are  replaced by unitary operators. There is no need anymore for considering canonical transformations nor, for that matter, for sorting quantities into pairs of conjugate variables. Jordan's reliance on canonical transformations and conjugate variables became even more strained in {\it Neue Begr\"undung} II, when he tried to extend his approach to quantities with partly or wholly discrete spectra. He had a particularly hard time dealing with the purely discrete spectrum of the recently introduced spin observable. Minding his $p$'s and $q$'s, Jordan ended up putting himself in a straitjacket.

\subsection{Von Neumann's alternative to the Dirac-Jordan theory}

At the end of their exposition of Jordan's theory in {\it Neue Begr\"undung} I, 
\citet[p.\ 30]{Hilbert-von Neumann-Nordheim} emphasized the mathematical difficulties with Jordan's approach (some of which they had caught, some of which they too had missed), announced that they might return to these on another occasion, and made a tantalizing reference to the first of  three papers on quantum mechanics that von Neumann would publish in 1927 in the Proceedings of the G\"ottingen Academy \citep{von Neumann 1927a, von Neumann 1927b, von Neumann 1927c}. This trilogy formed the basis for his famous book \citep{von Neumann 1932}. The first of these papers, ``Mathematical foundations ({\it Mathematische Begr\"undung}) of quantum mechanics," is the one in which \citet{von Neumann 1927a} introduced the Hilbert space formalism and the spectral theorem (he only published a rigorous proof of the latter two years later). One might therefore expect at this juncture that von Neumann would simply make the observations that we made in Section 1.2, namely that the Hilbert space formalism provides the natural implementation of Jordan's axiomatic scheme and that the spectral theorem can be used to address the most glaring mathematical problems with this implementation. Von Neumann, however, did nothing of the sort.\footnote{Von Neumann's formulation of quantum mechanics is nonetheless often referred to as transformation theory. In fact, \citet[p.\ 1]{von Neumann 1932} himself, in the introduction of his famous book, used that term to describe both his own theory and the theory of Dirac and Jordan. And the title of the section in which \citet[pp.\ 307--322]{Jammer 1966} covers von Neumann's formulation is ``The statistical transformation theory in Hilbert space." At the beginning of the preceding section on Jordan and Dirac, \citet[p.\ 293]{Jammer 1966} warns his readers that ``a clear-cut or universally accepted definition of the subject matter of the transformation theory is hardly found in the literature." After giving a couple of examples of widely different usages, he offers a possible definition: ``the study of those transformations in quantum theory which leave the results of empirically significant formulae invariant" (ibid., p.\ 294). With this definition, there can be many different transformation theories depending ``on the kind of space with respect to which the performance of transformations is to be considered" (ibid.). Jammer then characterizes the progression from Jordan and Dirac to von Neumann in a way that fits well with our analysis: ``while the early phases of the development were characterized by transformations within the configuration and momentum space, its later elaboration led to the conception of abstract Hilbert spaces as the arena underlying the transformations. In this course the formerly important notion of canonical transformations in quantum mechanics gradually lost its peculiarity and importance" (ibid.).\label{nomenclature}}

Von Neumann was sharply critical of the Dirac-Jordan transformation theory. As he put it in the introduction of his 1932 book: ``Dirac's method does not meet the demands of mathematical rigor in any way---not even when it is reduced in the natural and cheap way to the level that is common in theoretical physics" \citep[p.\ 2; our emphasis]{von Neumann 1932}. He went on to say that ``the correct formulation is  {\it not just a matter of making Dirac's method mathematically precise and explicit} but right from the start calls for a different approach related to Hilbert's spectral theory of operators" (ibid., our emphasis).\footnote{As L\'eon \citet[pp.\ 95--96]{van Hove 1958}, noted in a concise and lucid summary of von Neumann's contribution to quantum mechanics for a special issue of the {\it Bulletin of the American Mathematical Society} dedicated to his memory: ``Although von Neumann himself attempted at first, in collaboration with Hilbert and Nordheim, to edify the quantum-mechanical formalism along similar lines [i.e., those of the transformation theory of Dirac and Jordan], he soon realized that a much more natural framework was provided by the abstract, axiomatic theory of Hilbert spaces and their linear operators." Unfortunately, there are many highly misleading statements in the historical literature about the relation between the Dirac-Jordan transformation theory and von Neumann's Hilbert space formalism. \citet[pp.\ 46--47]{Kragh 1990} calls the latter ``a mathematically advanced development of the Dirac-Jordan transformation theory." \citet[p.\ 301]{Lacki 2000} writes that ``von Neumann appears to follow exactly the program as set out in his previous paper  with Hilbert and Nordheim." And in our paper on the path to {\it Neue Begr\"undung}, we wrote: ``In the process of providing sound mathematical underpinnings of Jordan's transformation theory, von Neumann introduced the idea of representing quantum-mechanical states by vectors or rays in Hilbert space" \citep[p.\ 361]{Duncan and Janssen 2009}.} Von Neumann only referred to Dirac in this passage, but as co-author of the paper with Hilbert and Nordheim mentioned above, he was thoroughly familiar with Jordan's closely related work as well. He also clearly appreciated the difference in emphasis between Dirac and Jordan. Talking about the Schr\"odinger wave function in the introduction of the second paper of his 1927 trilogy, he wrote: ``Dirac interprets it as a row of a certain transformation matrix, Jordan calls it a probability amplitude" \citep[p.\ 246]{von Neumann 1927b}.\footnote{Discussing Jordan's approach in his  first paper, \citet{von Neumann 1927a} referred to ``transformation operators (the integral kernels of which are the ``probability amplitudes")" (p.\ 3).} In the opening paragraph of this article, von Neumann contrasted wave mechanics with ``transformation theory" or ``statistical theory," once again reflecting the difference in emphasis between Dirac and Jordan. Yet, despite his thorough understanding of it, von Neumann did not care for the Dirac-Jordan approach.

Von Neumann's best-known objection concerns the inevitable use of delta functions in the Dirac-Jordan approach. However, von Neumann also objected to the use of probability amplitudes. Jordan's basic amplitude, $\rho(p,q) = e^{-ipq/\hbar}$, is not in the space $L^2$ of  square-integrable functions that forms one instantiation of abstract Hilbert space. Moreover, probability amplitudes are only determined up to a phase factor, which von Neumann thought particularly unsatisfactory. ``It is true that the probabilities appearing as end results are invariant," he granted  in the introduction of his paper, ``but it is unsatisfactory and unclear why this detour through the unobservable and non-invariant is necessary" \citep[p.\ 3]{von Neumann 1927a}. So, rather than following the Jordan-Dirac approach and looking for ways to mend its mathematical shortcomings, 
von Neumann, as  indicated in the passage from his 1932 book quoted above, adopted an entirely new approach. He generalized Hilbert's spectral theory of operators\footnote{See \citet{Steen 1973} for a brief history of spectral theory.}  to provide a  formalism for quantum mechanics that is very different from the one proposed by Jordan and Dirac.

The only elements that von Neumann took from transformation theory---more specifically Jordan's version of it---were, first, Jordan's basic idea that quantum mechanics is ultimately a set of rules for conditional probabilities ${\rm Pr}(a|b)$, and second, the fundamental assumption that such probabilities are given by the absolute square of the corresponding probability amplitudes, which essentially boils down to the Born rule.\footnote{Interestingly, \citet[pp.\ 43--44]{von Neumann 1927a} mentioned Pauli, Dirac, and Jordan in this context, but not  Born. Unlike Heisenberg (see note \ref{Teil-Ganze}), von Neumann had no reason to dislike Born's work. And he did cite Born in the second paper of his 1927 trilogy on quantum mechanics \citep[p.\ 245]{von Neumann 1927b}.} Von Neumann derived a new expression for conditional probabilities in quantum mechanics that avoids probability amplitudes altogether and instead sets them equal to the trace of products of projection operators, as they are now called. Instead of the term `projection operator', Von Neumann used the term  {\it Einzeloperator} (or {\it E.Op.\ }for short; cf.\ note \ref{E.Op.}). The probability ${\rm Pr}(a|b)$, e.g., is given by ${\rm Tr}(\hat{E}(a) \hat{F}(b))$, where $\hat{E}(a)$ and $\hat{F}(b)$ are  projection operators onto, in Dirac notation, the `eigenvectors' $| a \rangle$ and $| b \rangle$ of the operators $\hat{a}$ and $\hat{b}$, respectively. Unlike probability amplitudes, these projection operators do not have any phase ambiguity. This is easily seen in Dirac notation. The projection operator $\hat{E}(a) = |a\rangle \langle a|$  does not change if the ket $|a \rangle$ is replaced by $e^{i \vartheta} |a \rangle$ and the bra $\langle a|$ accordingly by $e^{-i\vartheta}\langle a|$. We should emphasize, however, that, just as Jordan and Dirac with their probability amplitudes/transformation functions $\langle a | b \rangle$, von Neumann did not think of his projection operators as constructed out of bras and kets, thus avoiding the problem that many of these bras and kets are not in Hilbert space.

Toward the end of his paper, \citet[pp.\ 46--47]{von Neumann 1927a} noted that his trace expression for conditional probabilities is invariant under ``canonical transformations." What von Neumann called canonical transformations, however, are not Jordan's canonical transformations but simply, in modern terms, unitary transformations. Such transformations automatically preserve Hermiticity and the need for something like Jordan's {\it Erg\"anzungsamplitude} simply never arises. Von Neumann  noted that his trace expression for conditional probabilities does not change if the projection operators $\hat{E}$ and $\hat{F}$ are replaced by $\hat{U} \hat{E} \hat{U}^\dagger$ and $\hat{U} \hat{F} \hat{U}^\dagger$, where $\hat{U}$ is an arbitrary unitary operator ($\hat{U}^\dagger = \hat{U}^{-1}$). In von Neumann's approach, as becomes particularly clear in his second paper of 1927 (see below), one also does not have to worry about sorting variables into sets of mutually conjugate ones. This then is what the  `never mind your $p$'s and $q$'s' part of the title of our paper refers to. By avoiding conjugate variables and canonical transformations, von Neumann completely steered clear of the problem that ultimately defeated Jordan's attempt to derive all of quantum mechanics from his set of axioms, namely that canonical transformations can never get us from $\hat{p}$'s and $\hat{q}$'s with fully continuous spectra to quantities with wholly or partly discrete spectra, such as the Hamiltonian.

In {\it Mathematische Begr\"undung}, von Neumann not only provided an alternative to Jordan's analysis of probabilities in quantum mechanics, he also provided an alternative to the Dirac-Jordan transformation-theory approach to proving the equivalence of matrix mechanics and wave mechanics \citep[p.\ 14]{von Neumann 1927a}. This is where von Neumann put the abstract notion of Hilbert space that he introduced in his paper to good use. He showed that  matrix mechanics and wave mechanics correspond to two instantiations of abstract Hilbert space, the space $l^2$ of square-summable sequences and the space $L^2$ of square-integrable functions, respectively \citep[p.\ 172]{Dieudonne 1981}.\footnote{Not usually given to hyperbole, \citet[p.\ 316]{Jammer 1966} was moved to comment on this feat: ``In von Neumann's formulation of quantum mechanics was found the ultimate fulfillment---as far as the theory of quanta is concerned---of Hipparchus's insistence on the usefulness of investigating ``why on two hypotheses so different from one another \ldots\ the same results appear to follow," which, as we have seen, had already characterized Jordan's line of research." The references to Hipparchus and Jordan are to p.\ 307 of Jammer's book (see also p.\ 293, where Dirac is mentioned as well).} As von Neumann reminded his readers, well-known theorems due to Parseval and Riesz and Fisher had established that $l^2$ and $L^2$ are isomorphic.\footnote{In 1907--1908, Erhard Schmidt, a student of Hilbert who got his Ph.D. in 1905, fully worked out the theory of $l^2$ and called it `Hilbert space' \citep[p.\ 364]{Steen 1973}. In a paper on canonical transformations, Fritz \citet[p.\ 197]{London 1926b} used the term `Hilbert space' for $L^2$ 
(Jammer, 1966, p.\ 298; Duncan and Janssen 2009, p.\ 356, note 12).\label{Erhard}}

In his second 1927 paper, ``Probability-theoretical construction ({\it Wahr\-schein\-lich\-keits\-theoretischer Aufbau}) of quantum mechanics," \citet{von Neumann 1927b} freed himself even further from relying on the Dirac-Jordan approach. In {\it Ma\-the\-ma\-tische Begr\"undung} he had accepted the Born rule and recast it in the form of his trace formula. In {\it Wahr\-schein\-lich\-keits\-theoretischer Aufbau} he sought to derive this trace formula, and thereby the Born rule, from more fundamental assumptions about probability. Drawing on ideas of von Mises (see note \ref{von mises}), von Neumann started by introducing probabilities in terms of selecting members from large ensembles of systems. He then made two very general and {\it prima facie} perfectly plausible assumptions about expectation values of quantities defined on such ensembles \citep[pp.\ 246--250]{von Neumann 1927b}. From those assumptions, some assumptions about the repeatability of measurements \citep[p.\ 271, p.\ 262, cf.\ note \ref{neumann heisenberg}]{von Neumann 1927b}, and key features of his Hilbert space formalism (especially some assumptions about the association of observables with Hermitian operators), von Neumann did indeed manage to recover the Born rule. Admittedly, the assumptions needed for this result are not as innocuous as they look at first sight. They are essentially the same as those that go into von Neumann's infamous no-hidden-variable proof \citep{Bell 1966, Bacciagaluppi and Crull 2009, Bub 2010}.

Along the way \citet[p.\ 253]{von Neumann 1927b} introduced what we now call a density operator to characterize the ensemble of systems he considered. He found that the expectation value of an observable represented by some operator $\hat{a}$ in an ensemble characterized by $\hat{\rho}$ is given by ${\rm Tr}(\hat{\rho} \, \hat{a})$, where we used the modern notation $\hat{\rho}$ for the density operator (von Neumann used the letter $U$). This result holds both for what \citet{von Neumann 1927b} called a ``pure" ({\it rein}) or ``uniform" ({\it einheitlich}) ensemble (p.\ 255), consisting of identical systems in identical states,  and for what he called a ``mixture" ({\it Gemisch}) (p.\ 265). So the result is more general than the Born rule, which obtains only in the former case. Von Neumann went on to show that the density operator for a uniform ensemble is just the projection operator onto the ray in Hilbert space corresponding to the state of all systems in this ensemble. However, he found it unsatisfactory to characterize the state of a physical system by specifying a ray in Hilbert space. ``Our knowledge of a system," \citet[p.\ 260]{von Neumann 1927b} wrote, ``is never described by the specification of a state \ldots\ but, as a rule, by the results of experiments performed on the system." In this spirit, he considered the simultaneous measurement of a maximal set of commuting operators and constructed the density operator for an ensemble where what is known is that the corresponding quantities have values in certain intervals. He showed that such measurements can fully determine the state and that the density operator in that case is once again the projection operator onto the corresponding ray in Hilbert space.

Von Neumann thus arrived at the typical quantum-mechanical way of conceiving of a physical problem nowadays, which is very different from the classical way to which Jordan was still wedded in {\it Neue Begr\"undung}. In classical mechanics, as well as in Jordan's version of transformation theory, the full description of a physical system requires the specification of a complete set of $p$'s and $q$'s. In quantum mechanics, as was first made clear in von Neumann's {\it Wahr\-schein\-lich\-keits\-theoretischer Aufbau}, it requires the specification of the eigenvalues of all the operators in a maximal set of commuting operators for the system. In other words, the `never mind your $p$'s and $q$'s' part of the title of our paper carried the day.

\subsection{Outline of our paper}

In the balance of this paper we  cover the contributions of Jordan and von Neumann (initially with Hilbert and Nordheim) to the developments sketched above in greater detail. We give largely self-contained reconstructions of the central arguments and derivations in five key papers written in G\"ottingen and, in one case ({\it Neue Begr\"undung} II), in Copenhagen over the span of just one year, from late 1926 to late 1927 \citep{Jordan 1927b, Jordan 1927f, Hilbert-von Neumann-Nordheim, von Neumann 1927a, von Neumann 1927b}. To make the arguments and derivations in these papers easier to follow for a modern reader, we translate them all into the kind of modern notation introduced in Sections 1.2 and 1.3. To make it easier for the reader to check our claims against the primary sources, we provide detailed references to the latter, including where necessary equation numbers and legends for the original notation. We will not cover Dirac (cf.\ note \ref{dirac}), although we will occasionally refer to his work, both his original paper on transformation theory \citep{Dirac 1927} and the book based on it \citep{Dirac 1930}. We will also freely avail ourselves of his bra and ket notation. Our focus, however, will be on Jordan and von Neumann. 

We begin, in Section 2, with {\it Neue Begr\"undung} I \citep{Jordan 1927b}. In this paper Jordan only dealt with quantities with completely continuous spectra,  suggesting that the generalization to ones with partly or wholly discrete spectra would be straightforward \citep[p.\ 811, p.\ 816]{Jordan 1927b}. We cover Jordan's postulates for his probability amplitudes (Section 2.1) and his construction of a realization of these postulates, especially his use of canonical transformations between pairs of conjugate variables to derive the differential equations for these amplitudes (Section 2.3). In Section 2.2, drawing on an earlier paper \citep{Duncan and Janssen 2009}, we remind the reader of the role of canonical transformations in matrix mechanics. In Section 2.4, we take a closer look at Jordan's notion of a supplementary amplitude [{\it Erg\"anzungsamplitude}].

In Section 3, we discuss the paper by \citet{Hilbert-von Neumann-Nordheim}, submitted in April 1927, that grew out of the exposition of Jordan's approach in Hilbert's 1926/1927 course on quantum mechanics \citep[pp.\ 698--706]{Sauer and Majer 2009}. Hilbert and his co-authors had the advantage of having read the paper in which \citet{Dirac 1927} presented his version of transformation theory. Jordan only read Dirac's paper when he was correcting the page proofs of {\it Neue Begr\"undung} I \citep[p.\ 809; note added in proof]{Jordan 1927b}. 
 
In Section 4, we consider {\it Neue Begr\"undung} II \citep{Jordan 1927f}, received by {\it Zeitschrift f\"ur Physik} in early June 1927 and written in part in response to criticism of {\it Neue Begr\"undung} I by \citet{Hilbert-von Neumann-Nordheim} and by \citet{von Neumann 1927a} in {\it Mathematische Begr\"undung}. Since von Neumann introduced an entirely new approach, we deviate slightly from the chronological order of these papers, and discuss {\it Mathematische Begr\"undung} after {\it Neue Begr\"undung} II. In the abstract of the latter, \citet[p.\ 1]{Jordan 1927f} promised ``a simplified and generalized presentation of the theory developed in [{\it Neue Begr\"undung}] I."  Drawing on \citet{Dirac 1927}, Jordan simplified his notation somewhat, although he also added some new  and redundant elements to it. Most importantly, however, the crucial generalization to quantities with partly or wholly discrete spectra turned out to be far more problematic than he had suggested in {\it Neue Begr\"undung} I. Rather than covering {\it Neue Begr\"undung} II in detail, we highlight the problems Jordan ran into, especially in his attempt to deal with spin in his new formalism. 

In Sections 5 and 6, we turn to the first two papers of von Neumann's trilogy on quantum mechanics of 1927. In Section 5, on {\it Ma\-the\-ma\-ti\-sche Begr\"undung} \citep{von Neumann 1927a}, we focus on von Neumann's criticism of the Dirac-Jordan transformation theory, his proof of the equivalence of wave mechanics and matrix mechanics based on the isomorphism between $L^2$ and $l^2$, and his derivation of the trace formula for probabilities in quantum mechanics. We do not cover the introduction of his Hilbert space formalism, which takes up a large portion of his paper. This material is covered in any number of modern books on functional analysis.\footnote{See, e.g., \citet{Prugovecki 1981}, or, for a more elementary treatment, which will be more than sufficient to follow our paper, \citet[Ch.\ 3]{Dennery and Krzywicki 1996}.\label{Prugovecki}} In Section 6, on {\it Wahr\-schein\-lich\-keits\-theoretischer Aufbau} \citep{von Neumann 1927c}, we likewise focus on the overall argument of the paper, covering the derivation of the trace formula from some basic assumptions about the expectation value of observables in an ensemble of identical systems, the introduction of density operators, and the specification of pure states through the values of a maximal set of commuting operators.

In Section 7, we summarize the transition from Jordan's quantum-mechanical formalism rooted in classical mechanics (mind your $p$'s and $q$'s) to von Neumann's quantum-mechanical formalism which no longer depends on classical mechanics for its formulation (never mind your $p$'s and $q$'s). 

As a coda to our story, we draw attention to the reemergence of the canonical formalism, its generalized coordinates and conjugate momenta, even for spin-$\frac{1}{2}$ particles, in quantum field theory. 

\section{Jordan's {\it Neue Begr\"undung} I (December 1926)}
 
{\it Neue Begr\"undung} I was received by {\it Zeitschrift f\"ur Physik} on December 18, 1926 and published  January 18, 1927  \citep{Jordan 1927b}.  It  consists of two parts. In Part One ({\it I.\ Teil}\,), consisting of secs. 1--2 (pp.\ 809--816), Jordan laid down the postulates of his theory. In Part Two ({\it II.\ Teil}\,), consisting of secs.\ 3--7 (pp.\ 816--838), he presented the formalism realizing these postulates. In the abstract of the paper, Jordan announced that his new theory would unify all earlier formulations of quantum theory:
\begin{quotation}
The four forms of quantum mechanics that have been developed so far---matrix theory, the theory of Born and Wiener, wave mechanics, and $q$-number theory---are contained in a more general formal theory. Following one of Pauli's ideas, one can base this new theory on a few simple fundamental postulates ({\it Grundpostulate}) of a statistical nature \citep[p.\ 809]{Jordan 1927b}.
\end{quotation}
As we  mentioned in Section 1.2, Jordan claimed that he could recover both the time-dependent and the time-independent Schr\"odinger equation as special cases of the differential equations he derived for the probability amplitudes central to his formalism. This is the basis for his claim that wave mechanics can be subsumed under his new formalism. Nowhere in the paper did he  show explicitly how matrix mechanics is to be subsumed under the new formalism. Perhaps Jordan felt that this did not require a special argument as the new formalism had grown naturally  out of matrix mechanics and his own contributions to it \citep{Jordan 1926a, Jordan 1926b}. However, as emphasized repeatedly already, \citet{Jordan 1927b} restricted himself to quantities with purely continuous spectra in {\it Neue Begr\"undung} I, so the formalism as it stands is not applicable to matrix mechanics. 
Like Dirac's (1927) own version of statistical transformation theory, Jordan's version can be seen as a natural extension of Dirac's (1925) $q$-number theory. It is only toward the end of his paper (sec.\ 6)
that Jordan turned to the operator theory of \citet{Born and Wiener 1926}. In our discussion of {\it Neue Begr\"undung} I, we omit this section along with some mathematically intricate parts of secs.\ 3 and 5 that are not necessary for understanding the paper's overall argument. We do not cover the concluding sec.\ 7 of Jordan's paper either, which deals with quantum jumps  (recall his earlier paper on this topic [Jordan, 1927a], which we briefly discussed in Section 1.1).

Although we will not cover Jordan's unification of the various forms of quantum theory in any detail, we will cover (in Section 5) von Neumann's criticism of the Dirac-Jordan way of proving the equivalence of matrix mechanics and wave mechanics as a prelude to his own proof based on the isomorphism of $l^2$ and $L^2$ \citep{von Neumann 1927a}.
In our discussion of {\it Neue Begr\"undung} I in this section, we focus on the portion of Jordan's paper that corresponds to the last sentence of the abstract, which promises a statistical foundation of quantum mechanics. Laying this foundation actually takes up most of the paper (secs.\ 1--2, 4--5).

\subsection{Jordan's postulates for probability amplitudes}

The central quantities in {\it Neue Begr\"undung} I are generalizations of Schr\"odinger energy eigenfunctions which Jordan called ``probability amplitudes." He attributed both the generalization and the term to Pauli. Jordan referred to a footnote in a forthcoming paper by \citet[p.\ 83, note]{Pauli 1927a} proposing, in Jordan's terms, the following interpretation of the energy eigenfunctions $\varphi_n(q)$ (where $n$ labels the different energy eigenvalues) of a system (in one dimension): ``If $\varphi_n(q)$ is normalized, then $|\varphi_n(q)|^2 dq$ gives the probability that, if the system is in the state $n$, the coordinate [$\hat{q}$] has a value between $q$ and $q+dq$" \citep[p.\ 811]{Jordan 1927b}. A probability amplitude such as this one for position and energy can be introduced for any two quantities.

In  {\it Neue Begr\"undung} I, Jordan  focused on quantities with completely continuous spectra. He only tried to extend his approach, with severely limited success, to partly or wholly discrete spectra in {\it Neue Begr\"undung} II (see Section 4). For two quantities $\hat{x}$ and $\hat{y}$ that can take on a continuous range of values $x$ and $y$, respectively,\footnote{Recall that this is our notation (cf.\ note \ref{notation1}): Jordan used different letters for quantities and their numerical values. For instance, he used $q$ (with value $x$) and $\beta$ (with value $y$) for what we called $\hat{x}$ and $\hat{y}$, respectively \citep[p.\ 813]{Jordan 1927b}\label{NB1-notation}} there is a complex probability amplitude $\varphi(x,y)$ such that $|\varphi(x,y)|^2 \, dx$ gives the probability that $\hat{x}$ has a value between $x$ and $x+dx$ given that $\hat{y}$ has the value $y$.

In modern Dirac notation $\varphi(x,y)$ would be written as $\langle x | y \rangle$ (cf.\ our discussion in Section 1.2). Upon translation into this modern notation, many of Jordan's expressions turn into instantly recognizable expressions in modern quantum mechanics and we will frequently provide such translations to make it easier to read Jordan's text. We must be careful, however, not to read too much into it. First of all, von Neumann had not yet introduced the abstract notion of Hilbert space when Jordan and Dirac published their theories in early 1927, so neither one thought of probability amplitudes as `inner products' of `vectors' in Hilbert space at the time. More importantly, for quantities $\hat{x}$'s and $\hat{y}$'s with purely continuous spectra (e.g.,  position or momentum of a particle in an infinitely extended region), the `vectors'  $| x \rangle$ and $| y \rangle$ are {\em not} elements of Hilbert space, although an inner product $\langle x | y \rangle$ can be defined in a generalized sense (as a distribution) as an integral of products of continuum normalized wave functions, as is routinely done in elementary quantum mechanics. That continuum eigenstates can be treated {\it as though} they are indeed states in a linear space satisfying completeness and orthogonality relations which are continuum analogs of the discrete ones which hold rigorously in a Hilbert space is, as we will see later, just the von Neumann spectral theorem for self-adjoint operators with a (partly or wholly) continuous spectrum.

In the introductory section of {\it Neue Begr\"undung} I, \citet[p.\ 811]{Jordan 1927b} listed two postulates, labeled I and II. Only two pages later, in sec.\ 2, entitled ``Statistical foundation of quantum mechanics," these two postulates are superseded by a new set of  four postulates, labeled A through D.\footnote{In the short version of {\it Neue Begr\"undung} I presented to the G\"ottingen Academy on January 14, 1927, \citet[p.\ 162]{Jordan 1927c} only introduced postulates I and II. In this short version, \citet[p.\ 163]{Jordan 1927c} referred to ``a soon to appear extensive paper in {\it Zeitschrift f\"ur Physik}" (i.e., Jordan, 1927b).} In {\it Neue Begr\"undung} II, \citet[p.\ 6]{Jordan 1927f} presented yet another set of postulates, three this time, labeled I through III (see Section 4).\footnote{In his overview of recent developments in  quantum mechanics in {\it Die Naturwissenschaften}, \citet[Pt.\ 2, p.\ 648]{Jordan 1927i}, after explaining the basic notion of a probability amplitude (cf.\ Postulate A below), listed only two postulates, or axioms as he now called them, namely ``the assumption of probability interference" (cf. Postulate C below) and the requirement that there is a canonically conjugate quantity $\hat{p}$ for every quantum-mechanical quantity $\hat{q}$ (cf.\ Postulate D below).} The exposition of Jordan's theory by  \citet{Hilbert-von Neumann-Nordheim},
written in between {\it Neue Begr\"undung} I and II, starts from six ``physical axioms" (pp.\ 4--5), labeled I through VI (see Section 3). We will start from Jordan's four postulates of  {\it Neue Begr\"undung} I, which we paraphrase and comment on below, staying close to Jordan's own text but using the notation introduced above to distinguish between quantities and their numerical values.

{\it Postulate A}. For two mechanical quantities $\hat{q}$ and $\hat{\beta}$ that stand in a definite kinematical relation to one another there are two complex-valued functions, $\varphi(q, \beta)$ and $\psi(q, \beta)$, such that $\varphi(q, \beta) \psi^*(q, \beta) dq$ gives the probability of finding a value between $q$ and $q+dq$ for $\hat{q}$ given that $\hat{\beta}$ has the value $\beta$. The function $\varphi(q, \beta)$ is called the probability amplitude, the function $\psi(q, \beta)$ is called the ``supplementary amplitude" ({\it Erg\"anzungsamplitude}).

{\it Comments:} 
As becomes clear later on in the paper, ``mechanical quantities that stand in a definite kinematical relation to one another" are quantities that can be written as functions of some set of generalized coordinates and their conjugate momenta. In his original postulate I, \citet[p.\ 162]{Jordan 1927b} wrote that ``$\varphi(q, \beta)$ is independent of the mechanical nature (the Hamiltonian) of the system and is determined only by the kinematical relation between $\hat{q}$ and $\hat{\beta}$" (hats added). Hilbert et al. made this into a separate postulate, their axiom V: ``A further physical requirement is that the probabilities only depend on the functional nature of the quantities $F_1(pq)$ and $F_2(pq)$, i.e., on their kinematical connection [{\it Verkn\"upfung}], and not for instance on additional special properties of the mechanical system under consideration, such as, for example, its Hamiltonian" \citep[p.\ 5]{Hilbert-von Neumann-Nordheim}. With $\varphi(q, \beta) = \langle q|\beta \rangle$, the statement about the kinematical nature of probability amplitudes  translates into the observation that they depend only on the inner-product structure of  Hilbert space 
and not on the Hamiltonian governing the time evolution of the system under consideration.\footnote{In the AHQP interview with Jordan, Kuhn emphasized the importance of this aspect of Jordan's formalism: ``The terribly important step here is throwing the particular Hamiltonian function away and saying that the relationship is only in the kinematics" (session 3, p.\ 15).} 

It turns out that for all quantities represented, in modern terms, by Hermitian operators, the amplitudes $\psi(q, \beta)$ and $\varphi(q, \beta)$ are equal to one another. At this point, however, Jordan wanted to leave room for quantities represented by non-Hermitian operators. This is directly related to the central role of canonical transformations in his formalism. As Jordan (1926a,b) had found in a pair of papers published in 1926, canonical transformations need not be unitary and therefore do not always preserve the Hermiticity of the conjugate variables one starts from \citep{Duncan and Janssen 2009}. The {\it Erg\"anzungsamplitude} does not appear in the presentation of Jordan's formalism by \citet{Hilbert-von Neumann-Nordheim}.\footnote{Both $\psi$ and $\varphi$ are introduced in the lectures by Hilbert on which this paper is based but they are set equal to one another almost immediately  and without any further explanation \cite[p.\ 700]{Sauer and Majer 2009}.} In {\it Neue Begr\"undung} II, \citet[p.\ 3]{Jordan 1927f}  restricted himself to Hermitian quantities  and silently dropped the {\it Er\-g\"an\-zungs\-am\-pli\-tude}. We return to the {\it Erg\"anzungsamplitude} in Section 2.4 below, but until then we will simply set $\psi(q, \beta) = \varphi(q, \beta)$ everywhere. 

{\it Postulate B}. The probability amplitude $\bar{\varphi}(\beta,q)$ is the complex conjugate of the probability amplitude $\varphi(q, \beta)$. In other words, $\bar{\varphi}(\beta,q) = \varphi^*(q, \beta)$. This implies a symmetry property of the probabilities themselves: the probability density $|\bar{\varphi}(\beta, q)|^2$ for finding the value $\beta$ for $\hat{\beta}$ given the value $q$ for $\hat{q}$ is equal to the probability  density $|\varphi(q, \beta)|^2$ for finding the value $q$ for $\hat{q}$ given the value $\beta$ for $\hat{\beta}$.

{\it Comment.} This property is immediately obvious once we write $\varphi(q,\beta)$ as $\langle q|\beta\rangle$
with the interpretation of $\langle q|\beta\rangle$ as an `inner product' in Hilbert space (but recall that one has to be cautious when dealing with quantities with continuous spectra).

{\it Postulate C}. The probabilities combine through interference. In sec.\ 1,  \citet[p.\ 812]{Jordan 1927b} already introduced the phrase ``interference of probabilities" to capture the striking feature in his quantum formalism that the probability {\it amplitudes} rather than the probabilities themselves follow the usual composition rules for probabilities.\footnote{Recall, however, Heisenberg's criticism of this aspect of Jordan's work in his uncertainty paper \citep[pp.\ 183--184; cf.\ note \ref{HvJ0}]{Heisenberg 1927b}.} Let $F_1$ and $F_2$ be two outcomes [{\it Tatsachen}] for which the amplitudes are $\varphi_1$ and $\varphi_2$. If $F_1$ and $F_2$ are mutually exclusive,  $\varphi_1 + \varphi_2$ is the amplitude for the outcome `$F_1$ or $F_2$'. If $F_1$ and $F_2$ are independent,  $\varphi_1 \varphi_2$ is the amplitude for the outcome `$F_1$ and $F_2$'.

{\it Consequence.} Let $\varphi(q, \beta)$ be the probability amplitude for the outcome $F_1$ of finding the value $q$ for $\hat{q}$ given the value $\beta$ for $\hat{\beta}$.  Let $\chi(Q, q)$ be the probability amplitude for the outcome $F_2$ of finding the value $Q$ for $\hat{Q}$ given the value $q$ for $\hat{q}$.  Since $F_1$ and $F_2$ are independent, Jordan's multiplication rule tells us that the probability amplitude for `$F_1$ and $F_2$' is given by the product $\chi(Q, q) \, \varphi(q, \beta)$.  Now let  $\Phi(Q, \beta)$ be the probability amplitude for the outcome $F_3$ of finding the value $Q$ for $\hat{Q}$ given the value $\beta$ for $\hat{\beta}$. According to Jordan's addition rule, this amplitude is equal to the `sum' of the amplitudes for `$F_1$ and $F_2$'  for all different values of $q$. Since $\hat{q}$ has a continuous spectrum, this `sum' is actually an integral. The probability amplitude for $F_3$ is thus given by\footnote{We include the numbers of the more important equations in the original papers in square brackets. `NB1' refers to {\it Neue Begr\"undung} I \citep{Jordan 1927b}. Since the numbering of equations starts over a few times in this paper (see note \ref{numbering-NB1}), we will often include the section number as well.}
\begin{equation}
\Phi(Q, \beta) = \int \chi(Q, q) \, \varphi(q, \beta) \, dq. \quad [{\rm NB1, \, sec.\ 2, \, Eq.\ 14}]
\label{completeness}
\end{equation}

{\it Special case.} If $\hat{Q} = \hat{\beta}$, the amplitude $\Phi(\beta', \beta'')$ becomes the Dirac delta function. \citet[p.\ 814]{Jordan 1927b} introduced the notation $\delta_{\beta' \beta''}$ even though $\beta'$ and  $\beta''$ are continuous rather than discrete variables [NB1, sec.\ 2, Eq.\ 16]. In a footnote he conceded that this is mathematically dubious. In {\it Neue Begr\"un\-dung} II, \citet[p.\ 5]{Jordan 1927f}  used the delta function that  \citet[pp.\ 625--627]{Dirac 1927} had meanwhile introduced in his paper on transformation theory. Here and in what follows we will give Jordan the benefit of the doubt and assume the normal properties of the delta function.\footnote{For a brief history of the delta function that focuses on its role in quantum mechanics, see, e.g., \citet[pp.\ 301--302, pp.\ 313--314]{Jammer 1966}.} 

Using that the amplitude $\chi(\beta', q)$ is just the complex conjugate of the amplitude $\varphi(q, \beta')$, we arrive at the following expression for $\Phi(\beta', \beta'')$:
\begin{equation}
\Phi(\beta', \beta'') = \int \varphi^*(q, \beta') \, \varphi(q, \beta'') \,  dq = \delta_{\beta' \beta''}. \;\; [{\rm NB1, \, sec.\ 2, \, Eqs.\ 15,  16, 17}]
\label{orthogonality}
\end{equation}

{\it Comment.} Translating Eqs.\ (\ref{completeness})--(\ref{orthogonality}) above into  Dirac notation, we recognize them as familiar {\it completeness} and {\it orthogonality} relations:\footnote{The notation $\langle Q|\beta \rangle$ for $\Phi(Q, \beta)$ etc. obviates the need for different letters for different probability amplitudes that plagues Jordan's notation.}
\begin{equation}
\langle Q | \beta \rangle =  \int \langle Q | q \rangle \langle q | \beta \rangle \, dq, \quad
\langle \beta' | \beta'' \rangle = \int \langle  \beta' | q \rangle  \langle q | \beta'' \rangle \, dq = \delta(\beta' - \beta'').
\label{completeness/orthogonality}
\end{equation}
Since the eigenvectors $| q \rangle$ of the operator $\hat{q}$ are not in Hilbert space, the spectral theorem, first proven by \citet{von Neumann 1927a}, is required for the use of the resolution of the unit operator $\hat{1} = \int dq |q \rangle \langle q|$.

{\it Postulate D}. For every $\hat{q}$ there is a conjugate momentum $\hat{p}$. Before stating this postulate, Jordan offered a new definition of what it means for $\hat{p}$ to be the conjugate momentum of  $\hat{q}$. If the amplitude $\rho(p,q)$ of finding the value $p$ for $\hat{p}$ given the value $q$ for  $\hat{q}$ is given by 
\begin{equation}
\rho(p,q) = e^{-ipq/\hbar}, \quad [{\rm NB1, \, sec.\ 2, \, Eq.\ 18}]
\label{NB1-18}
\end{equation}
then $\hat{p}$ is the conjugate momentum of $\hat{q}$. 

Anticipating a special case of the uncertainty principle (cf.\ notes \ref{beller} and \ref{jordan uncertainty}), \citet[p.\ 814]{Jordan 1927b} noted that Eq.\ (\ref{NB1-18}) implies that ``[f]or a given value of  $\hat{q}$ all possible values of $\hat{p}$ are {\it equally probable}."

For $\hat{p}$'s and $\hat{q}$'s with completely continuous spectra, Jordan's definition of when $\hat{p}$ is conjugate to $\hat{q}$ is equivalent to the standard one that the operators $\hat{p}$ and $\hat{q}$ satisfy the commutation relation $[\hat{p}, \hat{q}]  \equiv \hat{p} \, \hat{q} - \hat{q} \, \hat{p} = \hbar/i$. This equivalence, however, presupposes the usual association of the differential operators $(\hbar/i) \,\partial/\partial q$ and `multiplication by $q$' with the quantities $\hat{p}$ and $\hat{q}$, respectively. As we emphasized in Section 1.2, Jordan did not think of these quantities as operators acting in an abstract Hilbert space, but he did associate them (as well as any other quantity obtained through adding and multiplying $\hat{p}$'s and $\hat{q}$'s) with the differential operators $(\hbar/i) \,\partial/\partial q$ and $q$  (and combinations of them). The manipulations in Eqs.\ (19ab)--(24) of {\it Neue Begr\"undung} I, presented under the subheading ``Consequences" ({\it Folgerungen}) immediately following postulate D,  are meant to show that this association follows from his postulates \citep[pp.\ 814--815]{Jordan 1927b}. Using modern notation, we reconstruct Jordan's rather convoluted argument. As we will see, the argument as it stands does not work, but a slightly amended version of it does.

The probability amplitude $\langle p| q\rangle = e^{-ipq/\hbar}$, Jordan's $\rho(p,q) $, trivially satisfies the following pair of equations:
\begin{equation}
 \left(p +\frac{\hbar}{i}\frac{\partial}{\partial q} \right)\langle p| q\rangle = 0, \quad \quad [{\rm NB1, \, sec.\ 2, \, Eq.\ 19a}]  \label{NB1-19a}
\end{equation}
\begin{equation}
   \left(\frac{\hbar}{i}\frac{\partial}{\partial p} + q \right)\langle p| q\rangle = 0. \quad \quad [{\rm NB1, \, sec.\ 2, \, Eq.\ 19b}]  \label{NB1-19b}
\end{equation}
Unless we explicitly say otherwise, expressions such as $\langle a | b \rangle$ are to be interpreted as our notation for Jordan's probability amplitudes $\varphi(a,b)$ and {\it not} as inner products of vectors $|a \rangle$ and $|b \rangle$ in Hilbert space.

Following Jordan (NB1,  sec.\ 2, Eqs.\ 20--22),
we define the map  $T$, which takes functions $f$ of $p$ and turns them into functions $Tf$ of $Q$ (the value of a new quantity $\hat{Q}$ with a fully continuous spectrum):
   \begin{equation}
        T:  \;\;  f(p) \;\; \rightarrow \;\; (Tf)(Q) \equiv \int \langle Q| p\rangle f(p)dp.
        \label{NB1-21}
        \end{equation} 
In other words, $T \ldots = \int  dp \, \langle Q | p \rangle \ldots$ [NB1, Eq.\ 21]. For the special case that $f(p) = \langle p | q \rangle$, we get:
       \begin{equation}
        \left(T\langle p | q \rangle \right)(Q) = \int \langle Q| p \rangle \langle p | q \rangle dp = \langle Q | q \rangle,
        \label{T(p,q)}
        \end{equation} 
where we used completeness, one of the consequences of Jordan's postulate C (cf.\ Eqs.\ (\ref{completeness})--(\ref{completeness/orthogonality})). In other words, $T$ maps $\langle p | q \rangle$  onto $\langle Q | q \rangle$:\footnote{At this point, Jordan's notation, $\varphi(x,y) = T . \rho(x,y)$ [NB1, sec.\ 2, Eq.\ 22], gets particularly confusing as the $x$ on the left-hand side and the $x$ on the right-hand side refer to values of different quantities. The same is true for the equations that follow [NB1, Eqs.\ 23ab, 24ab].}
\begin{equation}
\langle Q | q \rangle = T \, \langle p | q \rangle. \quad [{\rm NB1, \, sec.\ 2, \, Eq.\ 22}]
\label{NB1-22}
\end{equation}

Likewise, we define the inverse map $T^{-1}$, which takes functions $F$ of $Q$ and turns them into functions $T^{-1}F$ of $p$:\footnote{To verify that $T^{-1}$ is  indeed the inverse of $T$, we take $F(Q)$ in Eq.\ (\ref{T^-1F(p)}) to be $(Tf)(Q)$ in Eq.\ (\ref{NB1-21}). In that case we get:
        \begin{eqnarray}
(T^{-1}Tf)(p) & = & \int \langle p | Q \rangle \left( \int \langle Q| p' \rangle f(p')dp' \right) dQ \nonumber \\
 & = & \int \! \!\!  \int \langle p | Q \rangle \langle Q| p' \rangle f(p') \, dQ \, dp'   \nonumber \\
 & = &  \int \langle p | p' \rangle f(p') \,  dp' = f(p), \nonumber 
\end{eqnarray}
where we used the resolution of unity, $\hat{1} = \int dQ \, |Q \rangle \langle Q|$, and $\langle p | p' \rangle = \delta(p - p')$.}
        \begin{equation}
        T^{-1}:  \;\;  F(Q) \;\; \rightarrow \;\;   (T^{-1}F)(p) \equiv \int \langle p | Q \rangle F(Q) dQ.
        \label{T^-1F(p)}
        \end{equation} 
In other words, $T^{-1} \ldots = \int dQ \, \langle p | Q \rangle \ldots$\footnote{\citet[p.\ 815, note]{Jordan 1927b} used the {\it Erg\"anzungsamplitude} to represent $T^{-1}$ in this form.} For the special case that $F(Q) = \langle Q | q \rangle$ we get (again, by completeness): 
       \begin{equation}
       \left( T^{-1}\langle Q | q \rangle \right)(p)  = \int \langle p | Q \rangle \langle Q | q \rangle dQ = \langle p | q \rangle,
       \label{T^-1(Q,q)}
        \end{equation} 
or, more succinctly,
\begin{equation}
\langle p | q \rangle = T^{-1} \langle Q | q \rangle.
\label{pq}
\end{equation}

Applying $T$ to the left-hand side of Eq. (\ref{NB1-19a}) [NB1, Eq.\ 19a], we find:        
\begin{equation}
T \left( \left(p +\frac{\hbar}{i}\frac{\partial}{\partial q} \right) \langle p| q\rangle \right)
= T p \langle p| q\rangle + \frac{\hbar}{i}\frac{\partial}{\partial q} T \langle p| q\rangle = 0,
\label{towards23a}
\end{equation}
where we used that differentiation with respect to $q$ commutes with applying $T$ (which only affects the functional dependence on $p$). Using that $\langle p | q \rangle = T^{-1}\langle Q | q \rangle$ (Eq.\ (\ref{pq})) and $T\langle p | q \rangle = \langle Q | q \rangle$ (Eq.\ (\ref{NB1-22})),
we can rewrite Eq.\ (\ref{towards23a}) as:\footnote{There is a sign error in NB1, sec.\ 2, Eq.\ 23a: $- T x T^{-1}$ should be $T x T^{-1}$.}
\begin{equation}
\left( T p T^{-1} + \frac{\hbar}{i}\frac{\partial}{\partial q} \right) \langle Q | q\rangle = 0. \quad [{\rm NB1, \, sec.\ 2, \, Eq.\ 23a}]
\label{NB1-23a}
\end{equation}

Similarly, applying $T$ to the left-hand side of Eq. (\ref{NB1-19b}) [NB1, Eq.\ 19b], we find:
\begin{equation}
T \left(  \left( \frac{\hbar}{i}\frac{\partial}{\partial p} + q \right) \langle p| q\rangle \right)
= T \frac{\hbar}{i}\frac{\partial}{\partial p} \langle p| q\rangle 
+ q T \langle p| q\rangle = 0,
\label{towards23b}
\end{equation}
where we used that multiplication by $q$ commutes with applying $T$. Once again using that $\langle p | q \rangle = T^{-1}\langle Q | q \rangle$ and $T\langle p | q \rangle = \langle Q | q \rangle$, 
we can rewrite 
this as:\footnote{There is a sign error in NB1, sec.\ 2, Eq.\ 23b: $- y$ should be $y$. The sign errors in Eqs.\ \eqref{NB1-23a}-\eqref{NB1-23b} confused Heisenberg, who was relying on {\it Neue Begr\"undung} I for the mathematical part of his uncertainty paper (see note \ref{beller}). He wrote to Jordan to ask for clarification:  
\begin{quotation}
Today just a quick question, since I have been trying in vain, enduring persistent fits of rage, to derive your Eq.\ (23a and b) from (19a and b) in your transformation paper. According to my certainly not authoritative opinion it should be $\rho = e^{+ xy/\varepsilon}$ and not  $\rho = e^{- xy/\varepsilon}$, for out of $(x + \varepsilon \, \partial/\partial y) \, \rho(x,y)$ [NB1, Eq.\ 19a, our Eq.\ \eqref{NB1-19a}] I always get ---God be darned---$(+TxT^{-1} + \varepsilon \, \partial/\partial y) \, \varphi(x,y) =0$ [NB1, Eq.\ 23a, our Eq.\ \eqref{NB1-23a}]. Now it's possible that I am doing something nonsensical with these constantly conjugated quantities ($\tilde{F}$, $F^*$, $F^\dagger$: read: $F$-blurry, $F$-ill, and $F$-deceased), but I don't understand anything anymore. Since, however, the quantity $\rho(x,y)$ forms the basis of my mathematics, I am kindly asking you for clarification of the sign (Heisenberg to Jordan, March 17, 1927, AHQP).
\end{quotation}
Jordan's response apparently has not been preserved but from another letter from Heisenberg to Jordan a week later, we can infer that Jordan wrote back that the expression for $\rho(x,y)$ in {\it Neue Begr\"undung} I is correct but that there are sign errors in Eqs.\ (23ab). In the meantime Heisenberg had submitted his uncertainty paper and replied: ``I now fully agree with your calculations and will change my calculations accordingly in the proofs" (Heisenberg to Jordan, March 24, 1927, AHQP).\label{F-verschwommen}}
\begin{equation}
\left( T \frac{\hbar}{i}\frac{\partial}{\partial p} T^{-1} + q  \right) \langle Q | q\rangle = 0. \quad [{\rm NB1, \, sec.\ 2, \, Eq.\ 23b}]
\label{NB1-23b}
\end{equation}

Eqs.\ (\ref{NB1-23a}) and (\ref{NB1-23b}) [NB1, Eqs.\ 23ab] gave Jordan a representation of the quantities $\hat{p}$ and $\hat{q}$ in the $Q$-basis. The identification of $\hat{p}$ in the $Q$-basis is straightforward. The quantity $p$ in Eq.\ (\ref{NB1-19a}) [NB1, Eq.\ 19a] turns into the quantity $T p T^{-1}$ in Eq.\ (\ref{NB1-23a}), [NB1, Eq.\ 23a]. This is just what Jordan had come to expect on the basis of his earlier use of canonical transformations (see Section 2.2 below). The identification of  $\hat{q}$ in the $Q$-basis is a little trickier. Eq.\ (\ref{NB1-19b}) [NB1, Eq.\ 19b] told Jordan that the position operator in the original $p$-basis is $- (\hbar/i) \, \partial/\partial p$ (note the minus sign). This quantity turns into $- T \, (\hbar/i) \, \partial/\partial p \, T^{-1}$ in Eq.\ (\ref{NB1-23b}) [NB1, Eq.\ 23b]. This then should be the representation of $\hat{q}$ in the new $Q$-basis, as Jordan stated right below this last equation: ``With respect to ({\it in Bezug auf}\,) the fixed chosen quantity [$\hat{Q}$] every other quantity [$\hat{q}$] corresponds to an operator [$- T \, (\hbar/i) \, \partial/\partial p \, T^{-1}$]" \citep[p.\ 815]{Jordan 1927b}.\footnote{Because of the sign error in Eq.\ (\ref{NB1-23b}) [NB1, Eq.\ 23b], Jordan set $\hat{q}$ in the $Q$-basis equal to $T ((\hbar/i) \, \partial/\partial p) T^{-1}$.}

With these representations of his quantum-mechanical quantities $\hat{p}$ and $\hat{q}$, Jordan could now define their addition and multiplication through the corresponding addition and multiplication of the differential operators representing these quantities.

Jordan next step was to work out what the differential operators $T p T^{-1}$ and $- T (\hbar/i) \, \partial/\partial p T^{-1}$, representing $\hat{p}$ and $\hat{q}$ in the $Q$-basis, are in the special case that $\hat{Q}=\hat{q}$. In that case, Eqs.\ (\ref{NB1-23a}) and (\ref{NB1-23b}) [NB1, Eqs.\ 23ab] turn into:
\begin{equation}
\left( T p T^{-1} + \frac{\hbar}{i}\frac{\partial}{\partial q} \right) \langle q' | q\rangle = 0,
\label{NB1-23a-Q=q}
\end{equation}
\begin{equation}
\left( T \frac{\hbar}{i}\frac{\partial}{\partial p} T^{-1} + q  \right) \langle q' | q\rangle = 0. 
\label{NB1-23b-Q=q}
\end{equation}
On the other hand, $\langle q'| q \rangle = \delta(q' - q)$. So $\langle q'| q \rangle$ trivially satisfies:
\begin{equation}
\left( \frac{\hbar}{i}\frac{\partial}{\partial q'} + \frac{\hbar}{i}\frac{\partial}{\partial q} \right) \langle q' | q\rangle  = 0, \quad \quad [{\rm NB1, sec.\ 2, Eq.\ 24a}] \label{NB1-24a}
\end{equation}
\begin{equation}
\left( - q' + q  \right) \langle q' | q\rangle  = 0.  \quad \quad \quad \quad [{\rm NB1, sec.\ 2, Eq.\ 24b}] \label{NB1-24b}
\end{equation}

Comparing Eqs.\ (\ref{NB1-24a})--(\ref{NB1-24b}) with Eqs.\ (\ref{NB1-23a-Q=q})--(\ref{NB1-23b-Q=q}), we arrive at
\begin{equation}
T p T^{-1} \langle q' | q\rangle = \frac{\hbar}{i}\frac{\partial}{\partial q'} \langle q' | q\rangle,
\label{IDmomentum0}
\end{equation}
\begin{equation}
- T \frac{\hbar}{i}\frac{\partial}{\partial p} T^{-1} \langle q' | q\rangle = q' \langle q' | q\rangle.
\label{IDposition0}
\end{equation}
Eq.\ (\ref{IDmomentum0}) suggests that $T p T^{-1}$, the momentum  $\hat{p}$ in the $q$-basis acting on the $q'$ variable, is just $(\hbar/i) \, \partial/\partial q'$. Likewise, Eq.\ (\ref{IDposition0}) suggests that $- T \, (\hbar/i) \, \partial/\partial p \, T^{-1}$, the position  $\hat{q}$ in the $q$-basis acting on the $q'$ variable, is just multiplication by $q'$. As Jordan put it in a passage that is hard to follow because of his confusing notation:
\begin{quotation}
Therefore, as a consequence of (24) [our Eqs.\ (\ref{NB1-24a})--(\ref{NB1-24b})], the operator $x$ [multiplying by $q'$ in our notation] is assigned  ({\it zugeordnet}) to the quantity [{\it Gr\"osse}] $Q$ itself  [$\hat{q}$ in our notation]. One sees furthermore that the operator $\varepsilon \, \partial/\partial x$ [$(\hbar/i) \, \partial/\partial q'$ in our notation] corresponds to the momentum $P$ [$\hat{p}$]  belonging to $Q$ [$\hat{q}$] \citep[p.\ 815]{Jordan 1927b}.\footnote{We remind the reader that Jordan used the term `operator' [{\it Operator}] {\it not} for an  operator acting in an abstract Hilbert space but for the differential operators $(\hbar/i) \, \partial/\partial x$ and (multiplying by) $x$ and for combinations of them.}
\end{quotation}
It is by this circuitous route that Jordan arrived at the usual functional interpretation of coordinate and momentum operators in the Schr\"odinger formalism. \citet[pp.\ 815--816]{Jordan 1927b} emphasized that the association of $(\hbar/i) \, \partial/\partial q$ and $q$ with $\hat{p}$ and $\hat{q}$ can easily be generalized. Any quantity ({\it Gr\"osse}) obtained through multiplication and addition of $\hat{q}$ and $\hat{p}$ is associated with the corresponding combination of differential operators $q$ and $(\hbar/i) \, \partial/\partial q$.

Jordan's argument as it stands fails. We cannot conclude that  two operations are identical from noting that they give the same result when applied to one special case, here the delta function  $\langle q'| q \rangle = \delta(q' - q)$ (cf.\ Eqs.\ (\ref{IDmomentum0})--(\ref{IDposition0})). We need to show that they give identical results when applied to an {\it arbitrary} function. We can easily remedy this flaw in Jordan's argument, using only the kind of manipulations he himself used at this point (though we will do so in modern notation). We contrast this proof in the spirit of Jordan with a modern proof showing that Eqs.\ (\ref{NB1-23a}) and (\ref{NB1-23b}) imply that $\hat{p}$ and $\hat{q}$, now understood in the spirit of von Neumann as operators acting in an abstract Hilbert space, are represented by $(\hbar/i) \, \partial/\partial q$ and $q$, respectively, in the $q$-basis. The input for the proof  {\it \`{a} la} Jordan are his postulates and the identification of the differential operators representing momentum and position  in the $Q$-basis as $T p T^{-1}$ and $- T \, (\hbar/i) \, \partial/\partial p \, T^{-1}$, respectively (cf.\ our comments following Eq.\ (\ref{NB1-23b})). The input for the proof {\it \`{a} la} von Neumann are the inner-product structure of Hilbert space and the spectral decomposition of the operator $\hat{p}$. Of  course, \citet{von Neumann 1927a} only introduced these elements {\it after} Jordan's {\it Neue Begr\"undung} I.

Closely following Jordan's approach, we can show that Eqs.\ (\ref{NB1-23a}) and (\ref{NB1-23b}) [NB1, Eqs.\ 23ab] imply that, for arbitrary functions $F(Q)$, if $Q$ is set equal to $q$,
\begin{equation}
(T p T^{-1}F)(q)  =  \frac{\hbar}{i} \frac{\partial}{\partial q}F(q), 
\label{IDmomentum}
\end{equation}
\begin{equation}
\left(- T \frac{\hbar}{i} \frac{\partial}{\partial p} T^{-1}F \right)(q)  =  q F(q). 
\label{IDposition}
\end{equation}
Since $F$ is an arbitrary function, the problem we noted with Eqs. (\ref{IDmomentum0})--(\ref{IDposition0}) is solved. Jordan's identification of the differential operators representing momentum and position in the $q$-basis does follow from Eqs.\ (\ref{IDmomentum})--(\ref{IDposition}). 

To derive Eq.\ (\ref{IDmomentum}), we apply $T$, defined in  Eq.\ (\ref{NB1-21}), to $p \, (T^{-1}F)(p)$. We then use the definition  of $T^{-1}$ in  Eq.\ (\ref{T^-1F(p)}) to write $(T p T^{-1}F)(Q)$ as:
\begin{eqnarray}
(T p T^{-1}F)(Q) & = & \int \langle Q |  p  \rangle \;  p \; (T^{-1}F)(p) \; dp \nonumber \\
 & = & \int \langle Q | p \rangle \; p 
\left[ \int \langle p | Q' \rangle F(Q') dQ'  \right] dp \nonumber \\
 & = & \int \!\!\! \int \langle Q | p \rangle \; p \; \langle p | Q' \rangle F(Q') \; dp \, dQ'. \label{IDmomentum1}
\end{eqnarray}
We now set $\hat{Q}=\hat{q}$, use Eq.\ (\ref{NB1-19a}) to substitute $- (\hbar/i) \,\partial/\partial q' \, \langle p| q' \rangle$ for $p \langle p| q' \rangle$, and perform a partial integration:
\begin{eqnarray}
(T p T^{-1}F)(q) & = & \int \!\!\! \int \langle q | p \rangle \; p \; \langle p | q' \rangle F(q') \; dp \, dq'  \nonumber \\
 & = & \int \!\!\! \int \langle q | p \rangle \left( - \frac{\hbar}{i}\frac{\partial}{\partial q'} 
 \langle p | q' \rangle \right)  F(q') \; dp \, dq'   \nonumber \\
 & = & \int \!\!\! \int \langle q | p \rangle  
 \langle p | q' \rangle \frac{\hbar}{i}\frac{dF(q')}{dq'}   \; dp \, dq'. \label{IDmomentum2}
\end{eqnarray}
On account of completeness and orthogonality (see Eq.\ (\ref{completeness/orthogonality}) [NB1, Eqs.\ 14--17]), the right-hand side reduces to $(\hbar/i) \, F'(q)$. This concludes the proof of Eq.\ (\ref{IDmomentum}). 

To derive Eq.\ (\ref{IDposition}), we similarly apply $T$ to  $- (\hbar/i) \, \partial/\partial p \, (T^{-1}F)(p)$:
\begin{eqnarray}
\left(- T \frac{\hbar}{i} \frac{\partial}{\partial p} T^{-1}F \right)(Q) & = & - \int \langle Q |  p  \rangle \;  \frac{\hbar}{i} \frac{\partial}{\partial p} \,  (T^{-1}F)(p)  \, dp \nonumber \\
 & = & - \int \langle Q | p \rangle \; \frac{\hbar}{i} \frac{\partial}{\partial p} 
\left[ \int \langle p | Q' \rangle F(Q') dQ'  \right] dp \nonumber \\
 & = & - \int \!\!\! \int \langle Q | p \rangle \; \frac{\hbar}{i} \frac{\partial}{\partial p} \; \langle p | Q' \rangle F(Q') \; dp \, dQ'. \label{IDposition1}
\end{eqnarray}
We now set $\hat{Q}=\hat{q}$ and use Eq.\ (\ref{NB1-19b}) to substitute $q'  \langle p| q' \rangle$ for 
$- (\hbar/i) \, \partial/\partial p \, \langle p| q' \rangle$:
\begin{equation}
\left(- T \frac{\hbar}{i} \frac{\partial}{\partial p} T^{-1}F \right)(q)
= \int \!\!\! \int \langle q | p \rangle \, q'  \,  \langle p | q' \rangle \, F(q') \, dp \, dq' = q \, F(q),
\label{IDposition2}
\end{equation}
where in the last step we once again used completeness and orthogonality. This concludes the proof of Eq.\ (\ref{IDposition}). 

We now turn to the modern proofs. It is trivial to show that the representation of the position operator $\hat{q}$ in the $q$-basis is simply multiplication by the eigenvalues $q$. Consider an arbitrary eigenstate $|q \rangle$ of position with eigenvalue $q$, i.e., $\hat{q} \, |q \rangle = q \, |q \rangle$. It follows that  $\langle Q |  \, \hat{q} \, | q \rangle = q \langle Q | q \rangle$, where $| Q \rangle$ is an arbitrary eigenvector of an arbitrary Hermitian operator $\hat{Q} = \hat{Q}^\dagger$ with eigenvalue $Q$. The complex conjugate of this last relation,
\begin{equation}
\langle q |  \, \hat{q} \, | Q \rangle = q \, \langle q  | Q \rangle,
\end{equation}
is just the result we wanted prove.

It takes a little more work to show that Eq.\ (\ref{NB1-23a}) [NB1, Eq.\ 23a] implies that the representation of the momentum operator $\hat{p}$ in the $q$-basis is $(\hbar/i) \, \partial/\partial q$. Consider Eq.\ (\ref{IDmomentum1}) for the special case $F(Q) = \langle Q | q \rangle$:
\begin{equation}
T p T^{-1} \langle Q | q \rangle = \int \!\!\! \int \langle Q | p \rangle \; p \; \langle p | Q' \rangle \langle Q' | q \rangle \; dp \, dQ'.
\end{equation}
Recognizing the spectral decomposition $\int dp \, p \, |p\rangle\langle p|$ of $\hat{p}$ in this equation, we can rewrite it as:
\begin{equation}
T p T^{-1} \langle Q | q \rangle =
 \int \langle Q | \, \hat{p} \, | Q' \rangle \langle Q' | q \rangle \, dQ' = \langle Q | \, \hat{p} \, | q \rangle,
\end{equation}
where in the last step we used the resolution of unity, $\hat{1} = \int dQ' \, | Q' \rangle \langle Q' |$.
Eq.\ (\ref{NB1-23a}) tells us that
\begin{equation}
T p T^{-1} \langle Q | q \rangle = - \frac{\hbar}{i} \frac{\partial}{\partial q} \langle Q | q \rangle.
\end{equation}
Setting the complex conjugates of the right-hand sides of these last two equations equal to one another, we arrive at:
\begin{equation}
\langle q | \, \hat{p} \, | Q \rangle = \frac{\hbar}{i} \frac{\partial}{\partial q} \langle q | Q \rangle,
\label{p in the q basis}
\end{equation}
which is the result we wanted to prove. Once again, the operator $\hat{Q}$ with eigenvectors $|Q \rangle$ is arbitrary. If $\hat{Q}$ is the Hamiltonian and $\hat{q}$ is a Cartesian coordinate, $\langle q | Q \rangle$ is just a Schr\"odinger energy eigenfunction.

With these identifications of $\hat{p}$ and $\hat{q}$ in the $q$-basis we can finally show that Jordan's new definition of conjugate variables in Eq.\ (\ref{NB1-18}) [NB1, Eq.\ 18] reduces to the standard definition, $[\hat{p}, \hat{q}] = \hbar/i$, at least for quantities with completely continuous spectra. Letting $[(\hbar/i) \,\partial/\partial q, q]$ act on an arbitrary function $f(q)$, one readily verifies that the result is $(\hbar/i) \, f(q)$. Given the association of $(\hbar/i) \, \partial/\partial q$ and $q$ with the quantities $\hat{p}$ and $\hat{q}$ that has meanwhile been established, it follows that these quantities indeed satisfy the usual commutation relation
\begin{equation}
[\hat{p}, \hat{q}] = \hat{p} \, \hat{q} - \hat{q} \, \hat{p} = \frac{\hbar}{i}. \quad [{\rm NB1, Eq.\ 25}]
\label{NB1-25}
\end{equation}

This concludes Part I (consisting of secs.\ 1--2) of {\it Neue Begr\"undung} I. Jordan wrote:
\begin{quotation}
This is the content of the new theory. The rest of the paper will be devoted, through a mathematical discussion of these differential equations [NB1, Eqs.\ 23ab, our Eqs.\ (\ref{NB1-23a})--(\ref{NB1-23b}), and similar equations for  other quantities], on the one hand, to proving that our postulates are mathematically consistent [{\it widerspruchsfrei}\,] and, on the other hand, to showing that the earlier forms [{\it Darstellungen}] of quantum mechanics  are contained in our theory \citep[p.\ 816]{Jordan 1927b}.
\end{quotation}
In this paper we focus on the first of these tasks, which amounts to providing a realization of the postulates discussed in this secton. 

\subsection{Canonical transformations in classical mechanics, the old quantum theory and matrix mechanics}

Given the central role of canonical transformations in {\it Neue Begr\"undung}, we insert a brief subsection to review the use of canonical transformations in the developments leading up to it.\footnote{This subsection is based on \citet[sec.\ 2]{Duncan and Janssen 2009}.} Canonical transformations in classical physics are transformations of the position and conjugate momentum variables $(q,p)$ that preserve the form of Hamilton's equations,
\begin{equation}
\dot{q} = \frac{\partial H(p,q)}{\partial p}, \;\;\;\; \dot{p} = - \frac{\partial H(p,q)}{\partial q}.
\label{Ham eqns}
\end{equation}
Following \citet[p.\ 810]{Jordan 1927b} in {\it Neue Begr\"undung} I, we assume that the system is one-dimensional. For convenience, we assume that the Hamiltonian $H(p,q)$ does not explicitly depend on time. The canonical transformation to new coordinates and momenta $(Q, P)$ is given through a {\it generating function}, which is a function of one of the old and one of the new variables. For a generating function of the form $F(q, P)$, for instance,\footnote{In the classification of \citet[p.\ 373, table 9.1]{Goldstein}, this corresponds to a generating function of type 2, $F_2(q,P)$. The other types depend on $(q,Q)$, $(p,Q)$, or $(p,P)$. Which type one chooses is purely a matter of convenience and does not affect the physical content.\label{types}} we find the equations for the canonical transformation 
$(q,p) \rightarrow (Q,P)$ by solving the equations
\begin{equation}
p = \frac{\partial F(q,P)}{\partial q}, \;\;\;\; Q = \frac{\partial F(q,P)}{\partial P}
\label{gen func}
\end{equation}
for $Q(q,p)$ and $P(q,p)$. This transformation preserves the form of Hamilton's equations:\footnote{For elementary discussion, see, e.g., \citet[Pt.\ 2, sec.\ 5.1]{Duncan and Janssen 2007}.}
\begin{equation}
\dot{Q} = \frac{\partial \overline{H}(P,Q)}{\partial P}, \;\;\;\; \dot{P} = - \frac{\partial \overline{H}(P,Q)}{\partial Q},
\end{equation}
where the Hamiltonians $H(p,q)$ and $\overline{H}(P,Q)$ are numerically equal to one another but given by different functions of their respective arguments. One way to solve the equations of motion is to find a canonical transformation such that, in terms of the new variables, the Hamiltonian depends only on momentum, $\overline{H}(P,Q) = \overline{H}(P)$. Such variables are called {\it action-angle variables} and the standard notation for them is $(J, w)$. The basic quantization condition of the old quantum theory of Bohr and Sommerfeld restricts the value of a set of action variables for the system under consideration to integral multiples of Planck's constant, $J=nh$ ($n = 0, 1, 2, \ldots$). Canonical transformations to action-angle variables thus played a central role in the old quantum theory. With the help of them, the energy spectrum of the system under consideration could be found. 

In classical mechanics, canonical transformations preserve the so-called Poisson bracket, $\{ p, q \} = 1$. For any two  phase-space functions $G(p,q), H(p,q)$ of the pair of canonical variables $(p,q)$, the Poisson bracket is defined as
   \begin{equation}
   \{ G(p,q), H(p,q) \} \equiv \frac{\partial G(p,q)}{\partial p}\frac{\partial H(p,q)}{\partial q} - \frac{\partial H(p,q)}{\partial p}\frac{\partial G(p,q)}{\partial q}.
   \end{equation}
For $G(p,q) = p$ and $H(p,q) = q$, this reduces to $\{ p, q \} = 1$. We now compute the Poisson bracket $\{ P, Q \}$ of a new pair of canonical variables related to $(p, q)$ by the generating function $F(q,P)$ as in Eq.\ (\ref{gen func}):  
  \begin{equation}
  \{ P(p,q), Q(p,q) \} = \frac{\partial P(p,q)}{\partial p}\frac{\partial Q(p,q)}{\partial q} - \frac{\partial Q(p,q)}{\partial p}\frac{\partial P(p,q)}{\partial q}.
\label{poisson brackets invariance}
\end{equation}

By the usual chain rules of partial differentiation, we have
\begin{equation}
 \left.  \frac{\partial Q}{\partial p}\right|_{q} = \left. \frac{\partial^{2}F}{\partial P^{2}}\right|_{q}
   \left. \frac{\partial P}{\partial p} \right|_{q},
\end{equation}
\begin{equation}
 \left.  \frac{\partial Q}{\partial q}\right|_{p} =  \frac{\partial^{2}F}{\partial q\partial P} +  \left. \frac{\partial^{2}F}{\partial P^{2}}\right|_{q}  \left. \frac{\partial P}{\partial q}\right|_{p}.
\end{equation}
Substituting these two expressions into Eq.\ (\ref{poisson brackets invariance}), we find
\begin{eqnarray}
 \{ P(p,q), Q(p,q) \} &=& \left. \frac{\partial P}{\partial p}\right|_{q}  \left(
  \frac{\partial^{2}F}{\partial q\partial P} +  \left. \frac{\partial^{2}F}{\partial P^{2}}\right|_{q}  \left. \frac{\partial P}{\partial q}\right|_{p}
 \right) \nonumber \\
& & 
 \;\;\;\; \;\;\;\;  \;\;\;\;  \;\;\;\;  \;\;\;\;  \;\;\;\;  \;\;\;\;  -  \left(
  \left. \frac{\partial^{2}F}{\partial P^{2}}\right|_{q}
   \left. \frac{\partial P}{\partial p} \right|_{q}
 \right) \left. \frac{\partial P}{\partial q}\right|_{p} \nonumber \\
 &=& \left. \frac{\partial^{2}F}{\partial q\partial P}\frac{\partial P}{\partial p}\right|_{q}.
   \end{eqnarray}
The final line is identically equal to 1, as
   \begin{equation}
   \frac{\partial^{2}F}{\partial q\partial P} = \left. \frac{\partial p}{\partial P}\right|_{q} = \left( \left. \frac{\partial P}{\partial p}\right|_{q} \right)^{-1}.
   \end{equation}
This shows that the Poisson bracket $\{ p, q \} = 1$ is indeed invariant under canonical transformations.

In matrix mechanics a canonical transformation is a transformation of the matrices $(q,p)$ to new matrices  $(Q,P)$ preserving the canonical commutation relations 
\begin{equation}
[p, q] \equiv pq - qp = \frac{\hbar}{i}
\label{commutator}
\end{equation}
that replace the Poisson bracket $\{ p, q \} = 1$ in quantum mechanics. Such transformations are of the form
\begin{equation}
P = TpT^{-1}, \;\;\; Q = TqT^{-1},  \;\;\; \overline{H} = THT^{-1},
\end{equation}
where $\overline{H}$ is obtained by substituting $TpT^{-1}$ for $p$ and $TqT^{-1}$ for $q$ in the operator $H$ given as a function $p$ and $q$. One easily recognizes that this transformation does indeed preserve the form of the commutation relations (\ref{commutator}): $[P, Q] = \hbar/i$.
Solving the equations of motion in matrix mechanics boils down to finding a transformation matrix $T$ such that the new Hamiltonian  $\overline{H}$ is diagonal. The diagonal elements, $\overline{H}_{mm}$, then give the (discrete) energy spectrum.

In two papers before {\it Neue Begr\"undung}, \citet{Jordan 1926a, Jordan 1926b} investigated the relation between the matrices $T$ implementing canonical transformations in matrix mechanics and generating functions in classical mechanics. He showed that the matrix $T$ corresponding to a generating function of the form\footnote{In the classification of Goldstein {\it et al.} (cf.\ note \ref{types}), this corresponds to a generating function of type 3, $F_3(p,Q)$.}
\begin{equation}
F(p,Q) = \sum_n f_n(p) g_n(Q),
\label{gen func 1}
\end{equation}
is given by 
\begin{equation}
T(q,p) = \exp{\frac{i}{\hbar} \left\{  (p,q) - \sum_n (f_n(p), \, g_n(q)) \right\}},
\label{Jordan T}
\end{equation}
where the notation $(.\,,.)$ in the exponential signals an ordering such that, when the exponential is expanded, all $p$'s are put to the left of all $q$'s in every term of the expansion.\footnote{See \citet[pp.\ 355--356]{Duncan and Janssen 2009} for a reconstruction of Jordan's proof of this result.} 

When he wrote {\it Neue Begr\"undung} I in late 1926, Jordan was thus steeped in the use of canonical transformations, both in classical and in quantum physics. 
When Kuhn asked Jordan about his two papers on the topic \citep{Jordan 1926a, Jordan 1926b} in an interview for the AHQP project, Jordan told him:  
\begin{quotation}
Canonical transformations in the sense of Hamilton-Jacobi were \ldots\ our daily
bread in the preceding years, so to tie in the new results with those as closely as
possible---that was something very natural for us to try (AHQP interview with Jordan, session 4, p.\ 11).
\end{quotation}

\subsection{The realization of Jordan's postulates: probability amplitudes and canonical transformations}

At the beginning of sec.\ 4 of {\it Neue Begr\"undung} I, ``General comments on the differential equations for the amplitudes," Jordan announced:
\begin{quotation}
To prove that our postulates are mathematically consistent, we want to give a new foundation of the theory---independently from the considerations in sec.\ 2---based on the differential equations which appeared as end results there \citep[p.\ 821]{Jordan 1927b}.
\end{quotation}

He began by introducing the canonically conjugate variables $\hat{\alpha}$ and $\hat{\beta}$, satisfying, by definition, the commutation relation $[\hat{\alpha}, \hat{\beta}] = \hbar/i$. They are related to the basic variables $\hat{p}$ and $\hat{q}$, for which the probability amplitude, according to Jordan's postulates, is $\langle p | q \rangle = e^{-ipq/\hbar}$ (see Eq.\ (\ref{NB1-18})), via
\begin{equation}
\hat{\alpha} = f(\hat{p},\hat{q}) =  T\hat{p}T^{-1},
\label{NB1-alpha}
\end{equation}
\begin{equation}
 \hat{\beta}  = g(\hat{p},\hat{q}) =  T\hat{q}T^{-1}.
 \label{NB1-beta}
\end{equation}
with $T = T(\hat{p}, \hat{q})$ [NB1, sec.\ 4, Eq.\ 1].\footnote{The numbering of equations in {\it Neue Begr\"undung} I starts over in sec.\ 3, the first section of Part Two ({\it II. Teil}\,), and then again in sec.\ 4. Secs.\ 6 and 7, finally, have their own set of equation numbers.\label{numbering-NB1}} Note that the operator $T(\hat{p}, \hat{q})$ defined here is different from the operator $T \ldots = \int  dp \, \langle Q | p \rangle \ldots$ defined in sec.\ 2 (see Eq.\ (\ref{NB1-21}), Jordan's Eq.\ (21)). The $T(\hat{p}, \hat{q})$ operator defined  in sec.\ 4 is a similarity transformation operator implementing the canonical transformation from the pair ($\hat{p},\hat{q}$) to the pair ($\hat{\alpha},\hat{\beta}$). We will see later that there is an important relation between the $T$ operators defined in secs.\ 2 and 4.

Jordan now posited the fundamental differential equations for the probability amplitude $\langle q| \beta\rangle$ in his theory:\footnote{He introduced separate equations for the {\it Erg\"anzungsamplitude} [NB1, sec.\ 4, Eqs.\ 3ab] (see Eqs. (\ref{NB1-4-2a})--(\ref{NB1-4-2b}) below). We  ignore these additional equations for the moment but  will examine them for some special cases in sec.\  2.4.}
\begin{equation}
\left\{ f\left(\frac{\hbar}{i}\frac{\partial}{\partial q},q\right)+\frac{\hbar}{i}\frac{\partial}{\partial\beta}  \right\}
\langle q| \beta\rangle = 0,
 \quad \quad [{\rm NB1, \, sec.\ 4, \, Eq.\  2a}]
\label{NB1-2a}
\end{equation}
\begin{equation}
\left\{ g\left(\frac{\hbar}{i}\frac{\partial}{\partial q},q\right)-\beta \right\}
\langle q| \beta\rangle = 0.
\quad \quad [{\rm NB1,  \, sec.\ 4, \, Eq.\  2b}]
\label{NB1-2b}
\end{equation}
These equations have the exact same form as Eqs.\ (\ref{NB1-23a})--(\ref{NB1-23b}) [NB1, sec.\ 2, Eqs.\ 23ab], with the understanding that the operator $T$ is defined differently. As Jordan put it in the passage quoted above, he took the equations that were the end result in sec.\ 2 as his starting point in sec.\ 4.

Before turning to Jordan's discussion of these equations, we show that they are easily recovered in the modern Hilbert space formalism. The result of the momentum operator $\hat{\alpha}$ in Eq.\ (\ref{NB1-alpha}) acting on eigenvectors $| \beta\rangle$ of its conjugate operator $\hat{\beta}$ in Eq.\  (\ref{NB1-beta}) is, as we saw in Section 2.1:\footnote{The complex conjugate of Eq.\ (\ref{p in the q basis}) can be written as 
$$\langle Q | \, \hat{p} \, | q \rangle = - \frac{\hbar}{i} \frac{\partial}{\partial q} \langle Q | q \rangle 
= \langle Q | \left(- \frac{\hbar}{i} \frac{\partial}{\partial q} \right) | q \rangle.
$$
Since this holds for arbitrary $|Q \rangle$, it follows that $\hat{p} \, | q \rangle = 
- (\hbar/i) (\partial/\partial q) | q \rangle$. This will be true for any pair of conjugate variables.}
     \begin{equation}
     \hat{\alpha}|\beta\rangle =
     -\frac{\hbar}{i}\frac{\partial}{\partial\beta}|\beta\rangle.
     \end{equation}
Taking the inner product of these expressions with $|q \rangle$ and using that $\hat{\alpha} = f(\hat{p},\hat{q})$, we find that
\begin{equation}
- \frac{\hbar}{i}\frac{\partial}{\partial\beta}\langle q|\beta\rangle
=  \langle q|\hat{\alpha}|\beta\rangle
= \langle q| f(\hat{p},\hat{q})|\beta\rangle.
\label{2a-1}
\end{equation}
Since $\hat{p}$ and $\hat{q}$ are represented by the differential operators $(\hbar/i) \, \partial/\partial q$ and $q$, respectively, in the $q$-basis, we can rewrite this as
\begin{equation}
\langle q| f(\hat{p},\hat{q})|\beta\rangle = 
f \left(\frac{\hbar}{i}\frac{\partial}{\partial q},q \right)\langle q|\beta\rangle.
\label{2a-2}
\end{equation}
Combining these last two equations, we arrive at Eq.\ (\ref{NB1-2a}).
Likewise, using that $\hat{\beta}|\beta\rangle =  \beta |\beta\rangle$ and that $\hat{\beta} = g(\hat{p}, \hat{q})$, we can write the inner product $\langle q |  \hat{\beta}  |\beta\rangle$ as 
\begin{equation} 
\langle q |  \hat{\beta}  |\beta\rangle
= \beta \langle q | \beta \rangle
=  g \left(\frac{\hbar}{i}\frac{\partial}{\partial q},q \right) \langle q  |\beta\rangle,
\label{2b-1}
\end{equation}
where in the last step we used the representation of $\hat{p}$ and $\hat{q}$ in the $q$-basis. From this equation we can read off Eq.\ (\ref{NB1-2b}). 

We turn to Jordan's discussion of Eqs.\ (\ref{NB1-2a})--(\ref{NB1-2b}) [NB1, sec.\ 4, Eqs.\ 2ab]. As he pointed out:
\begin{quotation}
As is well-known, of course, one cannot in general simultaneously impose two partial differential equations on one function of two variables. We will prove, however, in sec.\ 5: the presupposition---which we already made---that $\hat{a}$ and $\hat{\beta}$ are connected to $\hat{p}$ and $\hat{q}$ via a canonical transformation (1) [our Eqs.\ (\ref{NB1-alpha})--(\ref{NB1-beta})] is the necessary and sufficient condition for (2) [our Eqs.\ (\ref{NB1-2a})--(\ref{NB1-2b})] to be solvable \citep[p.\ 822; hats added]{Jordan 1927b}.
\end{quotation}
In sec.\ 5, ``Mathematical theory of the amplitude equations," \citet[pp.\ 824--828]{Jordan 1927b} made good on this promise. To prove that the ``presupposition" is sufficient, he used canonical transformations to explicitly construct a simultaneous solution of the pair of  differential equations (\ref{NB1-2a})--(\ref{NB1-2b}) for probability amplitudes (ibid., pp.\ 824--825, Eqs.\ 9--17). He did this in two steps. 
\begin{enumerate}
\item He showed that the sufficient condition for $\langle Q | \beta \rangle$ to be a solution of the amplitude equations in the $Q$-basis, given that $\langle q | \beta \rangle$ is a solution of these equations in the $q$-basis, is that $(\hat{p}, \hat{q})$ and $(\hat{P}, \hat{Q})$ are related by a canonical transformation. 
\item He established a starting point for generating such solutions by showing that a very simple canonical transformation (basically switching $\hat{p}$ and $\hat{q}$) turns the amplitude equations  (\ref{NB1-2a})--(\ref{NB1-2b}) into a pair of equations immediately seen to be satisfied by the amplitude $\langle q|\beta \rangle = e^{iq\beta/\hbar}$. 
\end{enumerate}
With these two steps Jordan had shown that the assumption that $\hat{P}$ and $\hat{Q}$ are related to $\hat{p}$ and $\hat{q}$ through a canonical transformation is indeed a sufficient condition for the amplitude equations (\ref{NB1-2a})--(\ref{NB1-2b}) [NB1, sec.\ 4, Eqs.\ 2ab] to be simultaneously solvable. We will cover this part of Jordan's argument in detail. 

 
The proof that this assumption is necessary as well as sufficient is much more complicated \citep[pp.\ 825--828, Eqs.\ 18--34]{Jordan 1927b}. The mathematical preliminaries presented in sec.\ 3 of {\it Neue Begr\"undung}  I (ibid., pp.\ 816--821) are needed only for this part of the proof in sec.\ 5. We will cover neither this part of sec.\ 5 nor sec.\ 3. 

However, we do need to explain an important result that Jordan derived in sec.\ 5 as a consequence of this part of his proof (ibid., p.\ 828, Eqs.\ 35--40): Canonical transformations $T(\hat{p}, \hat{q})$ as defined above (see Eqs.\ (\ref{NB1-alpha})--(\ref{NB1-beta}) [NB1, sec.\ 4, Eq. 1]), which are differential operators once $\hat{p}$ and $\hat{q}$ have been replaced by their representations $(\hbar/i) \, \partial/\partial q$ and $q$ in the $q$-basis, can be written as integral operators $T$ as defined in sec.\ 2 (see Eq.\  (\ref{NB1-21}) [NB1, sec.\ 2, Eq.\ 21]). 

This result is central to the basic structure of Jordan's theory and to the logic of his {\it Neue Begr\"undung} papers. It shows that Jordan's probability amplitudes do double duty as integral kernels of the operators implementing canonical transformations. As such, Jordan showed, they satisfy the completeness and orthogonality relations required by postulate C (see Eqs. (\ref{completeness})--(\ref{completeness/orthogonality}) [NB1, sec.\ 2, Eqs. 14--17]). To paraphrase the characterization of Jordan's project by Hilbert et al. that we already quoted in Section 1.1, Jordan postulated certain relations between his probability amplitudes in Part One of his paper and then, in Part Two, presented ``a simple analytical apparatus in which quantities occur that satisfy these relations exactly" \citep[p.\ 2]{Hilbert-von Neumann-Nordheim}. These quantities, it turns out, are the integral kernels of canonical transformations. Rather than following Jordan's own proof of this key result, which turns on properties of canonical transformations, we present a modern proof, which turns on properties of Hilbert space and the spectral theorem.

But first we show, closely following Jordan's own argument in sec.\ 5 of {\it Neue Begr\"undung} I, how to construct a simultaneous solution of the differential equations (\ref{NB1-2a})--(\ref{NB1-2b}) [NB1, sec.\ 4, Eqs.\ 2ab] for the amplitudes. Suppose we can exhibit just one case of a canonical transformation $({\hat{p},\hat{q}})\rightarrow (\hat{\alpha},\hat{\beta})$ (Eqs. (\ref{NB1-alpha})--(\ref{NB1-beta}) [NB1, sec.\ 4, Eq. 1]) where the amplitude equations manifestly have a unique simultaneous solution. According to Jordan,\footnote{We will see below that this assumption is problematic.} any other canonical pair can be arrived at from the pair (${\hat{p},\hat{q}}$) via a new transformation function $S({\hat{P},\hat{Q}})$, in the usual way
\begin{equation}
        \hat{p} = S\hat{P}S^{-1},\quad \hat{q} = S\hat{Q}S^{-1}.
\end{equation}
with $S = S(\hat{P}, \hat{Q})$ [NB1, sec.\ 5, Eq.\ 10]. The connection between the original pair ($\hat{\alpha},\hat{\beta}$) and the new pair (${\hat{P},\hat{Q}}$) involves the composite of two canonical transformations  [NB1, sec.\ 5, Eq.\ 11]:
\begin{equation}
\hat{\alpha} = f(\hat{p},\hat{q}) = f(S\hat{P}S^{-1},S\hat{Q}S^{-1}) \equiv F(\hat{P},\hat{Q}),
\label{NB1-5-11a}
\end{equation}
\begin{equation}
 \hat{\beta} = g(\hat{q},\hat{q})  = g(S\hat{P}S^{-1},S\hat{Q}S^{-1})\equiv G(\hat{P},\hat{Q}).
 \label{NB1-5-11b}
\end{equation}

In the new $Q$-basis, the  differential equations (\ref{NB1-2a})--(\ref{NB1-2b}) [NB1, sec.\ 4, Eqs.\ 2ab] for probability amplitudes take the form
\begin{equation}
 \left\{ F\left(\frac{\hbar}{i}\frac{\partial}{\partial Q},Q\right) +  \frac{\hbar}{i}\frac{\partial}{\partial \beta} \right\}  \langle Q|\beta\rangle = 0, \quad [{\rm NB1, \, sec.\ 5, \, Eq.\ 12a}]
 \label{NB1-12a}
\end{equation}
\begin{equation}
\left\{ G\left(\frac{\hbar}{i}\frac{\partial}{\partial Q},Q\right) - \beta \right\}
 \langle Q|\beta\rangle =  0. \quad [{\rm NB1, \, sec.\ 5, \, Eq.\ 12b}]
 \label{NB1-12b}
\end{equation}
Jordan  now showed that
\begin{equation}
\langle Q| \beta\rangle = \left\{S\left(\frac{\hbar}{i}\frac{\partial}{\partial q},q\right)\langle q|\beta\rangle\right\}_{q=Q} \quad [{\rm NB1, \, sec.\ 5, \, Eq.\ 13}]
 \label{NB1-13}
\end{equation}
is a simultaneous solution of the amplitude equations (\ref{NB1-12a})--(\ref{NB1-12b}) in the $Q$-basis if  $\langle q|\beta\rangle$ is a simultaneous solution of the amplitude equations (\ref{NB1-2a})--(\ref{NB1-2b}) in the $q$-basis. Using the operator $S$ and its inverse $S^{-1}$, we can rewrite the latter as\footnote{This step is formally the same as the one that got us from Eqs.\ (\ref{NB1-19a})--(\ref{NB1-19b}) [NB1, sec.\ 2, Eqs.\ 19ab] to Eqs.\ (\ref{NB1-23a})--(\ref{NB1-23b}) [NB1, sec.\ 2, Eqs.\ 23ab].}
\begin{equation}
S \, \left\{ f\left(\frac{\hbar}{i}\frac{\partial}{\partial q},q\right)+\frac{\hbar}{i}\frac{\partial}{\partial\beta}  \right\} \, S^{-1} S \,
\langle q| \beta\rangle = 0,
 \quad \quad [{\rm NB1, \, sec.\ 5, \, Eq.\  14a}]
\label{NB1-14a}
\end{equation}
\begin{equation}
S \, \left\{ g\left(\frac{\hbar}{i}\frac{\partial}{\partial q},q\right)-\beta \right\} \, S^{-1} S \,
\langle q| \beta\rangle = 0,
\quad \quad [{\rm NB1,  \, sec.\ 5, \, Eq.\  14b}]
\label{NB1-14b}
\end{equation}
both taken, as in Eq.\ (\ref{NB1-13}), at $q=Q$. Written more carefully, the first term in curly brackets in  Eq.\ (\ref{NB1-14a}), sandwiched between $S$ and $S^{-1}$, is
\begin{equation}
S \, f\left(\frac{\hbar}{i}\frac{\partial}{\partial q},q\right) \, S^{-1}
= \left\{ S\left(\frac{\hbar}{i}\frac{\partial}{\partial q},q\right) \;
 f\left(\frac{\hbar}{i}\frac{\partial}{\partial q},q\right) \;
 S^{-1}\left(\frac{\hbar}{i}\frac{\partial}{\partial q},q\right) \right\}_{q=Q}.
 \label{14a-1}
\end{equation}
With the help of Eq.\ (\ref{NB1-5-11a}), this can further be rewritten as 
\begin{equation}
 S(P,Q) \, f(P,Q) \, S(P,Q)^{-1}|_{P=\frac{\hbar}{i}\frac{\partial}{\partial Q}} = 
 F\left(\frac{\hbar}{i}\frac{\partial}{\partial Q},Q\right).
 \label{14a-2}
\end{equation}
The second term in curly brackets in  Eq.\ (\ref{NB1-14a}), sandwiched between $S$ and $S^{-1}$, is simply equal to
\begin{equation}
S \, \frac{\hbar}{i}\frac{\partial}{\partial\beta} \, S^{-1} = \frac{\hbar}{i}\frac{\partial}{\partial\beta},
\label{14a-3}
\end{equation}
as $S$ does not involve $\beta$. Using Eqs.\ (\ref{NB1-13}) and (\ref{14a-1})--(\ref{14a-3}), we can rewrite Eq.\ (\ref{NB1-14a}) [NB1, sec.\ 5, Eq.\ 14a] as
\begin{equation}
\left\{ F\left(\frac{\hbar}{i}\frac{\partial}{\partial Q},Q\right) +  \frac{\hbar}{i}\frac{\partial}{\partial \beta} \right\}  \langle Q|\beta\rangle = 0,
\end{equation}
which is just Eq.\ (\ref{NB1-12a}) [NB1, sec.\ 5, Eq.\ 12a]. A completely analogous argument establishes that Eq.\ (\ref{NB1-14b}) [NB1, sec.\ 5, Eq.\ 14b] reduces to Eq.\ (\ref{NB1-12b}) [NB1, sec.\ 5, Eq.\ 12b]. This concludes the proof that $\langle Q|\beta \rangle$ is a solution of the amplitude equations in the new $Q$-basis, if $\langle q|\beta \rangle$, out of which $\langle Q|\beta \rangle$ was constructed with the help of the operator $S$ implementing a canonical transformation, is a solution of the amplitude equations in the old $q$-basis.

As $S$ is completely general, we need only  exhibit a single valid starting point, i.e., a pair ($f,g$)  and an amplitude $\langle q|\beta\rangle$ satisfying the amplitude equations in the $q$-basis (Eqs.\  (\ref{NB1-2a})--(\ref{NB1-2b}) [NB1, sec.\ 4, Eqs.\ 2ab]), to construct general solutions of the amplitude equations in some new $Q$-basis (Eqs.\ (\ref{NB1-12a})--(\ref{NB1-12b}) [NB1, sec.\ 5, Eqs.\ 12ab]). The trivial example of a canonical transformation switching the roles of coordinate and momentum does the trick (cf.\ Eqs.\ (\ref{NB1-5-11a})--(\ref{NB1-5-11b}) [NB1, sec.\ 5, Eq.\ 11]):
\begin{equation}
\hat{\alpha} = f(\hat{p},\hat{q}) = -\hat{q},\;\;\; \beta= g(\hat{p},\hat{q}) = \hat{p}. \quad [{\rm NB1, \, sec.\ 5, \, Eq.\ 15}]
\label{NB1-15}         
\end{equation}
In that case, Eqs.\ (\ref{NB1-2a})--(\ref{NB1-2b}) 
become [{\rm NB1, sec.\ 5, Eq.\ 16}]
\begin{equation}
\left\{ q - \frac{\hbar}{i} \frac{\partial}{\partial \beta}  \right\} \langle q|\beta\rangle = 0,
\label{NB1-16a}         
\end{equation}
\begin{equation}
\left\{ \frac{\hbar}{i} \frac{\partial}{\partial q} - \beta  \right\} \langle q|\beta\rangle = 0.
\label{NB1-16b}         
\end{equation}
Except for the minus signs, these equations are of the same form as the trivial equations (\ref{NB1-19a})--(\ref{NB1-19b})  [NB1, sec.\ 2, Eqs. 19ab] for $\langle p| q \rangle$,  satisfied by the basic amplitude $\langle p| q \rangle = e^{-ipq/\hbar}$. In the case of Eqs.\ (\ref{NB1-16a})--(\ref{NB1-16b}), the solution is:
\begin{equation}
\langle q|\beta\rangle = e^{i\beta q/\hbar}. \quad [{\rm NB1, \, sec.\ 5, \, Eq.\ 17}]
\label{NB1-17}         
\end{equation}
This establishes that  the canonical nature of the transformation to the new variables is a sufficient condition for the consistency (i.e. simultaneous solvability) of the pair of differential equations (\ref{NB1-12a})--(\ref{NB1-12b}) [NB1, sec.\ 5, Eq.\ 12ab] for the probability amplitudes. 

\citet[pp.\ 825--828]{Jordan 1927b} went on to prove the converse, i.e., that the canonical connection is also a necessary  condition for the consistency of Eqs.\ (\ref{NB1-12a})--(\ref{NB1-12b}). This is done, as  Jordan explained at the top of p.\ 827 of his paper, by explicit construction of the operator $S$ (in Eq.\ (\ref{NB1-13})), given the validity of Eqs.\ (\ref{NB1-12a})--(\ref{NB1-12b}). We skip this part of the proof.

\citet[p.\ 828]{Jordan 1927b} then used some of the same techniques to prove a key result in his theory. As mentioned above, we will appeal to the modern Hilbert space formalism and the spectral theorem to obtain this result. Once again consider Eq.\ (\ref{NB1-13}):
\begin{equation}
\langle Q| \beta\rangle = \left\{S\left(\frac{\hbar}{i}\frac{\partial}{\partial q},q\right)\langle q|\beta\rangle\right\}_{q=Q}. \quad [{\rm NB1, \, sec.\ 5, \, Eq.\ 13}]
\label{NB1-13-copy}
\end{equation}
This equation tells us that the differential operator $S((\hbar/i) \, \partial/\partial q, q)$ maps arbitrary states $\langle q|\beta \rangle$ in the $q$-basis (recall that $\hat{\beta}$ can be any operator) onto the corresponding states $\langle Q|\beta \rangle$ in the $Q$-basis. The spectral theorem, which gives us the resolution  $\int dq |q \rangle \langle q|$ of the unit operator, tells us that this mapping can also be written as
\begin{equation}
\langle Q| \beta\rangle = \int dq \, \langle Q| q \rangle  \langle q| \beta\rangle.
\label{completeness-again}
\end{equation}
Schematically, we can write
\begin{equation}
S\left(\frac{\hbar}{i}\frac{\partial}{\partial q},q\right) \ldots = \int dq \, \langle Q| q \rangle \ldots
\label{amplitude/kernel}
\end{equation}
In other words, the probability amplitude $\langle Q| q \rangle$ is the integral kernel for the integral representation of the canonical transformation operator $S((\hbar/i) \, \partial/\partial q, q)$. Using nothing but the properties of canonical transformations and his differential equations for probability amplitudes (Eqs.\ (\ref{NB1-2a})--(\ref{NB1-2b}) [NB1, sec.\ 4, Eqs.\ 2ab]), \citet[p.\ 828]{Jordan 1927b} derived an equation of exactly the same form as Eq.\ (\ref{amplitude/kernel}), which we give here in its original notation:
\begin{equation}
T\left( \varepsilon \, \frac{\partial}{\partial x}, x \right) = \int dx . \varphi(y, x). \ldots \quad [{\rm NB1, \, sec.\ 5, \, Eq.\ 40}]
\label{NB1-40} 
\end{equation}


Jordan claimed that Eqs.\ (\ref{NB1-2a})--(\ref{NB1-2b}) [NB1, sec.\ 4, Eqs.\ 2ab] contain both the time-independent and the time-dependent Schr\"odinger equations as special cases.\footnote{Cf.\ \citet[sec.\ 10, pp.\ 27--29, ``The Schr\"odinger differential equations"]{Hilbert-von Neumann-Nordheim}, discussed briefly by \citet[p.\ 311]{Jammer 1966}.} The time-independent Schr\"odinger equation is a special case of  Eq.\ (\ref{NB1-2b}): 
\begin{quotation}
If in (2b) we take $\beta$ to be the energy $W$, and $g$ to be the Hamiltonian function $H(p, q)$ of the system, we obtain the Schr\"odinger wave equation, which corresponds to the classical Hamilton-Jacobi equation. With (2b) comes (2a) as a second equation. In this equation we need to consider $f$ to be the time $t$ (as a function of $p$ and $q$) \citep[p.\ 822]{Jordan 1927b}.
\end{quotation}
Actually, the variable conjugate to $\hat{H}$ would have to be {\it minus} $\hat{t}$. For $\hat{\alpha} = f(\hat{p}, \hat{q}) = - \hat{t}$ and $\hat{\beta} = g(\hat{p}, \hat{q}) = \hat{H}$ (with eigenvalues $E$), Eqs.\ (\ref{NB1-2a})--(\ref{NB1-2b}) become:
\begin{equation}
\left\{ \hat{t} - \frac{\hbar}{i}\frac{\partial}{\partial E}  \right\}
\langle q| E \rangle = 0,
 \label{NB1-2a'}
\end{equation}
\begin{equation}
\left\{ \hat{H} - E \right\}
\langle q| E \rangle = 0.
\label{NB1-2b'}
\end{equation}
If $\langle q| E \rangle$ is set equal to $\psi_E(q)$, Eq.\ (\ref{NB1-2b'}) is indeed just the time-independent Schr\"odinger equation. 

Jordan likewise claimed that the time-dependent Schr\"o\-din\-ger equation is a special case of Eq.\ (\ref{NB1-2a}) 
\begin{quotation}
if for $\beta$ we choose the time $t$ [this, once again, should be $-t$], for $g$ [minus] the time $t(p,q)$ as function of $p, q$, and, correspondingly, for $f$ the Hamiltonian function $H(p, q)$ \citep[p.\ 823]{Jordan 1927b}.
\end{quotation}
This claim is more problematic. For $\hat{\alpha} = f(\hat{p}, \hat{q}) = \hat{H}$ (eigenvalues $E$) and $\hat{\beta} = g(\hat{p}, \hat{q}) = - \hat{t}$, Eqs.\ (\ref{NB1-2a})--(\ref{NB1-2b}) [NB1, sec.\ 4, Eqs.\ 2ab] become:
\begin{equation}
\left\{ \hat{H} - \frac{\hbar}{i}\frac{\partial}{\partial t}  \right\}
\langle q| t \rangle = 0,
 \label{NB1-2a''}
\end{equation}
\begin{equation}
\left\{ \hat{t} - t \right\}
\langle q| t \rangle = 0.
\label{NB1-2b''}
\end{equation}
If $\langle q| t \rangle$ is set equal to $\psi(q,t)$, Eq.\ (\ref{NB1-2a''}) turns into the time-dependent Schr\"o\-din\-ger equation. However, time is a parameter  in quantum mechanics and {\it not} an operator $\hat{t}$ with eigenvalues $t$ and eigenstates $| t \rangle$.\footnote{There is an extensive literature on this subject. For an introduction to this issue, see, e.g., \citet{Hilgevoord 2002} and the references therein.} 

This also makes  Eqs.\ (\ref{NB1-2a'}) and (\ref{NB1-2b''}) problematic. Consider the former. For a free particle, the Hamiltonian is $\hat{H} = \hat{p}^2/2m$, represented by $((\hbar/i) \partial/\partial q)^2/2m$  in the $q$-basis. The solution of Eq.\ (\ref{NB1-2b'}),
\begin{equation}
\langle q | E \rangle = e^{i \sqrt{2mE} q/\hbar},
\end{equation}
is also a solution of Eq.\ (\ref{NB1-2a'}) as long as we define $\hat{t} \equiv m\hat{q}\hat{p}^{-1}$, as suggested by the relation $q=(p/m)t$, Note, however, that we rather arbitrarily decided on this particular ordering of the non-commuting operators $\hat{p}$ and $\hat{q}$. Using that 
\begin{equation}
\frac{\hbar}{i} \frac{\partial}{\partial q} \, e^{i \sqrt{2mE} q/\hbar} = \sqrt{2mE} \, e^{i \sqrt{2mE} q/\hbar},
\end{equation}
we find that $\hat{t} \, \langle q| E \rangle = m\hat{q}\hat{p}^{-1} \langle q| E \rangle$ is given by:
\begin{equation}
m\hat{q} \left( \frac{\hbar}{i} \frac{\partial}{\partial q} \right)^{-1} e^{i \sqrt{2mE} q/\hbar} = \frac{mq}{\sqrt{2mE}} \, e^{i \sqrt{2mE} q/\hbar}.
\end{equation}
This is indeed equal to $(\hbar/i) \partial/\partial E \,  \langle q| E \rangle$ as required by Eq.\ (\ref{NB1-2a'}):
\begin{equation}
\frac{\hbar}{i} \frac{\partial}{\partial E} \, e^{i \sqrt{2mE} q/\hbar} = \frac{mq}{\sqrt{2mE}} \, e^{i \sqrt{2mE} q/\hbar}.
\end{equation}
So with $\hat{t} \equiv m\hat{q}\hat{p}^{-1}$, both Eq.\ (\ref{NB1-2a}) and Eq.\ (\ref{NB1-2b}) [NB1, sec.\ 4, Eqs.\ 2ab] hold in the special case of a free particle. It is not at all clear, however, whether this will  be true in general.

It is probably no coincidence that we can get Jordan's formalism to work, albeit with difficulty, for a free particle where the energy spectrum is fully continuous. Recall that, in {\it Neue Begr\"undung} I, Jordan restricted himself to quantities with completely continuous spectra. As he discovered when he tried to generalize his formalism to quantities with partly or wholly discrete spectra in {\it Neue Begr\"undung} II, this restriction is not nearly as innocuous as he made it sound in {\it Neue Begr\"undung} I. 

Consider the canonical transformation $\hat{\alpha} = T \hat{p} T^{-1}$ (Eq.\ (\ref{NB1-alpha}) [NB1, sec.\ 4, Eq.\ 1]) that plays a key role in Jordan's construction of the model realizing his postulates. Consider (in modern terms) an arbitrary eigenstate $|p \rangle$ of the operator $\hat{p}$ with eigenvalue $p$, i.e., $\hat{p} |p \rangle = p | p \rangle$. It only takes one line to show that then $T |p \rangle$ is an eigenstate of $\hat{\alpha}$ with the same eigenvalue $p$:
\begin{equation}
\hat{\alpha} \, T |p \rangle =  T \hat{p} T^{-1} T |p \rangle = T \hat{p} |p \rangle = p \, T |p \rangle.
\label{trigger}
\end{equation}
In other words, the operators $\hat{\alpha}$ and $\hat{p}$ connected by the canonical transformation $\hat{\alpha} = T \hat{p} T^{-1}$ have the same spectrum. This simple observation, more than anything else, reveals the limitations of Jordan's formalism. It is true, as Eq.\ (\ref{NB1-2b'}) demonstrates, that his differential equations Eqs.\ (\ref{NB1-2a})--(\ref{NB1-2b}) [NB1, sec.\ 4, Eqs.\ 2ab] for probability amplitudes contain the time-independent Schr\"odinger equation as a special case. However, since the energy spectrum is bounded from below and, in many interesting cases, at least partially discrete, it is impossible to arrive at the time-independent Schr\"odinger equation starting from the trivial equations (\ref{NB1-16a})--(\ref{NB1-16b}) [NB1, sec.\ 5, Eq.\ 16] for the probability amplitude $e^{iq\beta/\hbar}$ between $\hat{q}$ and $\hat{\beta}$---recall that $\hat{\beta} = \hat{p}$ in this case (see Eq.\ (\ref{NB1-15}))---and performing some canonical transformation. As Eq.\ (\ref{trigger}) shows, a canonical transformation cannot get us from  $\hat{p}$'s and $\hat{q}$'s with completely continuous spectra to $\hat{\alpha}$'s and $\hat{\beta}$'s with partly discrete spectra. This, in turn, means that, in many interesting cases (i.e., for Hamiltonians with at least partly discrete spectra), the time-independent Schr\"odinger equation does {\it not} follow from Jordan's postulates. In Jordan's defense one could note at this point that this criticism is unfair as he explicitly restricted himself to quantities with fully continuous spectra in {\it Neue Begr\"undung} I. However, as we will see when we turn to {\it Neue Begr\"undung} II in Section 4, Jordan had to accept in this second paper that the extension of his general formalism to quantities with wholly or partly discrete spectra only served to drive home the problem and did nothing to alleviate it. 

\subsection{The confusing matter of the {\em Erg\"anzungsamplitude}}

In this subsection, we examine the ``supplementary amplitude" ({\it Er\-g\"an\-zungs\-amplitude}) $\psi(x,y)$ that Jordan introduced in {\it Neue Begr\"undung} I in addition to the probability amplitude $\varphi(x,y)$.\footnote{This subsection falls somewhat outside the main line of argument of our paper and can be skipped by the reader without loss of continuity.} Jordan's (1927b, p.\ 813) postulate I 
sets the conditional probability ${\rm Pr}(x|y)$ that $\hat{x}$ has a value between $x$ and $x+dx$ given that $\hat{y}$ has the value $y$ equal to:
   \begin{equation}
   \label{phipsistar}
     \varphi(x,y)\psi^{*}(x,y)dx. \quad \quad [{\rm NB1, \, sec.\ 2, \, Eq.\ 10}]
     \end{equation}
Jordan allowed the eigenvalues $x$ and $y$ to be complex. He stipulated that the ``star" in $\psi^{*}(x,y)$ is to be interpreted in such a way that, when taking the complex conjugate of $\psi(x,y)$, one should retain $x$ and $y$ and {\it not} replace them, as the ``star" would naturally suggest, by their complex conjugates. The rationale for this peculiar rule will become clear below. 

For general complex amplitudes, Eq.\ (\ref{phipsistar}) only makes sense as a positive real probability if the phases of $\varphi(x,y)$ and $\psi^*(x,y)$ exactly compensate, leaving only their positive absolute magnitudes (times the interval $dx$, as we are dealing with continuous quantities). Jordan certainly realized that in cases where the mechanical quantities considered were represented by self-adjoint operators, this duplication was unnecessary.\footnote{For instance, if $\hat{x}$ is a Cartesian coordinate and $\hat{y}$ is the Hamiltonian, the amplitudes $\varphi(x,y) = \psi(x,y)$ are just the Schr\"odinger energy eigenfunctions of the system in coordinate space.} He seems to have felt the need, however, to advance a more general formalism, capable of dealing with the not uncommon circumstance that a canonical transformation of perfectly real (read ``self-adjoint" in the quantum-mechanical case)  mechanical quantities actually leads to a new canonically conjugate, but {\em complex} (read ``non-self-adjoint")  pair of quantities. An early example of this can be found in London's (1926b) solution of the quantum harmonic oscillator by canonical transformation from the initial ($\hat{q}, \hat{p}$) coordinate-momentum pair to raising and lowering operators, which are obviously not self-adjoint \citep[sec.\ 6.2, pp.\ 357--358]{Duncan and Janssen 2009}. 
     
Jordan  could hardly have been aware at this stage of the complete absence of  ``nice" spectral properties in the general case of a non-self-adjoint operator, with the exception of a very special subclass to be discussed shortly. In contrast to the self-adjoint case,  such operators  may lack a complete set of eigenfunctions spanning the Hilbert space, or there may be an overabundance
of eigenfunctions which form an ``over-complete" set, in the sense that proper subsets of eigenfunctions may suffice to construct an arbitrary state.  To the extent that eigenfunctions exist, the associated eigenvalues are in general complex, occupying some domain---of possibly very complicated structure---in the complex plane. In the case of the lowering operator in the simple harmonic oscillator, the spectrum occupies the entire complex plane! Instead,   \citet[p.\ 812]{Jordan 1927b} seems to have thought of the eigenvalue spectrum as lying on a curve even in the general case of arbitrary non-self-adjoint quantities. 
     
There is one subclass of non-self-adjoint operators for which Jordan's attempt to deal with complex mechanical quantities can be given at least a limited validity. 
The spectral theorem usually associated with self-adjoint and unitary operators (existence and completeness of eigenfunctions) actually extends with full force to the larger class of normal operators $\hat{N}$, defined as satisfying the commutation relation $[\hat{N}, \hat{N}^{\dagger}]=0$, which obviously holds for self-adjoint ($\hat{N}=\hat{N}^{\dagger}$) and unitary ($\hat{N}^{\dagger} = \hat{N}^{-1}$) operators.\footnote{For discussion of the special case of finite Hermitian matrices, see \citet[sec.\ 24.3, pp.\ 177--178]{Dennery and Krzywicki 1996}. For a more general and more rigorous discussion, see \citet[Ch.\ II, sec.\ 10]{von Neumann 1932}.} The reason that the spectral theorem holds for such operators is very simple: given a normal operator $\hat{N}$, we may easily construct a pair of {\em commuting} self-adjoint operators:
\begin{equation}
\hat{A} \equiv \frac{1}{2}(\hat{N} + \hat{N}^{\dagger}), \quad \hat{B} \equiv \frac{1}{2i}(\hat{N} - \hat{N}^{\dagger}).
\end{equation}
It follows that $\hat{A} = \hat{A}^{\dagger}$, $\hat{B} = \hat{B}^{\dagger}$, and that $ [\hat{A}, \hat{B}] = (i/2) \, [N, N^\dagger] = 0$.
 A well-known theorem assures us that a complete set of simultaneous eigenstates $|\lambda\rangle$ of $\hat{A}$ and $\hat{B}$ exist, where the parameter $\lambda$ is chosen to label uniquely the state (we ignore the possibility of degeneracies here), with
 \begin{equation}
 \hat{A} \, |\lambda\rangle = \alpha(\lambda) \, |\lambda\rangle, \quad 
  \hat{B} \, |\lambda\rangle =  \beta(\lambda) \, |\lambda\rangle, \quad
   \hat{N} \, |\lambda\rangle = \zeta(\lambda) \, |\lambda\rangle,
\end{equation}
where $\zeta(\lambda) \equiv \alpha(\lambda)+i\beta(\lambda)$ are the eigenvalues of $\hat{N} = \hat{A} + i \hat{B}$.
Of course, there is no guarantee that $\alpha(\lambda)$ and $\beta(\lambda)$ are continuously connected (once we eliminate the parameter $\lambda$), so the spectrum of $\hat{N}$ (the set of points $\zeta(\lambda)$ in the complex plane) may have a very complicated structure. For a normal operator, there at least exists the possibility though that the spectrum indeed lies on a simple curve, as assumed by Jordan. In fact, it is quite easy to construct an example along these lines, and to show that Jordan's two amplitudes, $\varphi(x,y)$ and $\psi(x,y)$, do exactly the right job in producing the correct probability density in the (self-adjoint) $\hat{x}$ variable for a given {\it complex} value of the quantity $\hat{y}$, in this case associated with a normal operator with a complex spectrum. 
        
Our example is a simple generalization of one that \citet[sec.\ 5, pp.\ 830--831]{Jordan 1927b} himself gave (in the self-adjoint case). For linear canonical transformations, the differential equations specifying the amplitudes $\varphi(x,y)$ and $\psi(x,y)$ [NB1, sec.\ 4, Eqs.\ 2ab and 3ab] are readily solved analytically. Thus, suppose that the canonical transformation from a  self-adjoint conjugate pair ($\hat{p}, \hat{q}$) to a new conjugate pair ($\hat{\alpha},\hat{\beta}$) is given by
\begin{equation}
\hat{\alpha} = a \, \hat{p} + b \, \hat{q}, \quad \hat{\beta} = c \, \hat{p} + d \, \hat{q},
\label{alphbet}
\end{equation}
where the coefficients $a,b,c,d$ must satisfy $ad-bc=1$ for the transformation to be canonical, but may otherwise be complex numbers [cf.\ NB1, sec.\ 5, Eqs. 56--57]. The requirement that $\hat{\alpha}$ be a normal operator (i.e., $[\hat{\alpha}, \hat{\alpha}^{\dagger}]=0$) is easily seen to imply $a/a^{*}= b/b^{*}$.
Thus, $a$ and $b$ have the same complex phase (which we may call $e^{i\vartheta}$). Likewise, normality of $\hat{\beta}$ implies that $c$ and $d$ have equal phase (say, $e^{i\chi}$). Moreover, the canonical condition $ad-bc=1$ implies that the phases $e^{i\varphi}$ and $e^{i\chi}$ must cancel, so we henceforth set $\chi = -\vartheta$, and rewrite the basic canonical transformation as
\begin{equation}
 \hat{\alpha} =  \zeta (a \, \hat{p} + b \, \hat{q}), \quad  \hat{\beta} = \zeta^{*} (c \, \hat{p} + d \, \hat{q}),
\end{equation}
where $\zeta \equiv e^{i\vartheta}$, and $a,b,c,d$ are now real and satisfy $ad-bc=1$. We see that the spectrum of $\hat{\beta}$ lies along the straight line in the complex plane with phase $-\vartheta$ (as the operator $c \, \hat{p} + d \, \hat{q}$ is self-adjoint and therefore has purely real eigenvalues): the allowed values for $\hat{\beta}$ are such that $\zeta\beta$ is real.
     
 It is now a simple matter to solve the differential equations for the amplitude $\varphi(q,\beta)$ and the supplementary amplitude $\psi(q,\beta)$ in this case. The general equations are \citep[sec.\ 4, p.\ 821]{Jordan 1927b}:\footnote{\citet[p.\ 817]{Jordan 1927b} introduced the notation $F^\dagger$ for the adjoint of $F$ in sec.\ 3 of his paper.}
  \begin{eqnarray}
\left\{ f\left(\frac{\hbar}{i}\frac{\partial}{\partial q},q\right)+\frac{\hbar}{i}\frac{\partial}{\partial\beta}  \right\}
\varphi(q,\beta) & = & 0,  \quad \quad [{\rm NB1, \, sec.\ 4, \, Eq.\  2a}] \label{NB1-4-2a} \\
\left\{ g\left(\frac{\hbar}{i}\frac{\partial}{\partial q},q\right)-\beta \right\}
\varphi(q,\beta) & = & 0, \quad \quad [{\rm NB1,  \, sec.\ 4, \, Eq.\  2b}] \label{NB1-4-2b}\\
\left\{ f^\dagger\left(\frac{\hbar}{i}\frac{\partial}{\partial q},q\right)+\frac{\hbar}{i}\frac{\partial}{\partial\beta}  \right\}
\psi(q,\beta) & = & 0, \quad \quad [{\rm NB1, \, sec.\ 4, \, Eq.\  3a}] \label{NB1-4-3a} \\
\left\{ g^\dagger\left(\frac{\hbar}{i}\frac{\partial}{\partial q},q\right)-\beta \right\}
\psi(q,\beta) & = & 0. \quad \quad [{\rm NB1,  \, sec.\ 4, \, Eq.\  3b}] \label{NB1-4-3b}
\end{eqnarray}
The differential operators in this case are (recall that $\hat{\alpha} = f(\hat{p},\hat{q})$ and $\hat{\beta} = g(\hat{p},\hat{q})$ (see Eqs.\ (\ref{NB1-alpha})--(\ref{NB1-beta}) [NB1, sec.\ 4, Eq. 1]):
\begin{eqnarray}
    f = \zeta \left(a \, \frac{\hbar}{i}\frac{\partial}{\partial q}+b \, q \right), \quad  \quad  f^{\dagger} = \zeta^{*}\left(a \, \frac{\hbar}{i}\frac{\partial}{\partial q}+b \, q \right), \nonumber \\ 
      \label{fg and fg dagger}   \\
    g = \zeta^{*}\left(c \, \frac{\hbar}{i}\frac{\partial}{\partial q}+d \, q \right),  \quad  \quad
    g^{\dagger} = \zeta \left(c \, \frac{\hbar}{i}\frac{\partial}{\partial q}+d \, q \right), \nonumber
    \end{eqnarray}
and for $\varphi(q,\beta)$  and $\psi(q,\beta)$ we find (up to an overall constant factor):
    \begin{eqnarray}
    \varphi(q,\beta) &=& \exp{\left\{-\frac{i}{\hbar} \left(\frac{d}{2c}q^{2}-\frac{1}{c}\zeta\beta q+\frac{a}{2c}(\zeta\beta)^{2} \right)\right\}},
     \nonumber \\
    \label{phi psi sol 1} & & \\
     \psi(q,\beta) &=& \exp{\left\{-\frac{i}{\hbar} \left(\frac{d}{2c}q^{2}-\frac{1}{c}\zeta^{*}\beta q+\frac{a}{2c}(\zeta^{*}\beta)^{2} \right) \right\}}. 
     \nonumber
     \end{eqnarray}
We note that the basic amplitude $\varphi(q,\beta)$ is a pure oscillatory exponential, as the combinations $\zeta\beta$ and the constants $a,c,$ and $d$ appearing in the exponent are all real, so the exponent is overall purely imaginary, and the amplitude has unit absolute magnitude. This is not the case for $\psi(q,\beta)$, due to the appearance of $\zeta^{*}$, but at this point we recall that, according to Jordan's postulate A, the correct probability density is obtained by multiplying $\varphi(x,y)$ by $\psi^{*}(x,y)$, where the star symbol includes the instruction that the eigenvalue $y$ of $\hat{y}$ is {\em not to be conjugated} (cf.\ our comment following Eq.\ (\ref{phipsistar})). This rather strange prescription is essential if we are to maintain consistency of  the orthogonality property
     \begin{equation}
       \int \varphi(x,y^{\prime\prime})\psi^{*}(x,y^{\prime})dx = \delta_{y^{\prime}y^{\prime\prime}}
     \end{equation}
with the differential equations (\ref{NB1-4-2b}) and (\ref{NB1-4-3b}) for the amplitudes [NB1, sec.\ 4, Eqs.  2b and 3b]. With this proviso, we find (recalling again that $\zeta\beta$ is real):
     \begin{equation}
       \psi^{*}(q,\beta) = \exp{\left\{\frac{i}{\hbar} \left(\frac{d}{2c}q^{2}-\frac{1}{c}\zeta\beta q+\frac{a}{2c}(\zeta\beta)^{2} \right) \right\}} = \bar{\varphi}(q,\beta),
       \end{equation}
where the bar now denotes conventional complex conjugation, and we see that the product $\varphi(q,\beta)\psi^{*}(q,\beta) = \varphi(q,\beta) \bar{\varphi}(q,\beta)$ is indeed real, and in fact, equal to unity, as we might expect in the case of a purely oscillatory wave function.

That Jordan's prescription for the construction of conditional probabilities cannot generally be valid in the presence of classical complex (or quantum-mechanically non-self-adjoint) quantities is easily verified by relaxing the condition of normal operators in the preceding example. In particular, we consider the example of the raising and lowering operators for the simple harmonic oscillator, obtained again by a complex linear canonical transformation of the
 ($\hat{q},\hat{p}$) canonical pair.\footnote{As mentioned above, \citet{London 1926b} had looked at this example of a canonical transformation \citep[sec.\ 6.2, p.\ 358]{Duncan  and Janssen 2009}.} Now, as a special case of Eq.\ (\ref{alphbet}), we take
 \begin{equation}
 \hat{\alpha} = \frac{1}{\sqrt{2}}( \hat{p} + i\hat{q} ) = f(\hat{p},\hat{q}), \quad  
 \hat{\beta} = \frac{1}{\sqrt{2}}(i \hat{p} + \hat{q})= g(\hat{p},\hat{q}),
\label{alphbet'}  
\end{equation}
which, though canonical, clearly does not correspond to normal operators, as the coefficients $a,b$ (and likewise $c,d$) are now 90 degrees out of phase. Solving the differential equations for the amplitudes, Eqs.\ (\ref{NB1-4-2a})--(\ref{NB1-4-3b}) [NB1, sec.\ 4, Eqs.\ 2ab, 3ab], we now find (up to an overall constant factor [cf.\ Eq.\ (\ref{phi psi sol 1})]):
 \begin{eqnarray}
    \varphi(q,\beta) &=& \exp{\left\{-\frac{1}{2\hbar} \left(q^{2}-2\sqrt{2}\beta q+\beta^{2} \right) \right\}}, 
    \nonumber \\
    & & \label{phi psi sol 2} \\
    \psi(q,\beta) &=& \exp{\left\{\frac{1}{2\hbar} \left(q^{2}-2\sqrt{2}\beta q+\beta^{2} \right) \right\}} = \psi^{*}(q,\beta) = \frac{1}{\varphi(q,\beta)}, 
    \nonumber
    \end{eqnarray}
where $\beta$ is an {\em arbitrary} complex number. In fact, the wave function $\varphi(q,\beta)$ is a {\em square-integrable} eigenfunction of $\hat{\beta}$ for an arbitrary complex value of $\beta$: it corresponds to the well-known ``coherent eigenstates" of the harmonic oscillator, with the envelope (absolute magnitude) of the wave function executing simple harmonic motion about the center of the potential well with frequency $\omega$ (given the Hamiltonian $\hat{H}=\frac{1}{2}(\hat{p}^{2}+\omega^{2}\hat{q}^{2})$). The probability density in $q$ of such a state for fixed $\beta$ is surely given by the conventional prescription $|\varphi(q,\beta)|^{2}$. On the other hand, for complex $\beta$, the Jordan prescription requires us to form the combination (with the peculiar interpretation of the ``star" in $\psi^*(q,\beta)$, in which  $\beta$ is {\em not} conjugated):
    \begin{equation}
    \varphi(q,\beta)\psi^{*}(q,\beta) = 1,
    \end{equation}
which clearly makes no physical sense in this case, as the state in question is a localized, square-integrable one. If  Jordan's notion of an {\em Erg\"anzungsamplitude} is to have any nontrivial content, it would seem to require, at the very least, that the complex quantities considered fall into the very special category of normal operators after quantization. 

In fact, as we will see shortly, in the paper by \citet{Hilbert-von Neumann-Nordheim} on Jordan's version of  statistical transformation theory, the requirement of self-adjointness already acquires the status of a {\em sine qua non} for physical observables in quantum theory, and the concept of an {\em Erg\"anzungs\-am\-pli\-tude} disappears even from Jordan's own treatment of his theory after {\it Neue Begr\"undung} I.

\section{Hilbert, von Neumann, and Nordheim's  {\it Grundlagen} (April 1927)}

In the winter semester of 1926/27, Hilbert gave a course entitled ``Mathematical methods of quantum theory." The course consisted of two parts. The first part, ``The older quantum theory," was essentially a repeat of the course that Hilbert had given under the same title in 1922/23. The second part, ``The new quantum theory," covered the developments since 1925. As he had in 1922/23, Nordheim prepared the notes for this course, which have recently been published \citep[pp.\ 504--707; the second part takes up pp. 609--707]{Sauer and Majer 2009}. At the very end (ibid., pp.\  700--706), we find a concise exposition of the main line of reasoning of Jordan's {\it Neue Begr\"undung} I. 

This presentation served as the basis for a paper by \citet{Hilbert-von Neumann-Nordheim}. As the authors explained in the introduction (ibid., pp.\ 1--2),  ``important parts of the mathematical elaboration" were due to von Neumann, while Nordheim was responsible for the final text \citep[p.\ 361]{Duncan and Janssen 2009}. The paper was submitted to the {\it Mathematische Annalen} April 6, 1927, but, for reasons not clear to us, was only published at the beginning of the volume for 1928. It thus appeared after the trilogy by \citet{von Neumann 1927a, von Neumann 1927b, von Neumann 1927c} that rendered much of it obsolete. In this section we go over the main points of this three-man paper.\footnote{For other discussions of \citet{Hilbert-von Neumann-Nordheim}, see \citet[pp.\ 309--312; cf.\ note \ref{jammer blooper}]{Jammer 1966}, \citet[pp.\ 404--411]{Mehra Rechenberg}, and \citet[pp.\ 295--300, focusing mainly on the paper's axiomatic structure]{Lacki 2000}.}

In the lecture notes for Hilbert's course, Dirac is not mentioned at all, and even though in the paper it is acknowledged that \citet{Dirac 1927} had independently arrived at and published similar results, the focus continues to be on Jordan. There are only a handful of references to Dirac, most importantly in connection with the delta function and in the discussion of the Schr\"odinger equation for a Hamiltonian with a partly discrete spectrum \citep[p.\ 8 and p.\ 30, respectively]{Hilbert-von Neumann-Nordheim}. Both the lecture notes and the paper stay close to the relevant sections of {\it Neue Begr\"undung} I, but Hilbert and his collaborators did change Jordan's notation. Their notation is undoubtedly an improvement over his---not a high bar to clear---but the modern reader trying to follow the argument in these texts may still want to translate it into the kind of modern notation we introduced in Section 2. We will adopt the notation of Hilbert and his co-authors in this section, except that we will continue to use hats to distinguish (operators representing) mechanical quantities from their numerical values.

As we mentioned in Section 2.1, when we discussed postulates A through D of {\it Neue Begr\"un\-dung} I, \citet[pp.\ 4--5]{Hilbert-von Neumann-Nordheim} based their exposition of Jordan's theory on six ``physical axioms."\footnote{In the lecture notes we find four axioms that are essentially the same as Jordan's four postulates  \citep[pp.\ 700--701]{Sauer and Majer 2009}.} Axiom I introduces the basic idea of a probability amplitude. The amplitude for the probability that a mechanical quantity $\hat{F}_1(\hat{p} \, \hat{q})$ (some function of momentum $\hat{p}$  and coordinate $\hat{q}$) has the value $x$ given that another such quantity $\hat{F}_2(\hat{p} \, \hat{q})$ has the value $y$ is written as $\varphi(x \, y; \hat{F}_1 \, \hat{F}_2)$.

Jordan's {\it Erg\"anzungsamplitude} still made a brief appearance in the notes for Hilbert's course \citep[p.\ 700]{Sauer and Majer 2009} but is silently dropped in the paper. As we saw in Section 2, amplitude and supplementary amplitude are identical as long as we only consider quantities represented, in modern terms, by Hermitian operators. In that case, the probability $w(x \, y;  \hat{F}_1 \, \hat{F}_2)$ of finding the value $x$ for $\hat{F}_1$ given the value $y$ for $\hat{F}_2$ is given by the product of $\varphi(x \, y; \hat{F}_1 \,\hat{F}_2)$ and its complex conjugate, which, of course, will always be a real number. Although they did not explicitly point out that this eliminates the need for the {\it Erg\"anzungsamplitude}, \citet[p.\ 17--25]{Hilbert-von Neumann-Nordheim} put great emphasis on the restriction to Hermitian operators. Secs.\ 6--8 of their paper (``The reality conditions," ``Properties of Hermitian operators," and ``The physical meaning of the reality conditions") are devoted to this issue.

Axiom II corresponds to Jordan's postulate B and says that the amplitude for finding a value for $\hat{F}_2$ given the value of $\hat{F}_1$ is the complex conjugate of the amplitude of finding a value for $\hat{F}_1$ given the value of $\hat{F}_2$. This symmetry property entails that these two outcomes have the same probability. Axiom III is not among Jordan's postulates. It basically states the obvious demand that when $\hat{F}_1 = \hat{F}_2$, the probability $w(x \, y; \hat{F}_1 \, \hat{F}_2)$ be either 0 (if $x \neq y$) or 1 (if $x =y$). Axiom IV corresponds to Jordan's postulate C and states that the amplitudes rather than the probabilities themselves follow the usual composition rules for probabilities (cf.\  Eqs.\ (\ref{completeness}) and (\ref{completeness/orthogonality}) in Section 2.1):
\begin{equation}
\varphi(x \, z; \hat{F}_1 \, \hat{F}_3) = \int \varphi(x \, y; \hat{F}_1 \, \hat{F}_2) \, \varphi(y \, z; \hat{F}_2 \, \hat{F}_3) \, dy.
\label{HvNN-completeness}
\end{equation}
Though they did not use Jordan's phrase ``interference of probabilities," the authors emphasized the central importance of this particular axiom:\footnote{Recall, however, Heisenberg's criticism of this aspect of Jordan's work in the uncertainty paper \citep[pp.\ 183--184; cf.\ note \ref{HvJ0}]{Heisenberg 1927b}.\label{HvJ1}}

\begin{quotation}
This requirement [Eq.\ (\ref{HvNN-completeness})] is obviously analogous to the addition and multiplication theorems of ordinary probability calculus, except that in this case they hold for the amplitudes rather than for the probabilities themselves.

The characteristic difference to ordinary probability calculus lies herein that initially, instead of the probabilities themselves, amplitudes occur, which in general will be complex quantities and only give ordinary probabilities if their absolute value is taken and then squared \citep[p.\ 5]{Hilbert-von Neumann-Nordheim}
\end{quotation}
Axiom V, as we already mentioned in Section 2.1, makes part of Jordan's postulate A into a separate axiom. It demands that probability amplitudes for quantities  $\hat{F}_1$ and $\hat{F}_2$ depend only on the functional dependence of these quantities on $\hat{q}$ and $\hat{p}$ and not on ``special properties of the system under consideration, such as, for example, its Hamiltonian" (ibid., p.\ 5). Axiom VI, finally, adds another obvious requirement to the ones recognized by Jordan: that probabilities be independent of the choice of coordinate systems.

Before they introduced the axioms, \citet[p.\ 2]{Hilbert-von Neumann-Nordheim} had already explained, in a passage that we quoted in Section 1.1, that the task at hand was to find ``a simple analytical apparatus in which quantities occur that satisfy" axioms I--VI. As we know from {\it Neue Begr\"undung} I, the quantities that fit the bill are the integral kernels of certain canonical transformations, implemented as $T \hat{p} T^{-1}$ and $T \hat{q} T^{-1}$ (cf.\ Eqs.\ (\ref{NB1-alpha})--(\ref{NB1-beta})). After introducing this ``simple analytical apparatus" in secs.\ 3--4  (``Basic formulae of the operator calculus," ``Canonical operators and transformations"), the authors concluded in sec.\ 5 (``The physical interpretation of the operator calculus"):
\begin{quotation}
{\it The probability amplitude $\varphi(x \, y; \hat{q} \, \hat{F})$ between the coordinate $\hat{q}$ and an arbitrary mechanical quantity $\hat{F}(\hat{q} \, \hat{p})$---i.e., for the situation that for a given value $y$ of $\hat{F}$, the coordinate lies between $x$ and $x + dx$---is given by the kernel of the integral operator that canonically transforms the operator 
$\hat{q}$ into the operator corresponding to the mechanical quantity $\hat{F}(\hat{q} \, \hat{p})$} \citep[p.\ 14; emphasis in the original, hats added]{Hilbert-von Neumann-Nordheim}.
\end{quotation}
They immediately generalized this definition to cover the probability amplitude between two arbitrary quantities $\hat{F}_1$ and $\hat{F}_2$.
In sec.\ 3, the authors already derived differential equations for integral kernels $\varphi(x \, y)$ (ibid., pp.\ 10--11, Eqs.\ (19ab) and (21ab)).
Given the identification of these integral kernels with probability amplitudes in sec.\ 5, these equation are just Jordan's fundamental differential equations for the latter (NB1, sec.\  4, Eqs. (2ab); our Eqs.\ (\ref{NB1-2a})--(\ref{NB1-2b}) in Section 2.3).

In sec.\ 4, they also stated the key assumption that any quantity of interest can be obtained through a canonical transformation starting from some canonically conjugate pair of quantities $\hat{p}$ and $\hat{q}$:
\begin{quotation}
We will assume that every operator $\hat{F}$ can be generated out of the basic operator $\hat{q}$ by a canonical transformation. This statement can also be expressed in the following way, namely that, given $\hat{F}$, the operator equation $T\hat{q}T^{-1}$ has to be solvable.

The conditions that $\hat{F}$ has to satisfy for this to be possible will not be investigated here  \citep[p.\ 12; hats added]{Hilbert-von Neumann-Nordheim}.
\end{quotation}
What this passage suggests is that the authors, although they recognized the importance of this assumption, did not quite appreciate that, as we showed at the end of Section 2.3, it puts severe limits on the applicability of Jordan's formalism. In the simple examples of canonical transformations  ($\hat{F} = f(\hat{q})$ and $\hat{F}=\hat{p}$) that they considered in sec.\ 9 (``Application of the theory to special cases"), the assumption is obviously satisfied and the formalism works just fine (ibid., pp.\ 25--26). In sec.\ 10 (``The Schr\"odinger differential equations"), however, they set $\hat{F}$ equal to the Hamiltonian $\hat{H}$ and  claimed that one of the differential equations for the probability amplitude $\varphi(x\,W; \hat{q} \, \hat{H})$ (where $W$ is an energy eigenvalue) is the time-independent Schr\"odinger equation. As soon as the Hamiltonian has a wholly or partly discrete spectrum, however, there simply is no operator $T$ such that $\hat{H}=T\hat{q}T^{-1}$.

In secs.\ 6--8, which we already briefly mentioned above, \citet[pp.\ 17--25]{Hilbert-von Neumann-Nordheim} showed that the necessary and sufficient condition for the probability $w(x \, y;  \hat{F}_1 \, \hat{F}_2)$ to be real is that $\hat{F}_1$ and $\hat{F}_2$ are both represented by Hermitian operators. As we pointed out earlier, they implicitly rejected Jordan's attempt to accommodate $\hat{F}$'s represented by non-Hermitian operators through the introduction of the {\it Erg\"anzungsamplitude}. They also showed that the operator representing the canonical conjugate $\hat{G}$ of a quantity $\hat{F}$ represented by a Hermitian operator is itself Hermitian. 

The authors ended their paper on a cautionary note emphasizing its lack of mathematical rigor. They referred to von Neumann's (1927a) forthcoming paper, {\it Mathematische Begr\"undung}, for a more satisfactory treatment 
of the Schr\"odinger equation for Hamiltonians with a partly discrete spectrum. In the concluding paragraph, they warned the reader more generally:
\begin{quotation}
In our presentation the general theory receives such a perspicuous and formally simple form that we have carried it through in a mathematically still imperfect form, especially since a fully rigorous  presentation might well be considerably more tedious and circuitous \citep[p.\ 30]{Hilbert-von Neumann-Nordheim}.
\end{quotation}

\section{Jordan's {\it Neue Begr\"undung} II (June 1927)}

In April and May 1927, while at Bohr's institute in Copenhagen on an International Education Board fellowship, \citet{Jordan 1927f} wrote {\it Neue Begr\"undung} II, which was received by {\it Zeitschrift f\"ur Physik} June 3, 1927.\footnote{Not long after leaving for Copenhagen, Jordan wrote a long letter to Dirac, who was still in G\"ottingen, touching on some of the issues addressed in {\it Neue Begr\"undung} II. Jordan wrote: ``Following our conversation a few days ago I want to write you a little more. It's about a few things connected to those we discussed. Most of these considerations originate in the fall of last year. Back then, however, I didn't succeed in achieving a complete clarification of the question that interested me, and later I haven't really had the courage to take up the issue again. It would certainly please me if you could make some progress on these questions" (Jordan to Dirac, April 14, 1927, AHQP). Unfortunately, we do not have Dirac's reply. In between {\it Neue Begr\"undung} I and II, \citet{Jordan 1927e} published a short paper showing that his theory has the desirable feature that the conditional probability of finding a certain value for some quantity is independent of the scale used to measure that quantity.} In the abstract he announced a ``simplified and generalized" version of the theory presented in  {\it Neue Begr\"undung} I.\footnote{When Kuhn complained about the ``dreadful notation" of {\it Neue Begr\"undung} I in his interview with Jordan for the AHQP project, Jordan said that in {\it Neue Begr\"undung} II he just wanted to give a ``prettier and clearer" exposition of the same material (session 3, p.\ 17, quoted by Duncan and Janssen, 2009, p.\ 360).}

One simplification was that Jordan, like \citet{Hilbert-von Neumann-Nordheim}, dropped the {\it Erg\"anzungsamplitude} and restricted himself accordingly to physical quantities represented by Hermitian operators and to canonical transformations preserving Hermiticity. Another simplification was that he adopted Dirac's (1927) convention of consistently using the same letter for a mechanical quantity and its possible values, using primes to distinguish the latter from the former. When, for instance, the letter $\beta$ is used for some quantity, its values are denoted as $\beta', \beta''$, etc. We will continue to use the notation $\hat{\beta}$ for the quantity (and the operator representing that quantity) and the notation $\beta, \beta', \ldots$ for its values. While this new notation for quantities and their values was undoubtedly an improvement, the new notation for probability amplitudes and for transformation operators with those amplitudes as their integral kernels is actually more cumbersome than in {\it Neue Begr\"undung} I. 

In the end, however, these new notational complications only affect the cosmetics of the paper. What is more troublesome is that the generalization of the formalism promised in the abstract to handle cases with wholly or partly discrete spectra is much more problematic than Jordan suggested and, we argue, ultimately untenable.
By the end of {\it Neue Begr\"undung} II, Jordan is counting quantities nobody would think of as canonically conjugate (e.g., different components of spin) as pairs of conjugate variables and has abandoned the notion, central to the formalism of  {\it Neue Begr\"undung} I, that any quantity of interest (e.g, the Hamiltonian) is a member of a pair of conjugate variables connected to
some initial pair of $\hat{p}$'s and $\hat{q}$'s by a canonical transformation. It is fair to say that, although Jordan was still clinging to his $p$'s and $q$'s in {\it Neue Begr\"undung} II, they effectively ceased to play any significant role in his formalism.

As we showed at the end of Section 2.3, the canonical transformation 
\begin{equation}
\hat{\alpha} = T \hat{p} T^{-1}, \quad \hat{\beta} = T \hat{q} T^{-1}
\label{alpha-beta}
\end{equation}
(cf.\ Eqs.\ (\ref{NB1-alpha})--(\ref{NB1-beta}))
can never get us from a quantity  with a completely continuous spectrum (such as position or momentum) to a quantity with a wholly or partly discrete spectrum (such as the Hamiltonian). In {\it Neue Begr\"undung} II, \citet[pp.\ 16--17]{Jordan 1927f} evidently recognized this problem even though it is not clear that he realized the extent to which this undercuts his entire approach. 

The central problem is brought out somewhat indirectly in the paper. As \citet[pp.\ 1--2]{Jordan 1927f} already mentioned in the abstract and then demonstrated in the introduction, the commutation relation, $[\hat{p}, \hat{q}] = \hbar/i$, for two canonically conjugate quantities $\hat{p}$ and $\hat{q}$ cannot hold as soon as the spectrum of one of them is partly discrete. Specifically, this means that 
action-angle variables $\hat{J}$ and $\hat{w}$, where the eigenvalues of the action variable $\hat{J}$ are restricted to integral multiples of Planck's constant, cannot satisfy the canonical commutation relation. 

The proof of this claim is very simple.  \citet[p.\ 2]{Jordan 1927f} considered a pair of conjugate quantities $\hat{\alpha}$ and $\hat{\beta}$ where $\hat{\beta}$ is assumed to have a purely discrete spectrum. We will see that one runs into the same problem as soon as one  $\hat{\alpha}$ or $\hat{\beta}$ has a single discrete eigenvalue. Suppose $\hat{\alpha}$ and $\hat{\beta}$ satisfy the standard commutation relation:
\begin{equation}
[ \hat{\alpha},  \hat{\beta}]= \frac{\hbar}{i}.
\label{NB2-2}
\end{equation}
As Jordan pointed out, it then follows that an operator that is some function $F$ of $\hat{\beta}$ satisfies\footnote{If the function $F(\beta)$ is assumed to be a polynomial, $\sum_n c_n \, \beta^n$, which is all we need for what we want to prove, although \citet[p.\ 2]{Jordan 1927f} considered a ``fully transcendent function,"  Jordan's claim is a standard result in elementary quantum mechanics: 
$$
[\hat{\alpha},  F(\hat{\beta})] = [\hat{\alpha}, \sum_n c_n \, \hat{\beta}^n] = \sum_n  c_n \, n \, \frac{\hbar}{i} \, \hat{\beta}^{n-1} = \frac{\hbar}{i} \frac{d}{d\hat{\beta}} 
\left(\sum_n c_n \, \hat{\beta}^n\right) = \frac{\hbar}{i} \, F'(\hat{\beta}),
$$
where in the second step we repeatedly used that $[\hat{\alpha}, \hat{\beta}] = \hbar/i$ and that $[\hat{A}, \hat{B} \hat{C}] = [\hat{A}, \hat{B}] \hat{C} + \hat{B} [\hat{A}, \hat{C}]$  for any three operators $\hat{A}$,  $\hat{B}$,  and $\hat{C}$.}
\begin{equation}
[ \hat{\alpha}, F(\hat{\beta})] = \frac{\hbar}{i} F'(\hat{\beta}).
\label{NB2-3}
\end{equation}
Jordan now chose a function such that $F(\beta) = 0$ for all eigenvalues $\beta_1, \beta_2, \ldots$ of $\hat{\beta}$, while $F'(\beta) \neq 0$ at those same points. In that case, the left-hand side of Eq.\ (\ref{NB2-3}) vanishes at  all these points, whereas the right-hand side does not. Hence, Eq.\ (\ref{NB2-3}) and, by {\it modus tollens}, Eq.\ (\ref{NB2-2}) cannot hold. 

Since Eq.\ (\ref{NB2-3}) is an operator equation, we should, strictly speaking, compare the results of the left-hand side and the right-hand side acting on some state. To show that Eq.\ (\ref{NB2-3})---and thereby Eq.\ (\ref{NB2-2})---cannot hold, it suffices to show that it does not hold for one specific function $F$ and one specific state $|\psi\rangle$. Consider the simple function $F_1(\beta) = \beta - \beta_1$, for which $F_1(\beta_1)= 0$ and $F_1'(\beta_1) = 1$, and the discrete (and thus normalizable) eigenstate $|\beta_1\rangle$ of the operator corresponding to the quantity $\hat{\beta}$. Clearly, 
\begin{equation}
\langle \beta_1 | [\hat{\alpha}, F_1(\hat{\beta})] | \beta_1 \rangle = \langle \beta_1 | [\hat{\alpha}, \hat{\beta} - \beta_1] | \beta_1 \rangle = 0,
\end{equation}
as $\hat{\beta} | \beta_1 \rangle = \beta_1 |\beta_1 \rangle$, while 
\begin{equation}
\langle \beta_1 | (\hbar/i) F_1' (\hat{\beta})] | \beta_1 \rangle = \frac{\hbar}{i} \langle \beta_1 | \beta_1 \rangle = \frac{\hbar}{i},
\end{equation}
as $F'_1(\hat{\beta}) | \beta_1 \rangle = F'_1(\beta_1) | \beta_1 \rangle = | \beta_1 \rangle$. This shows that the relation, 
\begin{equation}
\langle \psi | [\hat{\alpha}, F(\hat{\beta})] | \psi \rangle = \langle \psi | \frac{\hbar}{i} F'(\hat{\beta}) | \psi \rangle,
\label{F=0 F'=1}
\end{equation}
and hence Eqs.\  (\ref{NB2-2})--(\ref{NB2-3}), cannot hold. The specific example $F_1(\beta) = \beta - \beta_1$ that we used above immediately makes it clear that the commutation relation $[\hat{\alpha}, \hat{\beta}] = \hbar/i$ cannot hold as soon as either one of the two operators has a single discrete eigenvalue.

Much later in the paper, in sec.\ 4 (``Canonical transformations"), \citet[p.\ 16]{Jordan 1927f} acknowledged that it follows directly from this result that no canonical transformation can ever get us from a pair of conjugate variables $\hat{p}$'s and $\hat{q}$'s with completely continuous spectra to $\hat{\alpha}$'s and $\hat{\beta}$'s with partly discrete spectra. It is, after all, an essential property of canonical transformations that they preserve canonical commutation relations. From Eq.\ (\ref{alpha-beta}) and $[\hat{p}, \hat{q}] = \hbar/i$ it follows that
\begin{equation}
[\hat{\alpha}, \hat{\beta}] = [T \hat{p} T^{-1}, T \hat{q} T^{-1}] = T [\hat{p}, \hat{q}] T^{-1} = \hbar/i.
\label{reductio}
\end{equation}
Since, as we just saw, only quantities with purely continuous spectra can satisfy this commutation relation, Eq. (\ref{reductio}) cannot hold for $\hat{\alpha}$'s and $\hat{\beta}$'s with partly discrete spectra and such $\hat{\alpha}$'s and $\hat{\beta}$'s cannot possibly be obtained from $\hat{p}$ and $\hat{q}$ through a canonical transformation of the form (\ref{alpha-beta}).

We will discuss below how this obstruction affects Jordan's general formalism. When Jordan, in the introduction of {\it Neue Begr\"undung} II, showed  that no quantity with a partly discrete spectrum can satisfy a canonical commutation relation, he presented it not as a serious problem for his formalism but as an argument for the superiority of his alternative definition of conjugate variables in {\it Neue Begr\"undung} I \citep[p.\ 814, cf.\ Eq.\ (\ref{NB1-18})]{Jordan 1927b}. In that definition $\hat{p}$ and $\hat{q}$ are considered canonically conjugate if the probability amplitude $\varphi(p, q)$ has the simple form $e^{-ipq/\hbar}$, which means that as soon as the value of one of the quantities $\hat{p}$ and $\hat{q}$ is known, all possible values of the other quantity are equiprobable. As we saw in Section 2.1, Jordan showed that for $\hat{p}$'s and $\hat{q}$'s with purely continuous spectra this implies that they satisfy $[\hat{p}, \hat{q}] = \hbar/i$ (cf.\ Eq.\ (\ref{NB1-25})), which is the standard definition of what it means for $\hat{p}$ and $\hat{q}$ to be conjugate variables. In {\it Neue Begr\"undung} II, \citet[p.\ 6, Eq.\ (C)]{Jordan 1927f} extended his alternative definition to quantities with wholly or partly discrete spectra, in which case the new definition, of course, no longer reduces to the standard one.

 As \citet{Jordan 1927f} wrote in the opening paragraph, his new paper only assumes a rough familiarity with the considerations of {\it Neue Begr\"undung} I. He thus had to redevelop much of the formalism of his earlier paper, while trying to both simplify and generalize it at the same time. In sec.\ 2 (``Basic properties of quantities and probability amplitudes"), Jordan began by restating the postulates to be satisfied by his probability amplitudes.
 
He introduced a new notation for these amplitudes. Instead of $\varphi(\beta, q)$ (cf.\ note \ref{NB1-notation}) he now wrote $\Phi_{\alpha p}(\beta', q')$. The primes, as we explained above, distinguish values of quantities from those quantities themselves. The subscripts $\alpha$ and $p$ denote which quantities are canonically conjugate to the quantities $\hat{\beta}$ and $\hat{q}$ for which the probability amplitude is being evaluated. As we will see below, one has a certain freedom in picking the $\hat{\alpha}$ and $\hat{p}$ conjugate to $\hat{\beta}$ and $\hat{q}$, respectively, and settling on a specific pair of $\hat{\alpha}$ and $\hat{p}$ is equivalent to fixing the phase ambiguity of the amplitude $\varphi(\beta, q)$ up to some constant factor. So for a given choice of  $\hat{\alpha}$ and $\hat{p}$, the amplitude $\Phi_{\alpha p}(\beta', q')$  is essentially unique. In this way, \citet[p.\ 20]{Jordan 1927f} could answer, at least formally, von Neumann's (1927a, p.\ 3) objection that probability amplitudes are not uniquely determined even though the resulting probabilities are. It is only made clear toward the end of the paper that this is the rationale behind these additional subscripts. Their only other role is to remind the reader that $\Phi_{\alpha p}(\beta', q')$ is determined not by one Schr\"odinger-type equation in Jordan's formalism but by a pair of such equations involving both canonically conjugate pairs of variables, $(\hat{p}, \hat{q})$ and $(\hat{\alpha}, \hat{\beta})$ \citep[p.\ 20]{Jordan 1927f}. As none of this is essential to the formalism, we will simply continue to use the notation $\langle \beta | q \rangle$ for the probability amplitude between the quantities $\hat{\beta}$ and $\hat{q}$.
 
Jordan also removed the restriction to systems of one degree of freedom that he had adopted for convenience in {\it Neue Begr\"undung} I \citep[p.\ 810]{Jordan 1927b}. So $\hat{q}$, in general, now stands for $(\hat{q}_1, \ldots \hat{q}_f)$, where $f$ is the number of degrees of freedom of the system under consideration. The same is true for other quantities.
\citet[pp.\ 4--5]{Jordan 1927f} spent a few paragraphs examining the different possible structures of the space of eigenvalues for such $f$-dimensional quantities depending on the nature of the spectrum of its various components---fully continuous, fully discrete, or combinations of both. He also introduced the notation $\delta(\beta' - \beta'')$ for a combination of the  Dirac delta function and the Kronecker delta (or, as Jordan called the latter, the ``Weierstrassian symbol"). 
 
In  {\it Neue Begr\"undung} II \citep[p.\ 6]{Jordan 1927f}, the four postulates of {\it Neue Begr\"undung} I (see our discussion in Section 2.1) are replaced by  three postulates or ``axioms,"  as Jordan now also called them, numbered with Roman numerals. This may have been in deference  to \citet{Hilbert-von Neumann-Nordheim}, although  they listed six such axioms (as we saw in Section 3).
Jordan's new postulates or axioms do not include the key portion of postulate A of  {\it Neue Begr\"undung} I stating the probability interpretation of the amplitudes. That is relegated to sec.\ 5, ``The physical meaning of the amplitudes" \citep[p.\ 19]{Jordan 1927f}. Right before listing the postulates, however, \citet[p.\ 5]{Jordan 1927f} did mention that he will only consider ``real (Hermitian) quantities," thereby obviating the need for the {\it Erg\"anzungsamplitude} and simplifying the relation between amplitudes and probabilities. There is no discussion of the {\it Erg\"anzungsamplitude} amplitude in the paper. Instead, following the lead of \citet{Hilbert-von Neumann-Nordheim}, Jordan silently dropped it. It is possible that this was not even a matter of principle for Jordan but only one of convenience. Right after listing the postulates, he wrote that the restriction to real quantities is made only ``on account of simplicity" ({\it der Einfachkeit halber}, ibid., p.\ 6). 
 
Other than the probability-interpretation part of postulate A, all four postulates of {\it Neue Begr\"undung} I return, generalized from one to $f$ degrees of freedom and from quantities with completely continuous spectra to quantities with wholly or partly discrete spectra. Axiom I corresponds to the old postulate D. It says that for every generalized coordinate there is a conjugate momentum. Axiom II consists of three parts, labeled (A), (B), and (C). Part (A) corresponds to the old postulate B, asserting the symmetry property, which, in the new notation, becomes:
\begin{equation}
\Phi_{\alpha p}(\beta', q') = \Phi^*_{p \alpha}(q', \beta').
\label{NB2-symmetry}
\end{equation}
Part (B) corresponds to the old postulate C, which gives the basic rule for the composition of probability amplitudes,
 \begin{equation}
\overline{\sum_{q'}} \, \Phi_{\alpha p}(\beta', q')  \Phi_{pP}(q', Q') =  \Phi_{\alpha P}(\beta', Q'),  
\label{NB2-composition}
\end{equation}
where the notation $\overline{\sum}_{q}$ indicates that, in general, we need a combination of  integrals over the continuous parts of the spectrum of a quantity and sums over its discrete parts. In Eq.\ (\ref{NB2-composition}), $\overline{\sum}_{q}$ refers to an ordinary integral as the coordinate $\hat{q}$ has a purely continuous spectrum. Adopting this  $\overline{\sum}$ notation, we can rewrite the composition rule (\ref{NB2-composition}) in the modern language introduced in Section 2 and immediately recognize it as a completeness relation (cf.\ Eqs.\ (\ref{completeness}) and (\ref{completeness/orthogonality})): 
\begin{equation}
\overline{\sum_{q}}\langle \, \beta | q \rangle \langle q | Q \rangle = \langle \beta | Q \rangle.
\label{NB2-completeness}
\end{equation}
We can likewise formulate orthogonality relations, as \citet[p.\ 7, Eq.\ (5)]{Jordan 1927f} did at the beginning of sec.\ 2 (``Consequences"):
\begin{equation}
\overline{\sum_{q}}\langle \, \beta | q \rangle \langle q | \beta' \rangle = \delta(\beta - \beta'), \quad
\overline{\sum_{\beta}}\langle \, q | \beta \rangle \langle \beta | q' \rangle = \delta(q - q').
\label{NB2-orthogonality}
\end{equation}
Recall that $\delta(\beta - \beta')$ can be either the Dirac delta function or the Kronecker delta, as $\hat{\beta}$ can have either a fully continuous or a partly or wholly discrete spectrum. The relations in Eq.\ (\ref{NB2-orthogonality}) can, of course, also be read as completeness relations, i.e., as giving two different resolutions,
\begin{equation}
\overline{\sum_{q}} | q \rangle \langle q |, \quad  \overline{\sum_{\beta}} | \beta \rangle \langle \beta |,
\end{equation}
 of the unit operator. Part (C) of axiom II is the generalization of the definition of conjugate variables familiar from {\it Neue Begr\"undung} I to $f$ degrees of freedom and to quantities with wholly or partly discrete spectra. Two quantities $\hat{\alpha} = (\hat{\alpha}_1, \ldots, \hat{\alpha}_f)$ and $\hat{\beta} = (\hat{\beta}_1, \ldots, \hat{\beta}_f)$ are canonically conjugate to one another if
\begin{equation}
\Phi_{\alpha, -\beta}(\beta, \alpha) = C \, e^{i \left( \sum_{k=1}^f \beta_k \alpha_k \right)/\hbar},
\label{NB2-C}
\end{equation}
where $C$ is a normalization constant.\footnote{Contrary to what \citet[p.\ 7]{Jordan 1927f} suggested, the sign of the exponent in Eq.\ (\ref{NB2-C}) agrees with the sign of the exponent in the corresponding formula in {\it Neue Begr\"undung} I \citep[p.\ 814, Eq.\ 18; our Eq.\ \eqref{NB1-18}]{Jordan 1927b}.} Axiom III, finally, is essentially axiom III of \citet[p.\ 4]{Hilbert-von Neumann-Nordheim}, which was not part of {\it Neue Begr\"undung} I and which says, in our notation, that $\langle \beta | \beta' \rangle = \delta(\beta - \beta')$, where, once again, $\delta(\beta - \beta')$ can be either the Dirac delta function or the Kronecker delta.

We need to explain one more aspect of Jordan's notation in {\it Neue Begr\"undung} II. As we have seen in Sections 2 and 3, probability amplitudes do double duty as integral kernels of  canonical transformations. \citet[p.\ 6, Eqs.\ (A$'$)-(B$'$)]{Jordan 1927f} introduced the special notation $\Phi_{\alpha p}^{\beta q}$ to indicate that the amplitude $\Phi_{\alpha p}(\beta', q')$ serves as such an integral kernel, thinking of $\Phi_{\alpha p}^{\beta q}$ as `matrices' with $\beta$ and $q$ as `indices' that, in general, will take on both discrete and continuous values. We will continue to use the modern notation $\langle \beta | q \rangle$ both when we want to think of this quantity as a probability amplitude and when we want to think of it as a transformation `matrix'. The notation of {\it Neue Begr\"undung} II, like the modern notation, clearly brings out the double role of this quantity. In {\it Neue Begr\"undung} I, we frequently encountered canonical transformations such as  $\hat{\alpha} = T\hat{p}T^{-1}$, $\hat{\beta} = T\hat{p}T^{-1}$ (cf.\ Eqs. (\ref{NB1-alpha})--(\ref{NB1-beta})). In {\it Neue Begr\"undung} II, such transformations are written with $\Phi_{\alpha p}^{\beta q}$'s instead of $T$'s. As we will explain in detail below, this conceals an important shift in Jordan's usage of such equations. This shift is only made explicit in sec.\ 4, which, as its title announces, deals specifically with ``Canonical transformations." Up to that point, and especially in sec.\ 3, ``The functional equations of the amplitudes," 
Jordan appears to be vacillating between two different interpretations of these canonical transformation equations, the one of {\it Neue Begr\"undung} I in which  $\hat{\alpha}$ and $\hat{\beta}$ are {\it new} conjugate variables different from the $\hat{p}$ and $\hat{q}$ we started from, and one, inspired by \citet{Dirac 1927}, as \citet[pp.\ 16--17]{Jordan 1927f} acknowledged in sec.\ 4, in which $\hat{\alpha}$ and  $\hat{\beta}$ are still the {\it same} $\hat{p}$ and $\hat{q}$ but expressed with respect to a {\it new basis}.\footnote{Cf.\ the letter from Dirac to Jordan quoted in Section 1.1 (cf.\ note \ref{MR2}).}

Before he got into any of this, \citet[sec.\ 2, pp.\ 8--10]{Jordan 1927f} examined five examples, labeled (a) through (e), of what he  considered to be pairs of conjugate quantities and convinced himself that they indeed qualify as such under his new definition (\ref{NB2-C}). More specifically,  he checked in these five cases that these purportedly conjugate pairs of quantities satisfy the completeness or orthogonality relations (\ref{NB2-orthogonality}). 
The examples include  familiar pairs of canonically conjugate variables, such as action-angle variables (Jordan's example (c)), but also quantities that we normally would not think of as conjugate variables, such as different components of spin (a special case of example (e)). We take a closer look at these two specific examples.

In example (c), the allegedly conjugate variables are the angle variable $\hat{w}$ with a purely continuous spectrum and eigenvalues $w \in [0, 1]$ (which means that the eigenvalues of a true angle variable $\hat{\vartheta} \equiv 2 \pi w$ are $\vartheta \in [0, 2 \pi]$) and the action variable $\hat{J}$ with a purely discrete spectrum and eigenvalues $J = C + nh$, where $C$ is an arbitrary (real) constant and $n$ is a positive integer. For convenience we set $C = -1$, so that $J = mh$ with $m = 0, 1, 2, \ldots$ The probability amplitude $\langle w | J \rangle$ has the form required by Jordan's definition (\ref{NB2-C}) of conjugate variables, with $\hat{\alpha}=\hat{J}$, $\hat{\beta}=\hat{w}$, and $f=1$:
\begin{equation}
\langle w | J \rangle = e^{iwJ/\hbar}.
\label{example-wJ-1}
\end{equation}
We now need to check whether $\langle w | J \rangle$ satisfies the two relations in Eq.\ (\ref{NB2-orthogonality}):
\begin{equation}
\int_0^1 \, dw \, \langle J_{n_1} | w \rangle \langle w | J_{n_2} \rangle = \delta_{n_1 n_2}, \quad
\sum_{n=0}^{\infty} \, \langle w | J_n \rangle \langle J_n | w' \rangle = \delta(w - w').
\label{example-wJ-2}
\end{equation}
Using Eq.\ (\ref{example-wJ-1}), we can write the integral in the first of these equations as:
\begin{equation}
\int_0^1 \, dw \,  \langle w | J_{n_2} \rangle \langle w | J_{n_1}  \rangle^* =
\int_0^1 \, dw \, e^{2 \pi i (n_2 - n_1)w} = \delta_{n_1 n_2}.
\label{example-wJ-2}
\end{equation}
Hence the first relation indeed holds. We can similarly write the sum in the second relation as
\begin{equation}
\sum_{n=0}^{\infty} \, \langle w | J_n \rangle \langle w' | J_n  \rangle^* =
\sum_{n=0}^{\infty} e^{2 \pi i n (w - w')}.
\label{example-wJ-3}
\end{equation}
Jordan set this equal to $\delta(w - w')$. However, for this to be true the sum over $n$ should have been from minus to plus infinity.\footnote{A quick way to see this is to consider the Fourier expansion of some periodic function $f(w)$:
$$
f(w) = \sum_{n= - \infty}^{\infty} c_n \, e^{2 \pi i n w}, \; {\rm with \; Fourier \; coefficients}\; c_n \equiv  \int_0^1 \, dw' \, e^{-2 \pi i n w'} \, f(w').
$$
Substituting the expression for $c_n$ back into the Fourier expansion of $f(w)$, we find
$$
f(w) = \int_0^1 \, dw'  \left( \sum_{n= - \infty}^{\infty}  e^{2 \pi i n (w - w')} \right) f(w'),
$$
which means that the expression in parentheses must be equal to $\delta(w - w')$. Note that the summation index does not run from $0$ to $\infty$, as in Eq. (\ref{example-wJ-3}), but from $- \infty$ to $+ \infty$.} If the action-angle variables are $(\hat{L}_z, \hat{\varphi})$, the $z$-component of angular momentum and the azimuthal angle, the eigenvalues of $\hat{J}$ are, in fact, $\pm m\hbar$ with $m = 0, 1, 2, \ldots$, but if the action variable is proportional to the energy, as it is in many applications in the old quantum theory \citep[Pt.\ 2, pp.\ 628--629]{Duncan and Janssen 2007}, the spectrum is bounded below. So even under Jordan's alternative definition of canonically conjugate quantities, action-angle variables do not always qualify. However, since action-angle variables do not play a central role in the {\it Neue Begr\"undung} papers, this is a  relatively minor problem.

We turn to Jordan's example (e), the other example of supposedly conjugate quantities that we want to examine.
Consider a quantity $\hat{\beta}$ with a purely discrete spectrum with $N$ eigenvalues $0, 1, 2, \ldots, N-1$. Jordan showed that,  if the completeness or orthogonality relations (\ref{NB2-orthogonality}) are to be satisfied, the quantity $\hat{\alpha}$ conjugate to $\hat{\beta}$ must also have a discrete spectrum with $N$ eigenvalues $hk/N$ where $k = 0, 1, 2, \ldots, N-1$. We will check this for the special case that $N=2$. As \citet[pp.\ 9--10]{Jordan 1927f} noted, this corresponds to the case of electron spin. In sec.\ 6, ``On the theory of the magnetic electron," he returned to the topic of spin, acknowledging a paper by \citet{Pauli 1927b} on the ``magnetic electron," which he had read in manuscript \citep[p.\ 21, note 2]{Jordan 1927f}. Pauli's discussion of spin had, in fact, been an important factor prompting Jordan to write {\it Neue Begr\"undung} II:
\begin{quotation}
But the magnetic electron truly provides a case where the older canonical commutation relations completely fail; the desire to fully understand the relations encountered in this case was an important reason for carrying out this investigation \citep[p.\ 22]{Jordan 1927f}.
\end{quotation}
The two conditions  on the amplitudes $\langle \beta | \alpha \rangle = C e^{i\beta\alpha/\hbar}$ (Eq.\ (\ref{NB2-C}) for $f=1$) that need to be verified in this case are (cf.\ Eq.\ \eqref{NB2-orthogonality}):
\begin{equation}
\sum_{k=1}^2 \langle \alpha_m | \beta_k \rangle \langle \beta_k | \alpha_n \rangle = \delta_{nm}, \quad
\sum_{k=1}^2 \langle \beta_m | \alpha_k \rangle \langle \alpha_k | \beta_n \rangle = \delta_{nm}.
\label{example-e-1}
\end{equation}
Inserting the expression for the amplitudes, we can write the first of these relations as
\begin{equation}
\sum_{k=1}^2  \langle \beta_k | \alpha_n \rangle \langle \beta_k | \alpha_m \rangle^*
= \sum_{k=1}^2 C^2 \, e^{i \beta_k (\alpha_n - \alpha_m)/\hbar} 
= C^2 \left( 1 + e^{i(\alpha_n - \alpha_m)/\hbar} \right),
\label{example-e-2}
\end{equation}
where in the last step we used that $\beta_1 = 0$ and $\beta_2 = 1$. The eigenvalues of $\hat{\alpha}$ in this case are $\alpha_1 = 0$ and $\alpha_2 = h/2$. For $m=n$, the right-hand side of Eq.\ (\ref{example-e-2}) is equal to $2C^2$.  For $m \neq n$,  $e^{i(\alpha_n - \alpha_m)/\hbar} = e^{\pm i \pi} = -1$ and the right-hand side of Eq.\ (\ref{example-e-2}) vanishes. Setting $C = 1/\sqrt{2}$, we  thus establish the first relation in Eq.\ (\ref{example-e-1}). A completely analogous argument establishes the second. 

This example is directly applicable to the treatment of two arbitrary components of the electron spin, $\hat{\bold{\sigma}} = (\hat{\sigma}_x, \hat{\sigma}_y, \hat{\sigma}_z)$, say the $x$- and the $z$-components, even though the eigenvalues of both $\hat{\sigma}_x$ and $\hat{\sigma}_z$ are $(\frac{1}{2}, -\frac{1}{2})$ rather than $(0, h/2)$ for $\hat{\alpha}$ and $(0,1)$ for $\hat{\beta}$ as in Jordan's example (e) for $N=2$. We can easily replace the pair of spin components $(\hat{\sigma}_x, \hat{\sigma}_z)$ by a pair of quantities $(\hat{\alpha}, \hat{\beta})$ that do have the exact same eigenvalues as in Jordan's example:
\begin{equation}
\hat{\alpha} \equiv \frac{h}{2} \left(\hat{\sigma}_x + \frac{1}{2} \right), \quad
\hat{\beta} \equiv \hat{\sigma}_z + \frac{1}{2}.
\label{example-e-3}
\end{equation}
The amplitudes $\langle \beta | \alpha \rangle = (1/\sqrt{2}) \, e^{i\beta\alpha/\hbar}$ now express that if the spin in one direction is known, the two possible values of the spin in the directions orthogonal to that direction are equiprobable. Moreover, as we just saw, amplitudes $\langle \beta | \alpha \rangle$ satisfy the completeness or orthogonality relations  (\ref{NB2-orthogonality}). It follows that {\it any two orthogonal components of spin are canonically conjugate to one another on Jordan's new definition}! One can thus legitimately wonder whether this definition is not getting much too permissive. However, as we will now show, the main problem with Jordan's formalism is not that it is asking too little of its conjugate variables, but rather that it is asking too much of its canonical transformations!

Canonical transformations enter into the formalism in sec. 3, where \citet[pp.\ 13--16]{Jordan 1927f} introduced a simplified yet at the same time generalized version of equations (2ab) of {\it Neue Begr\"undung} I for probability amplitudes \citep[p.\ 821]{Jordan 1927b}. They are simplified in that there are no longer additional equations for the {\it Erg\"anzungsamplitude} (ibid., Eqs. (3ab)). They are generalized in that they are no longer restricted to systems with only one degree of freedom and, much more importantly, in that they are no longer restricted to cases where all quantities involved have purely continuous spectra. Quantities with partly or wholly discrete spectra are now also allowed. 

Recall how Jordan built up his theory in {\it Neue Begr\"undung} I (cf. our discussion in Section 2.3). He posited a number of axioms to be satisfied by his  probability amplitudes. He then constructed a model for these postulates. To this end he identified probability amplitudes with the integral kernels for certain canonical transformations. Starting with differential equations trivially satisfied by the amplitude $\langle p | q \rangle = e^{-ipq/\hbar}$ for some initial pair of conjugate variables $\hat{p}$ and  $\hat{q}$, Jordan derived differential equations for amplitudes involving other quantities related to the initial ones through canonical transformations. As we already saw above, this approach breaks down as soon as we ask about the probability amplitudes for quantities with partly discrete spectra, such as, typically, the Hamiltonian.

Although \citet[p. 14]{Jordan 1927f} emphasized that one has to choose initial $\hat{p}$'s and  $\hat{q}$'s with ``fitting spectra" ({\it passende Spektren}) and that the equations for the amplitudes are solvable only ``if it is possible to find" such spectra, he did not state explicitly in sec.\ 3 that the construction of {\it Neue Begr\"undung} I fails for quantities with discrete spectra.\footnote{In his {\it Mathematische Begr\"undung}, \citet{von Neumann 1927a} had already put his finger on this problem: ``A special difficulty with [the approach of] Jordan is that one has to calculate not just the transforming operators (the integral kernels of which are the ``probability amplitudes"), but also the value-range onto which one is transforming (i.e., the spectrum of eigenvalues)" (p.\ 3).}  That admission is postponed until the discussion of canonical transformations in sec.\ 4. At the beginning of sec.\ 3, the general equations for probability amplitudes are given in the form \citep[p. 14, Eqs.\ (2ab)]{Jordan 1927f}:
\begin{equation}
\Phi^{\beta q}_{\alpha p} \hat{B}_k - \hat{\beta}_k \Phi^{\beta q}_{\alpha p} = 0, 
\label{NB2-3-2a}
\end{equation}
\begin{equation}
\Phi^{\beta q}_{\alpha p} \hat{A}_k - \hat{\alpha}_k \Phi^{\beta q}_{\alpha p} = 0,   
\label{NB2-3-2b}
\end{equation}
where $\hat{A}_k$ and $\hat{B}_k$ are defined as [NB2, sec.\ 3, Eq.\ (1)]:
\begin{equation}
\hat{B}_k = \left( \Phi^{\beta q}_{\alpha p} \right)^{-1} \hat{\beta}_k \, \Phi^{\beta q}_{\alpha p}, \quad
\hat{A}_k = \left( \Phi^{\beta q}_{\alpha p} \right)^{-1} \hat{\alpha}_k \,  \Phi^{\beta q}_{\alpha p}
\label{def A and B}
\end{equation}
\citet[pp.\ 14--15]{Jordan 1927f}  then showed that the differential equations of {\it Neue Begr\"undung} I are included in these new equations as a special case. Since there is only one degree of freedom in that case, we do not need the index $k$. We can also suppress all indices of $\Phi^{\beta q}_{\alpha p}$ as this is the only amplitude/transformation-matrix involved in the argument. So we have $\hat{A} = \Phi^{-1} \hat{\alpha} \, \Phi$ and $\hat{B} = \Phi^{-1} \hat{\beta} \, \Phi$. These transformations, however, are used very differently in the two installments of {\it Neue Begr\"undung}.  Although Jordan only discussed this change in sec.\ 4, he already alerted the reader to it in sec.\ 3, noting that ``$\hat{B}$, $\hat{A}$ are the operators for $\hat{\beta}$, $\hat{\alpha}$ with respect to $\hat{q}$, $\hat{p}$" \citep[p.\ 15]{Jordan 1927f} 

Suppressing all subscripts and superscripts, we can rewrite Eqs.\ (\ref{NB2-3-2a})--(\ref{NB2-3-2b}) as:
\begin{equation}
(\Phi \, \hat{B} \, \Phi^{-1} - \hat{\beta}) \, \Phi = 0, 
\label{NB2-3-2a-stripped}
\end{equation}
\begin{equation}
(\Phi \, \hat{A} \, \Phi^{-1} - \hat{\alpha}) \, \Phi = 0.
\label{NB2-3-2b-stripped}
\end{equation}
Using that 
\begin{equation}
\Phi \, \hat{A} \, \Phi^{-1} = \hat{\alpha} = f(\hat{p}, \hat{q}), \quad \Phi \, \hat{B} \, \Phi^{-1} = \hat{\beta} = g(\hat{p}, \hat{q})
\end{equation}
\citep[p.\ 15, Eq.\ 8]{Jordan 1927f}; that $\hat{p}$ and $\hat{q}$ in the $q$-basis are represented by $(\hbar/i) \partial/\partial q$ and multiplication by $q$, respectively; and that $\hat{\alpha}$ and $\hat{\beta}$ in Eqs. (\ref{NB2-3-2a-stripped})--(\ref{NB2-3-2b-stripped}) are represented by $- (\hbar/i) \partial/\partial \beta$ and multiplication by $\beta$, respectively, we see that in this special case Eqs.\ (\ref{NB2-3-2a})--(\ref{NB2-3-2b}) (or, equivalently, Eqs.\ (\ref{NB2-3-2a-stripped})--(\ref{NB2-3-2b-stripped})) reduce to [NB2, p. 15, Eqs.\ (9ab)] 
\begin{equation}
\left( g\left( \frac{\hbar}{i} \frac{\partial}{\partial q}, q \right) - \beta \right) \Phi = 0, 
\label{NB2-3-2a'}
\end{equation}
\begin{equation}
\left( f\left( \frac{\hbar}{i} \frac{\partial}{\partial q}, q \right) +  \frac{\hbar}{i} \frac{\partial}{\partial \beta} \right) \Phi = 0,   
\label{NB2-3-2b'}
\end{equation}
which are just Eqs. (2a) of {\it Neue Begr\"undung} I \citep[p.\ 821; cf.\ Eqs.\ (\ref{NB1-2a})--(\ref{NB1-2b}) with $\langle q | \beta \rangle$ written as $\Phi$]{Jordan 1927b}. This is the basis for Jordan's renewed claim that his general equations for probability amplitudes contain both the time-dependent and the time-independent Schr\"odinger equations as a special case (cf. our discussion at the end of Section 2.3).  It is certainly true that, if the quantity $\hat{B}$ in Eq.\ (\ref{def A and B}) is chosen to be the Hamiltonian, Eq. \ (\ref{NB2-3-2a'}) turns into the time-independent Schr\"odinger equation. However, there is no canonical transformation that connects this equation for $\psi_n(q) = \langle q | E \rangle$ to the equations trivially satisfied by $\langle p | q \rangle$ that formed the starting point for Jordan's construction of his formalism in {\it Neue Begr\"undung} I.\footnote{This problem does not affect Dirac's version of the theory (see our discussion toward the end of Section 1.1).} 

\citet[pp.\ 16--17]{Jordan 1927f} finally conceded this point in sec.\ 4 of {\it Neue Begr\"un\-dung} II. Following \citet{Dirac 1927}, Jordan switched to a new conception of canonical transformations. Whereas before, he saw canonical transformations such as $\hat{\alpha} = T \hat{p} T^{-1}, \hat{\beta} = T \hat{q} T^{-1}$, as taking us from one pair of conjugate variables ($\hat{p}, \hat{q}$) to a {\it different} pair ($\hat{\alpha}, \hat{\beta}$), he now saw them as taking us from one particular {\it representation} of a pair of conjugate variables to a {\it different representation} of those {\it same} variables.\footnote{Cf.\ the passage from the letter from Dirac to Jordan of December 24, 1916, that we quoted toward the end of Section 1.1 (cf.\ note \ref{MR2}).}
 The canonical transformation used in sec.\ 3, $\hat{A} = \Phi^{-1} \hat{\alpha} \, \Phi, \hat{B} = \Phi^{-1} \hat{\beta} \, \Phi$, is already an example of a canonical transformation in the new Dirac sense. By giving up on canonical transformations in the older sense, Jordan effectively abandoned the basic architecture of the formalism of {\it Neue Begr\"undung} I. 

This is how Jordan explained the problem at the beginning of sec.\ 4 of  {\it Neue Begr\"undung} II:
\begin{quotation}
Canonical transformations, the theory of which, as in classical mechanics, gives the natural generalization and the fundamental solution of the problem of the integration of the equations of motion, were originally  [footnote citing \citet{dreimaenner}] conceived of as follows: the canonical quantities $\hat{q}$, $\hat{p}$ should be represented as functions of certain other canonical quantities $\hat{\beta}$, $\hat{\alpha}$:
$$
\hat{q}_k = G_k(\hat{\beta}, \hat{\alpha}), \quad \hat{p}_k = F_k(\hat{\beta}, \hat{\alpha}). \quad \quad \quad \quad (1)
$$
On the assumption that canonical systems can be defined through the usual canonical commutation relations, a formal proof could be given [footnote referring to \citet{Jordan 1926a}] that for canonical $\hat{q}$, $\hat{p}$ and $\hat{\beta}$, $\hat{\alpha}$ equations (1), as was already suspected originally, can always be cast in the form
$$
\hat{q}_k = T \, \hat{\beta}_k \, T^{-1}, \quad \hat{p}_k = T \, \hat{\alpha}_k \, T^{-1}. \quad \quad \quad \quad (2)
$$
However, since, as we saw, the old canonical commutation relations are not valid [cf.\ Eqs.\ (\ref{NB2-2})--(\ref{F=0 F'=1})], this proof too loses its meaning; in general, one can {\it not} bring equations (1) in the form (2).

Now a modified conception of canonical transformation was developed by Dirac [footnote citing \citet{Dirac 1927} and \citet{Lanczos 1926}]. According to Dirac, [canonical transformations] are not about representing certain canonical quantities as functions of other canonical quantities, but rather about switching, {\it without a transformation of the quantities themselves}, to a different {\it matrix representation} \citep[pp.\ 16--17; emphasis in the original, hats added]{Jordan 1927f}.
\end{quotation} 

In modern terms, canonical transformations in the new Dirac sense transform the matrix elements of an operator in one basis to matrix elements of that same operator in another. This works whether or not the operator under consideration is part of a pair of operators corresponding to canonically conjugate quantities. The notation $\Phi^{\beta q}_{\alpha p} = \langle \beta | q \rangle$ introduced in {\it Neue Begr\"undung} II that replaces the notation $T$ in {\it Neue Begr\"undung} I for the operators implementing a canonical transformation nicely prepared us for this new way of interpreting such transformations. Consider the matrix elements of the position operator $\hat{q}$ (with a purely continuous spectrum) in the $q$-basis:
\begin{equation}
\langle q' | \hat{q} | q'' \rangle = q' \delta(q' - q'').
\end{equation}
Now let $\hat{\beta}$ be an arbitrary self-adjoint operator. In general, $\hat{\beta}$ will have a spectrum with both continuous and discrete parts. Von Neumann's spectral theorem tells us that
\begin{equation}
\hat{\beta} = \sum_n \beta_n \, | \beta_n \rangle \langle \beta_n | + \int  \beta \,  | \beta \rangle \langle \beta | \, d\beta,
\label{beta spectral decomposition}
\end{equation}
where sums and integrals extend over the discrete and continuous parts of the spectrum of $\hat{\beta}$, respectively. In Jordan's notation, the spectral decomposition of $\hat{\beta}$ can be written more compactly as: $\hat{\beta} = \overline{\sum} \, \beta \,  | \beta \rangle \langle \beta |$. We now want to find the relation between the matrix elements of $\hat{q}$ in the $\beta$-basis and its matrix elements in the $q$-basis. Using the spectral decomposition $\int  q \,  | q \rangle \langle q | \, dq$ of $\hat{q}$, we can write:
\begin{eqnarray}
\langle \beta' | \hat{q} | \beta'' \rangle &  = & \int dq' \langle \beta' | q' \rangle q' \langle q' | \beta'' \rangle \nonumber \\
& = & \int dq' dq'' \, \langle \beta' | q'' \rangle q'' \delta(q' - q'') \langle q'' | \beta'' \rangle  \\
& = & \int dq' dq'' \, \langle \beta' | q'' \rangle \langle q'' | \hat{q} | q' \rangle \langle q'' | \beta'' \rangle \nonumber 
\end{eqnarray}
In the notation of {\it Neue Begr\"undung} II, the last line would be the `matrix multiplication', $\left( \Phi^{q\beta}_{q \alpha} \right)^{-1} \hat{q} \, \Phi^{q\beta}_{q \alpha}$. This translation into modern notation shows that Jordan's formalism, even with a greatly reduced role for canonical transformations, implicitly relies on the spectral theorem, which \citet{von Neumann 1927a} published in {\it Mathematische Begr\"undung}, submitted just one month before {\it Neue Begr\"undung} II. 
We also note, however, that an explicit choice of quantities $\hat{p}$ and $\hat{\alpha}$ conjugate to $\hat{q}$ and $\hat{\beta}$, respectively, is completely irrelevant  for the application of the spectral theorem.

This gets us to the last aspect of {\it Neue Begr\"undung} II that we want to discuss in this section, namely Jordan's response to von Neumann's criticism of the Dirac-Jordan transformation theory. A point of criticism we already mentioned is that the probability amplitude $\varphi(\beta, q) = \langle \beta | q \rangle$ is determined only up to a phase factor. 

As we mentioned in Section 1.3, the projection operator $| a \rangle \langle a|$ does not change if the ket $|a \rangle$ is replaced by $e^{i\vartheta} | a \rangle$ and the bra $\langle a |$ accordingly by  $e^{-i\vartheta} \langle a |$, where $\vartheta$ can be an arbitrary real function of $a$. Hence, the spectral decomposition of $\hat{\beta}$ in Eq.\ (\ref{beta spectral decomposition}) does not change if $| \beta \rangle$ is replaced by $e^{-i \rho(\beta)/\hbar} | \beta \rangle$, where we have written the phase factor, in which $\rho$ is an arbitrary real function of $\beta$, in a way that corresponds to Jordan's notation for the resulting phase ambiguity in the amplitude $\varphi(\beta, q) = \langle \beta | q \rangle$. Similarly, the spectral decomposition of $\hat{q}$ does not change if we replace $| q \rangle$ by $e^{i \sigma(q)/\hbar} | q \rangle$, where $\sigma$ is an arbitrary real function of $q$. However, with the changes $| \beta \rangle  \rightarrow  e^{-i \rho(\beta)/\hbar} | \beta \rangle$ and $| q \rangle  \rightarrow  e^{i \sigma(q)/\hbar} | q \rangle$, the amplitude $\varphi(\beta, q) = \langle \beta | q \rangle$ changes \citep[p.\ 20]{Jordan 1927f}: 
\begin{equation}
 \langle \beta | q \rangle \quad \longrightarrow \quad e^{i(\rho(\beta) + \sigma(q))/\hbar} \langle \beta | q \rangle.
 \label{non-uniqueness 1}
\end{equation}
Unlike projection operators, as von Neumann pointed out, probability amplitudes, are determined only up to such phase factors. 

As we noted above, Jordan responded to this criticism by adding a dependence on quantities $\hat{\alpha}$ and $\hat{p}$ conjugate to $\hat{\beta}$ and $\hat{q}$, respectively, to the probability amplitude $\langle \beta | q \rangle$, thus arriving at the amplitudes $\Phi_{\alpha p}(\beta', q')$ of {\it Neue Begr\"undung} II. It turns out that the phase ambiguity of $\langle \beta | q \rangle$ is equivalent to a certain freedom we have in the definition of the quantities $\hat{\alpha}$ and $\hat{p}$ conjugate to $\hat{\beta}$ and $\hat{q}$, respectively. The phase ambiguity, as we saw, can be characterized by the arbitrary functions $\rho(\beta)$ and $\sigma(q)$. Following Jordan, we will show that our freedom in the definition of $\hat{\alpha}$ and $\hat{p}$ is determined by the derivatives $\rho'(\beta)$ and $\sigma'(q)$ of those same functions. By considering amplitudes $\Phi_{\alpha p}(\beta', q')$ with uniquely determined $\hat{\alpha}$ and $\hat{p}$, Jordan could thus eliminate the phase ambiguity that von Neumann found so objectionable.

Following \citet[p.\ 20]{Jordan 1927f}, we establish the relation between these two elements of arbitrariness for the special case that all quantities involved have fully continuous spectra. In sec.\ 6 on spin, \citet[pp.\ 21--25]{Jordan 1927f} tried to extend his argument to some special cases of discrete spectra. We will not discuss those efforts. 

Consider two complete sets of eigenstates of $\hat{q}$, $\{ |q \rangle_1 \}$ and $\{ |q \rangle_2 \}$, related to one another via
\begin{equation}
|q \rangle_1 = e^{i \sigma(q)/\hbar} \, | q \rangle_2.
\label{phase ambiguity}
\end{equation}
This translates into two different amplitudes, $\varphi_1(\beta, q) \equiv \langle \beta | q \rangle_1$ and $\varphi_2(\beta, q) \equiv \langle \beta | q \rangle_2$, that differ by that same phase factor: $\varphi_1(\beta, q) = e^{i \sigma(q)/\hbar}  \,\varphi_2 (\beta, q)$.\footnote{An argument completely analogous to one we give for the relation between the phase factor $e^{i \sigma(q)/\hbar}$ and the definition of $\hat{p}$ can be  be given for the relation between the phase factor $e^{i \rho(\beta)/\hbar}$ and the definition of $\hat{\alpha}$.} Suppose $\hat{p}_1$ is conjugate to $\hat{q}$ if we use the $| q \rangle_1$ set of eigenstates of $\hat{q}$. Since we restrict ourselves to quantities with fully continuous spectra, this means that $[\hat{p}_1, \hat{q}] = \hbar/i$. It also means, as we saw in Section 2.1, that 
\begin{equation}
\hat{p}_1 | q \rangle_1 = - \frac{\hbar}{i} \frac{\partial}{\partial q} | q \rangle_1.
\label{q1 states}
\end{equation}
Eigenstates $| p \rangle_1$ of $\hat{p}_1$ can be written as a Fourier series in terms of the $| q \rangle_1$ states:\footnote{One easily verifies that $| p \rangle_1$ is indeed an eigenstate of $\hat{p}_1$. The action of $\hat{p}_1$ on $| p \rangle_1$ can be written as:
$$
\hat{p}_1 | p \rangle_1 =   \int dq \, e^{ipq/\hbar} \hat{p}_1 | q \rangle_1
=  - \int dq \, e^{ipq/\hbar} \frac{\hbar}{i} \frac{\partial}{\partial q}  | q \rangle_1.
$$
Partial integration gives:
$$
\hat{p}_1 | p \rangle_1 = \int dq \, \frac{\hbar}{i} \frac{\partial}{\partial q} \left( e^{ipq/\hbar} \right) | q \rangle_1 
= p  \int dq \, e^{ipq/\hbar} | q \rangle_1 = p | p \rangle_1,
$$
which is what we wanted to prove.}
\begin{equation}
| p \rangle_1 = \int dq \, e^{ipq/\hbar} \, | q \rangle_1.
\label{p1 states}
\end{equation}
We can likewise construct a $\hat{p}_2$ conjugate to $\hat{q}$ if we use the $| q \rangle_2$ set of eigenstates of $\hat{q}$. Instead
of Eqs.\ (\ref{q1 states})--(\ref{p1 states}), we then have
\begin{equation}
\hat{p}_2 | q \rangle_2 = - \frac{\hbar}{i} \frac{\partial}{\partial q} | q \rangle_2,
\quad \quad
| p \rangle_2 = \int dq \, e^{ipq/\hbar} \, | q \rangle_2.
\label{q2 p2 states}
\end{equation}
The relation between these two different conjugate momenta, it turns out, is
\begin{equation}
\hat{p}_2 = \hat{p}_1 + \sigma'(\hat{q}).
\label{p ambiguity}
\end{equation}
Note that the commutator $[ \hat{p}_1, \hat{q}]$ does not change if we add an arbitrary function of $\hat{q}$ to $\hat{p}_1$.
To prove that Eq.\ (\ref{p ambiguity}) indeed gives the relation between $\hat{p}_1$ and $\hat{p}_2$, we show that $| p \rangle_2$ in Eq.\ (\ref{q2 p2 states}) is an eigenstate of $\hat{p}_2$ as defined in Eq.\ (\ref{p ambiguity}), using relation (\ref{phase ambiguity}) between $| q \rangle_1$ and $| q \rangle_2$:
\begin{eqnarray}
\hat{p}_2 | p \rangle_2 & =  & \int dq \, e^{ipq/\hbar} \, \hat{p}_2 \, | q \rangle_2 \nonumber \\
 & = &  \int dq \, e^{ipq/\hbar} \, (\hat{p}_1 + \sigma'(\hat{q})) \, e^{-i \sigma(q)/\hbar} \, | q \rangle_1 \nonumber \\
 & = &  \int dq \, e^{i(pq - \sigma(q))/\hbar} \, \left(- \frac{\hbar}{i} \frac{\partial}{\partial q} + \sigma'(q)\right) \, | q \rangle_1  
  \label{non-uniqueness 2}
 \\
 & = &  \int dq \, (p - \sigma'(q) + \sigma'(q)) \, e^{i(pq - \sigma(q))/\hbar}  | q \rangle_1 \nonumber \\
 & = & \; p \int dq \, e^{ipq/\hbar} \, | q \rangle_2 = p \, |p \rangle_2, \nonumber
\end{eqnarray}
where in the fourth step we performed a partial integration. This proves that the ambiguity (\ref{p ambiguity}) in the $\hat{p}$ conjugate to $\hat{q}$ corresponds directly to the phase ambiguity (\ref{phase ambiguity}) in the amplitude $\langle \beta | q\rangle$. Similarly, the ambiguity in the $\hat{\alpha}$ conjugate to $\hat{\beta}$, which is determined only up to a term $\rho'(\hat{\beta})$, corresponds directly to the phase ambiguity $e^{i\rho(\beta)/\hbar}$ in the amplitude $\langle \beta | q\rangle$. Hence, for specific conjugate variables $\hat{p}$ and $\hat{\alpha}$ conjugate to $\hat{q}$ and $\hat{\beta}$, the amplitude $\Phi_{\alpha p}(\beta', q')$ of {\it Neue Begr\"undung} II is unique up to a constant phase factor (i.e., one that is not a function of $q$ or $\beta$). 

In addition to responding to von Neumann's criticism of his approach,  \citet[p.\ 20]{Jordan 1927f} also offered some criticism of von Neumann's approach. In particular, he complained that von Neumann showed no interest in either canonical transformations or conjugate variables. As we will see when we cover von Neumann's {\it Mathematische Begr\"undung} in the next section, this is simply because von Neuman did not need either for his formulation of quantum mechanics. That formulation clearly did not convince Jordan. In fact, von Neumann's paper only seems to have increased Jordan's confidence in his own approach. After his brief discussion of {\it Mathematische Begr\"undung}, he concluded: ``It thus appears that the amplitudes themselves are to be considered the fundamental concept of quantum mechanics" \citep[pp.\ 20--21]{Jordan 1927f}. 

It is unclear whether Jordan ever came to appreciate the advantages of von Neumann's approach over his own. In the preface of his texbook on quantum mechanics, {\it Anschauliche Quantentheorie}, \citet{Jordan 1936} described the statistical transformation theory of Dirac and himself as ``the pinnacle of the development of quantum mechanics" (p.\ VI) and as the ``most comprehensive and profound version of the quantum laws" (ibid., p.\ 171; quoted by Duncan and Janssen, 2009, p.\ 361). He did not discuss any of von Neumann's contributions in this book. However, in the textbook he co-authored with Born, {\it Elementare Quantenmechanik}, we do find elaborate expositions \citep[Ch.\ 6, pp. 288--364]{Born and Jordan 1930} of the two papers by \citet{von Neumann 1927a, von Neumann 1927b} that will occupy us in the next two sections.

\section{Von Neumann's {\it Mathematische Begr\"undung} (May 1927)}

In the next two sections we turn our attention to the first two papers of the trilogy that \citet{von Neumann 1927a, von Neumann 1927b, von Neumann 1927c} published the same year as and partly in response to the papers by \citet{Dirac 1927} and \citet{Jordan 1927b} on transformation theory. This trilogy provided the backbone of his famous book published five years later \citep{von Neumann 1932}. The first paper in the trilogy, {\it Mathematische Begr\"undung}, was presented in the meeting of the G\"ottingen Academy of May 20, 1927. In this paper, von Neumann first introduced the Hilbert space formalism and the spectral theorem, at least for bounded operators, two contributions that have since become staples of graduate texts in quantum physics and functional analysis.\footnote{In the latter category we already mentioned \citet{Prugovecki 1981} and \citet[Ch.\ 3]{Dennery and Krzywicki 1996} (see note \ref{Prugovecki}).} In part because of this greater familiarity but also because of its intrinsic clarity, von Neumann's {\it Mathematische Begr\"undung} is much easier to follow for modern readers than Jordan's {\it Neue Begr\"undung}. There is  no need for us to cover it in as much detail as we did with Jordan's papers in Sections 2 and 4.\footnote{This paper is also discussed by \citet[Ch.\ III, sec.\ 2(d), pp.\ 411--417]{Mehra Rechenberg}.} 

{\it Mathematische Begr\"undung} is divided into nine parts, comprising 15 sections and two appendices:
\begin{quotation}
\begin{enumerate}
\item ``Introduction," sec.\ I, pp.\ 1--4;
\item ``The Hilbert space," secs.\ II--VI, pp.\ 4--22;
\item  ``Operator calculus," secs.\ VII--VIII, pp.\ 22--29;
\item ``The eigenvalue problem," sec.\ IX--X,  pp.\ 29--37;
\item ``The absolute value of an operator," sec.\ $\!\!$``IX" (a typo: this should be XI),  pp.\ 37--41;
\item ``The statistical assumption [{\it Ansatz}] 
of quantum mechanics," secs.\ XII--XIII,  pp.\ 42--47;
\item ``Applications," sec.\ XIV, pp.\ 47--50;
\item ``Summary," sec.\ XV, pp.\ 50--51;
\item ``Appendices," pp.\ 51--57.
\end{enumerate}
\end{quotation}
Abstract Hilbert space is introduced in secs.\ V--VI, the spectral theorem in secs.\ IX--X.
After going over the introduction of the paper, we focus on parts of sec.\ IV and secs.\ IX--XIII.

In Sec.\ IV, von Neumann criticized the way in which wave mechanics and matrix mechanics are unified in the approach of Dirac and Jordan and presented his superior alternative approach to this unification, based on the isomorphism of two concrete instantiations of abstract Hilbert space $\overline{\textfrak{H}}$, the space of square-summable sequences $\textfrak{H}_0$ and the space of square-integrable functions  $\textfrak{H}$  \citep[the designations $\textfrak{H}$, $\textfrak{H}_0$, and $\overline{\textfrak{H}}$ are introduced on pp.\ 14--15]{von Neumann 1927a}. In modern notation, this is the isomorphism between $l^2$ and $L^2$.  

Secs.\ IX--XIII contain von Neumann's criticism of Jordan's use of probability amplitudes and his derivation of an alternative formula for  conditional probabilities in quantum mechanics in terms of projection operators. Unlike von Neumann, we present this derivation in Dirac notation.

In the introduction of {\it Mathematische Begr\"undung}, \citet[pp.\ 1--3]{von Neumann 1927a} gave a list of seven points, labeled $\alpha$ through $\vartheta$ (there is no point $\eta$), in which he took stock of the current state of affairs in the new quantum theory and identified areas where it ran into mathematical difficulties. We paraphrase these points. ($\alpha$) Quantum theory describes the behavior of atomic systems in terms of certain eigenvalue problems. ($\beta$) This allows for a unified treatment of continuous and discontinuous elements in the atomic world.  ($\gamma$) The theory suggests that the laws of nature are stochastic.\footnote{Parenthetically, \citet[p.\ 1]{von Neumann 1927a} added an important qualification, ``(at least the quantum laws known to us)," leaving open the possibility that, at a deeper level, the laws would be deterministic again. Von Neumann's position at this point was thus basically the same as Jordan's (see note \ref{jordan indeterminism}). By the time of the second paper of his trilogy, von Neumann had read Heisenberg's (1927b) uncertainty paper and endorsed Heisenberg's position that the indeterminism of quantum mechanics is the result of the inevitable disturbance of quantum systems in our measurements \citep[p.\  273; cf.\ note \ref{neumann heisenberg}]{von Neumann 1927b}.} 
($\delta$) Returning to the formulation of the theory in terms of eigenvalue problems, von Neumann briefly characterized the different but equivalent ways in which such problems are posed in matrix mechanics and in wave mechanics. Both approaches have their difficulties. ($\varepsilon$) The application of matrix mechanics appears to be restricted to situations with purely discrete spectra. To deal with wholly or partly continuous spectra, one ends up using, side by side, matrices with indices taking on discrete values and ``continuous matrices," i.e., the integral kernels of the Dirac-Jordan transformation theory, with `indices'  taking on continuous values. It is ``very hard,"  \citet[p.\ 2]{von Neumann 1927a} warned, to do this in a mathematically rigorous way. ($\zeta$) These same problems start to plague 
wave mechanics as soon as wave functions are interpreted as probability amplitudes. Von Neumann credited Born, Pauli, and Jordan with transferring the probability concepts of matrix mechanics to wave mechanics and Jordan with developing these ideas into a ``closed system" (ibid.).\footnote{In this context \citet[p.\ 2]{von Neumann 1927a} referred to his forthcoming paper with Hilbert and Nordheim (1928). Oddly, von Neumann did not mention Dirac at this point, although Dirac is mentioned (alongside Pauli and Jordan) in sec.\ XII \citep[p.\ 43]{von Neumann 1927a} as well as in the second paper of the trilogy \citep[p.\ 245; see Section 6]{von Neumann 1927b}.\label{Dirac snub}} This system, however, faces serious mathematical objections because of the unavoidable use of improper eigenfunctions, such as the Dirac delta function, the properties of which von Neumann thought were simply ``absurd" (ibid., p.\ 3). His final objection seems mild by comparison but weighed heavily for von Neumann: ($\vartheta$) eigenfunctions in wave mechanics and probability amplitudes in transformation theory are determined only up to an arbitrary phase factor. The probabilities  one ultimately is after in quantum theory do not depend on these phase factors and von Neumann therefore wanted to avoid them altogether.\footnote{In {\it Neue Begr\"undung} II, as we saw in Section 5, \citet[p.\ 8]{Jordan 1927f} responded to this criticism by adding subscripts to the probability amplitudes for two quantities $\hat{\beta}$ and $\hat{q}$ indicating a specific choice of the canonically-conjugate quantities $\hat{\alpha}$ and $\hat{p}$ (see Eqs.\ \eqref{non-uniqueness 1}--\eqref{non-uniqueness 2}).}

In sec.\ II, von Neumann set the different guises in which the eigenvalue problems appear in matrix and in wave mechanics side by side. In matrix mechanics, the problem is to find square-summable infinite sequences of complex numbers $\bold{v} = (v_1, v_2, \ldots)$ such that 
\begin{equation}
\bold{H} \bold{v} = E  \bold{v},
\label{eigenvalues l2}
\end{equation}
where $\bold{H}$ is the matrix 
representing the Hamiltonian of the system in matrix mechanics and $E$ is an energy eigenvalue. In wave mechanics, the problem is to find square-integrable complex-valued functions $f(x)$ such that
\begin{equation}
\hat{H}  f(x) = E  f(x),
\label{eigenvalues L2}
\end{equation}
where $\hat{H}$ is the differential operator, involving multiplication by $x$ and differentiation with respect to $x$, that represents the Hamiltonian of the system in wave mechanics.

One way to unify these two approaches, \citet[pp.\ 10--11]{von Neumann 1927a} pointed out at the beginning of sec.\ IV, is to look upon the discrete set of values $1, 2, 3 \ldots$ of the index $i$ of the sequences $\{ v_i \}_{i=1}^\infty$ in matrix mechanics and the continuous (generally multi-dimensional) domain $\Omega$ of the functions $f(x)$ in wave mechanics as two particular realizations of some more general space, which von Neumann called $R$. 
Following the notation of his book \citep[ sec.\ 4, pp.\ 15--16]{von Neumann 1932}, we call the `space' of index values $Z$. Eq.\ \eqref{eigenvalues l2} can then be written as:\footnote{We  replaced von Neumann's (1927a, p.\ 10) $x_i$'s by $v_i$'s to avoid confusion with the argument(s) of the functions $f(x)$.\label{v for x}}
\begin{equation}
\sum_{j \in Z} H_{ij} v_j = E  v_i.
\label{eigenvalues l2 Z}
\end{equation}
`Summation over $Z$' can be seen as one instantiation of `integration over $R$;' `integration over $\Omega$' as another. In this way Eq.\ \eqref{eigenvalues L2} can, at least formally, be subsumed under matrix mechanics. One could represent the operator $\hat{H}$ in Eq.\ \eqref{eigenvalues L2} by the integral kernel $H(x, y)$ and write
\begin{equation}
\int_\Omega dy \, H(x,y) f(y) = E f(x).
\label{eigenvalues L2 Omega}
\end{equation}
Both the matrix $H_{ij}$ and the integral kernel $H(x,y)$ can be seen as `matrices' $H_{xy}$ with indices $x, y \in R$. For $H_{ij}$, $R = Z$; for $H(x,y)$, $R = \Omega$. Von Neumann identified this way of trying to unify matrix and wave mechanics as Dirac's way (and, one may add, although he is not mentioned by name at this point: Jordan's way). Von Neumann rejected this approach. He dismissed the analogy between $Z$ and $\Omega$ sketched above as ``very superficial, as long as one sticks to the usual measure of mathematical rigor" \citep[p.\ 11]{von Neumann 1927a}.\footnote{In Section 1.3, we already quoted some passages from the introduction of von Neumann's 1932 book in which he complained about the lack of mathematical rigor in Dirac's approach. After characterizing the approach in terms of the analogy between $Z$ and $\Omega$, he wrote: ``It is no wonder that this cannot succeed without some violence to formalism and mathematics: the spaces $Z$ and $\Omega$ are really very different and every attempt  to establish a relation between them must run into great difficulties" \citep[p.\ 15]{von Neumann 1932}.} He pointed out that even the simplest linear operator, the identity operator, does not have a proper integral-kernel representation. Its integral kernel is the improper Dirac delta function: $\int dy \, \delta(x-y) f(y) = f(x)$.

The appropiate analogy, \citet[pp.\ 11--14]{von Neumann 1927a} argued, is not between $Z$ and $\Omega$, but between the space of square-summable sequences {\it over} $Z$ and the space of square-integrable functions {\it over} $\Omega$. In his book, \citet[p.\ 16]{von Neumann 1932} used the notation $F_Z$ and $F_\Omega$ for these two spaces.\footnote{\citet[pp.\ 314--315]{Jammer 1966} also used this 1932 notation in his discussion of \citet{von Neumann 1927a}.} In 1927, as mentioned above, he used $\textfrak{H}_0$ and  $\textfrak{H}$, instead. Today they are called $l^2$ and $L^2$, respectively.\footnote{Earlier in his paper, \citet[p.\ 7]{von Neumann 1927a} remarked that what we now call $l^2$ was usually called ``(complex) Hilbert space." Recall, however, that \citet[p.\ 197]{London 1926b} used the term ``Hilbert space" for $L^2$ (note \ref{Erhard}).} 
Von Neumann (1927a, pp.\ 12--13) reminded his readers of the ``Parseval formula," which maps sequences in $l^2$ onto functions in $L^2$, and a ``theorem of Fischer and F.\ Riesz," which maps functions in $L^2$ onto sequences in $l^2$.\footnote{The paper cited by \citet[p.\ 13, note 15]{von Neumann 1927a} is \citet{Riess 1907a}. In his discussion of von Neumann's paper, \citet[pp.\ 314--315]{Jammer 1966} cited \citet{Riess 1907a, Riess 1907b} and \citet{Fischer 1907}.} The combination of these two results establishes that $l^2$ and $L^2$ are isomorphic. As \citet[p.\ 12]{von Neumann 1927a} emphasized, these ``mathematical facts that had long been known" could be used to unify matrix mechanics and wave mechanics in a mathematically impeccable manner. With a stroke of the pen, von Neumann thus definitively settled the issue of the equivalence of wave mechanics and matrix mechanics. {\it Anything that can be done in wave mechanics, i.e., in $L^2$, has a precise equivalent in matrix mechanics, i.e., in $l^2$}. This is true regardless of whether we are dealing with discrete spectra, continuous spectra, or a combination of the two. 

In sec.\ V,  \citet[pp.\ 14--18]{von Neumann 1927a} introduced abstract Hilbert space,  for which he used the notation $\overline{\textfrak{H}}$, carefully defining it in terms of five axioms labeled A through E.\footnote{In his book, \citet{von Neumann 1932} adopted the notation `H.\ R.' (shorthand for {\it Hilbertscher Raum}) for $\overline{\textfrak{H}}$.}  In sec.\ VI, he added a few more definitions and then stated and proved six theorems about Hilbert space, labeled 1 through 6 (ibid., pp. 18--22). In sec.\ VII, he turned to the discussion of operators acting in Hilbert space (ibid., pp.\ 25). This will be familiar terrain for the modern reader and need not be surveyed in any more detail.

The same goes for sec.\ VIII, in which von Neumann introduced a special class of Hermitian operators. Their defining property is that they are idempotent: $\hat{E}^2 = \hat{E}$. Von Neumann called an operator like this an {\it Ein\-zel\-operator} or {\it E.Op.} for short \citep[p.\ 25]{von Neumann 1927a}.\footnote{As he explains in a footnote, the term {\it Einzeloperator}  is based on Hilbert's term {\it Einzelform} \citep[p.\ 25, note 23]{von Neumann 1927a}. For historical discussion, see \citet[pp.\ 317--318]{Jammer 1966}.\label{E.Op.}} They are now known as projection operators. In a series of theorems, numbered 1 through 9, \citet[pp.\ 26--29]{von Neumann 1927a} proved some properties of such operators. For our purposes, it suffices to know that they are Hermitian and idempotent.

We do need to take a  closer look at sec.\ IX. In this section, \citet[pp.\ 29--33]{von Neumann 1927a} used projection operators to formulate the spectral theorem. 
Following \citet[p.\ 31]{von Neumann 1927a}, we start by considering a finite Hermitian operator $\hat{A}$ with a non-degenerate discrete spectrum. Order its real eigenvalues: $a_1 < a_2 < a_3 \ldots \;$ Let $|a_i \rangle$ be the associated normalized eigenvectors  ($\langle a_i | a_j \rangle = \delta_{ij}$). Now introduce the operator $\hat{E}(l)$:\footnote{Von Neumann initially defined this operator in terms of its matrix elements $\langle v | \hat{E}(l) | w \rangle$ for two arbitrary sequences $\{ v_i \}_{i=1}^\kappa$ and $\{ w_i \}_{i=1}^\kappa$ (where we replaced von Neumann's $x$ and $y$ by $v$ and $w$; cf.\ note \ref{v for x}). He defined (in our notation): $E(l; x|y) = \sum_{(i | a_i \leq l)} \langle v | a_i \rangle \langle a_i | w \rangle$ \citep[p.\ 31]{von Neumann 1927a}. \label{E Op matrix elements}}
\begin{equation}
\hat{E}(l) \equiv \sum_{(i | a_i \leq l)} |a_i \rangle \langle a_i |,
\label{E Op 1}
\end{equation}
where, unlike von Neumann, we used modern Dirac notation. As we already noted in Section 1.3, there is no phase ambiguity in $\hat{E}(l)$. The operator stays the same if we replace  $|a_i \rangle$ by $|a_i \rangle' = e^{i \varphi_i} |a_i \rangle$:
\begin{equation}
|a_i \rangle' \langle a_i |' = e^{i \varphi_i} |a_i \rangle \langle a_i | e^{- i \varphi_i} = | a_i \rangle \langle a_i |.
\label{E no phase}
\end{equation}
Of course, von Neumann did not think of an {\it E.Op.} as constructed out of bras and kets, just as Jordan did not think of a probability amplitude $\langle a|b \rangle$ as an inner product of $|a \rangle$ and $|b \rangle$.

The operator $\hat{E}(l)$ has the property:
\begin{equation}
\hat{E}(a_i) - \hat{E}(a_{i-1}) = |a_i \rangle \langle a_i |.
\label{E Op 2}
\end{equation}
It follows that:
\begin{equation}
\hat{A} = \sum_i a_i (\hat{E}(a_i) - \hat{E}(a_{i-1})) = \sum_i  a_i  | a_i \rangle \langle a_i |
\label{E Op 3}
\end{equation}
$\hat{E}(l)$ is piece-wise constant with jumps where $l$ equals an eigenvalue. Hence we can write $\hat{A}$ as a so-called Stieltjes integral, which von Neumann discussed and illustrated with some figures in appendix 3 of his paper \citep[pp.\ 55--57]{von Neumann 1927a}:
\begin{equation}
\hat{A} = \int l \, d\hat{E}(l).
\label{E Op 4}
\end{equation}
As \citet[p.\ 32]{von Neumann 1927a} noted, these results (Eqs.\ \eqref{E Op 1}--\eqref{E Op 4}) can easily be generalized from finite Hermitian matrices and finite sequences to bounded Hermitian operators and the space $\textfrak{H}_0$ or $l^2$ of infinite square-summable sequences. Since $\textfrak{H}_0$ is just a particular instantiation of the abstract Hilbert space $\overline{\textfrak{H}}$, it is clear that the same results hold for bounded Hermitian operators $\hat{T}$ in $\overline{\textfrak{H}}$. After listing the key properties of $\hat{E}(l)$ for $\hat{T}$,\footnote{As before (see note \ref{E Op matrix elements}), he first defined the matrix elements $\langle f | \hat{E}(l) | g \rangle$ for two arbitrary elements $f$ and $g$ of Hilbert space. So he started from the relation 
$$\langle f | \hat{T} | g \rangle = \int_{- \infty}^\infty l \, d\langle f | \hat{E}(l) | g \rangle,$$ 
and inferred from that, first, that $\hat{T} | g \rangle = \int_{- \infty}^\infty l \, d \{ \hat{E}(l) | g \rangle \}$, and, finally, that $\hat{T} = \int_{- \infty}^\infty l \, d \hat{E}(l)$ (cf.\ Eq. \eqref{E Op 4}). Instead of the notation $\langle f | g \rangle$, \citet[p.\ 12]{von Neumann 1927a} used the notation $Q(f, g)$ for the inner product of $f$ and $g$ (on p.\ 32, he also used $Q(f|g)$). So, in von Neumann's own notation, the relation he started from is written as  $Q(f, Tg) = \int_{- \infty}^\infty l \, dQ(f, E(l)g)$ \citep[p.\ 33]{von Neumann 1927a}. \label{inner product}}  he concluded sec.\ IX writing: ``We call $\hat{E}(l)$ the resolution of unity [{\it Zerlegung der Einheit}] belonging to $\hat{T}$" \citep[p.\ 33]{von Neumann 1927a}.

In sec.\ X, \citet[pp.\ 33--37]{von Neumann 1927a} further discussed the spectral theorem. Most importantly, he conceded that he had not yet been able to prove that it also holds for {\it unbounded} operators.\footnote{We remind the reader that a linear operator $\hat{A}$ in Hilbert space is bounded if there exists a positive real constant $C$ such that $|\hat{A} f| < C|f|$ for arbitrary vectors $f$ in the space (where $|...|$ indicates the norm of a vector, as
induced from the defining inner-product in the space). If this is {\em not} the case, then there exist vectors in the Hilbert
space on which the operator $\hat{A}$ is not well-defined, basically because the resultant vector has infinite norm. Instead,
such unbounded operators are only defined (i.e., yield finite-norm vectors) on a proper subset of the Hilbert space,
called the {\em domain} ${\mathcal D}(\hat{A})$ of the operator $\hat{A}$. The set of vectors obtained by applying $\hat{A}$ to all
elements of its domain is called the {\em range} ${\mathcal R}(\hat{A})$ of $\hat{A}$.  
Multiplication of two unbounded operators evidently becomes a delicate matter insofar as the domains and ranges of the respective operators may not coincide.
\label{bounded/unbounded}} 
He only published the proof of this generalization in a paper in {\it Mathematische Annalen} submitted on December 15, 1928 \citep{von Neumann 1929}.\footnote{For brief discussions, see \citet[p.\ 320]{Jammer 1966} and \citet[p.\ 415]{Mehra Rechenberg}.}
The key to the extension of the spectral theorem from bounded to unbounded operators is a so-called Cayley transformation \citep[p.\ 80]{von Neumann 1929}. Given an unbounded Hermitian operator $\hat{R}$, introduce the operator $\hat{U}$ and its adjoint 
\begin{equation}
\hat{U} = \frac{\hat{R} +i \hat{1}}{\hat{R} - i \hat{1}}, \quad \quad \hat{U}^\dagger = \frac{\hat{R} - i \hat{1}}{\hat{R} + i \hat{1}},
\label{Cayley}
\end{equation}
where $\hat{1}$ is the unit operator. Since $\hat{R}$ is Hermitian, it only has real eigenvalues, so $(\hat{R} - i \hat{1}) | \varphi \rangle \neq 0$ for any $|\varphi \rangle \in \overline{\textfrak{H}}$. Since $\hat{U}$ is unitary ($\hat{U}\hat{U}^\dagger = \hat{1}$), the absolute value of all its eigenvalues equals 1. $\hat{U}$ is thus a bounded operator for which the spectral theorem holds. If it holds for $\hat{U}$, however, it must also hold for the original unbounded operator $\hat{R}$. The spectral decomposition of $\hat{R}$ is essentially the same as that of $\hat{U}$. In his book, \citet[p.\ 80]{von Neumann 1932} gave Eq.\ \eqref{Cayley}, but he referred to his 1929 paper for a mathematically rigorous treatment of the spectral theorem for unbounded operators \citep[p.\ 75, p.\ 246, note 95, and p.\ 244, note 78]{von Neumann 1932}

Sec.\ XI concludes the purely mathematical part of the paper. In this section, \citet[pp.\ 37-41]{von Neumann 1927a} introduced the ``absolute value" of an operator, an important ingredient, as we will see, in his derivation of his formula for conditional probabilities in quantum mechanics  (see Eqs.\ \eqref{trace vN1}--\eqref{trace vN5} below). 

In sec.\ XII, \citet[pp.\ 42--45]{von Neumann 1927a} finally turned to the statistical interpretation of quantum mechanics. At the end of sec.\ I, he had already warned the reader that secs.\ II--XI would have a ``preparatory character" and that he would only get to the real subject matter of the paper in secs.\ XII--XIV. At the beginning of sec.\ XII, the first section of the sixth part of the paper (see our table of contents above), on the statistical interpretation of quantum mechanics, he wrote: ``We are now in a position to take up our real task, the mathematically unobjectionable unification of statistical quantum mechanics" \citep[p.\ 42]{von Neumann 1927a}. He then proceeded to use the spectral theorem and the projection operators $\hat{E}(l)$ of sec.\ IX to construct an alternative to Jordan's formula for conditional probabilities in quantum mechanics, which does not involve probability amplitudes. Recall von Neumann's objections to probability amplitudes (see Sections 1 and 4). First, Jordan's basic amplitudes, $\rho(p,q) = e^{-ipq/\hbar}$ (see Eq.\ \eqref{NB1-18}), which from the perspective of Schr\"odinger wave mechanics are eigenfunctions of momentum, are not square-integrable and hence not in Hilbert space \citep[p.\ 35]{von Neumann 1927a}. Second, they are only determined up to a phase factor \citep[p.\ 3, point $\vartheta$]{von Neumann 1927a}. Von Neumann avoided these two problems by deriving an alternative formula which expresses the conditional probability ${\rm Pr}(a|b)$ in terms of projection operators associated with the spectral decomposition of the operators for the observables $\hat{a}$ and $\hat{b}$.

Von Neumann took over Jordan's basic statistical {\it Ansatz}. Consider a one-particle system in one dimension with coordinate $q$. Von Neumann (1927a, p.\ 43)
considered the more general case with coordinates $q \equiv (q_1, \ldots, q_k)$. The probability of finding a particle in some region $K$ if we know that its energy is $E_n$ (i.e., if we know the particle is in the pure state $\psi_n(x)$ belonging to that eigenvalue), is given by (ibid.):\footnote{The left-hand side is short-hand for: ${\rm Pr}(\hat{q} {\rm \; has \; value} \; q \; {\rm in} \; K| \hat{H} {\rm \; has \; value} \; E_n)$. We remind the reader that the notation ${\rm Pr}(.\,| \,.)$ is ours and is not used in any of our sources.}
\begin{equation}
{\rm Pr}(q \; {\rm in} \; K| E_n) = \int_K |\psi_n(q)|^2 dq.
\label{prob K}
\end{equation}
Next, he considered the probability of finding the particle in some region $K$ if we know that its energy is in some interval $I$ that includes various eigenvalues of its energy, i.e., if the particle is in some mixed state where we only know that, with equal probability, its state is one of the pure states $\psi_n(x)$ associated with the eigenvalues within the interval $I$:\footnote{The left-hand side is short-hand for: ${\rm Pr}(\hat{q} {\rm \; has \; value} \; q \; {\rm in} \; K | \hat{H} {\rm \; has \; value} \; E_n \; {\rm in} \; I) $. On the right-hand side, $\sum_{(n|E_n \, {\rm in} \, I)}$ is the sum over all $n$ such that $E_n$ lies in the interval $I$.} 
\begin{equation}
{\rm Pr}(q \; {\rm in} \; K | E_n \; {\rm in} \; I) =  \sum_{(n|E_n \, {\rm in} \, I)} \int_K |\psi_n(q)|^2 dq.
 \label{prob K I}
\end{equation}
The distinction between pure states (in Eq.\ \eqref{prob K}) and mixed states (in Eq.\ \eqref{prob K I}) slipped in here was only made explicit in the second paper in the trilogy  \citep{von Neumann 1927b}. These conditional probabilities can be written in terms of the projection operators,
\begin{equation}
\hat{E}(I) \equiv \sum_{(n|E_n \, {\rm in} \, I)} |\psi_n \rangle \langle \psi_n|, \quad \hat{F}(K) \equiv \int_K |q \rangle \langle q| \, dq,
\label{E(I) and F(K)}
\end{equation}
that project arbitrary state vectors onto the subspaces of $\overline{\textfrak{H}}$ spanned by `eigenvectors' of the Hamiltonian $\hat{H}$ and of the position operator $\hat{q}$ with eigenvalues in the ranges $I$ and $K$, respectively. 
The right-hand side of Eq.\ \eqref{prob K I} can be rewritten as:
\begin{equation}
\sum_{(n|E_n \, {\rm in} \, I)} \int_K \langle \psi_n | q \rangle \langle q | \psi_n \rangle \, dq.
 \label{prob K I 2}
\end{equation}
We now choose an arbitrary orthonormal discrete basis $\{ | \alpha \rangle \}_{\alpha=1}^\infty$ of the Hilbert space $\overline{\textfrak{H}}$. Inserting the corresponding resolution of unity,  $\hat{1} = \sum_\alpha | \alpha \rangle \langle \alpha  |$, into Eq.\ \eqref{prob K I 2}, we find
\begin{equation}
\sum_\alpha \!\! \sum_{(n|E_n \, {\rm in} \, I)} \int_K \langle \psi_n | \alpha \rangle \langle \alpha | q \rangle \langle q | \psi_n \rangle \, dq.
\label{prob K I 3}
\end{equation}
This can be rewritten as:
\begin{equation}
\sum_\alpha \langle \alpha | \left( \int_K \! | q \rangle \langle q | \, dq \; \cdot \!\!\! \sum_{(n|E_n \, {\rm in} \, I)} \!\!\!\!  | \psi_n \rangle  \langle \psi_n | \right) | \alpha \rangle.
 \label{prob K I 4}
\end{equation}
This is nothing but the trace of the product of the projection operators $\hat{F}(K)$ and $\hat{E}(I)$ defined in Eq.\ \eqref{E(I) and F(K)}. The conditional probability in Eq.\ \eqref{prob K I} can thus be written as:
\begin{equation}
{\rm Pr}(x \; {\rm in} \; K | E_n \; {\rm in} \; I) = \sum_\alpha \langle \alpha | \hat{F}(K)\hat{E}(I)
| \alpha \rangle = {\rm Tr}(\hat{F}(K)\hat{E}(I)).
\label{trace formula}
\end{equation}

This is our notation for what  \citet[p.\ 45]{von Neumann 1927a} wrote as\footnote{Since von Neumann (ibid., p.\ 43) chose $K$ to be $k$-dimensional, he actually wrote: $[\hat{F}_1(J_1) \cdot \ldots \cdot \hat{F}_k(J_k), \hat{E}(I)]$ (ibid., p. 45; hats added). For the one-dimensional case we are considering, von Neumann's expression reduces to Eq.\ \eqref{trace vN1}. For other discussions of von Neumann's derivation of this key formula, see \citet[pp.\ 320--321]{Jammer 1966} and \citet[p.\ 414]{Mehra Rechenberg}.}
\begin{equation}
[\hat{F}(K), \hat{E}(I)].
\label{trace vN1}
\end{equation}
He  defined the quantity $[\hat{A}, \hat{B}]$---{\it not to be confused with a commutator}---as (ibid., p.\ 40):
\begin{equation}
[\hat{A}, \hat{B}] \equiv [\hat{A}^\dagger \hat{B}].
\label{trace vN2}
\end{equation}
For any operator $\hat{O}$, he defined the quantity $[\hat{O}]$, which he called the ``absolute value" of $\hat{O}$, as (ibid., pp.\ 37--38):\footnote{Using the notation $Q(.\,, .)$ for the inner product (see note \ref{inner product}) and using $A$ instead of $O$, \citet[p.\ 37]{von Neumann 1927a} wrote the right-hand side of Eq.\ \eqref{trace vN3} as $\sum_{\mu,\nu=1}^\infty | Q(\varphi_\mu, A\psi_\nu) |^2$.\label{Aphipsi}}
\begin{equation}
[\hat{O}] \equiv \sum_{\mu, \nu} | \langle \varphi_\mu | \hat{O} | \psi_\nu \rangle |^2,
\label{trace vN3}
\end{equation}
where $\{ | \varphi_\mu \rangle \}_{\mu =1}^\infty$ and $\{ | \psi_\nu \rangle \}_{\nu =1}^\infty$ are two arbitrary orthonormal bases of $\overline{\textfrak{H}}$. Eq.\ \eqref{trace vN3} can also be written as:
\begin{equation}
[\hat{O}] \equiv \sum_{\mu, \nu} \langle \varphi_\mu | \hat{O} | \psi_\nu \rangle  \langle \psi_\nu | \hat{O}^\dagger | \varphi_\mu \rangle 
= \sum_\mu \langle \varphi_\mu | \hat{O} \hat{O}^\dagger | \varphi_\mu \rangle 
= {\rm Tr}(\hat{O} \hat{O}^\dagger),
\label{trace vN4}
\end{equation}
where we used the resolution of unity, $\hat{1} = \sum_\nu  | \psi_\nu \rangle  \langle \psi_\nu |$, and the fact that ${\rm Tr}(\hat{O}) = \sum_\alpha \langle \alpha | \hat{O} | \alpha \rangle$ 
for any orthonormal basis $\{ | \alpha \rangle \}_{\alpha=1}^\infty$ of $\overline{\textfrak{H}}$.\footnote{Eq.\ \eqref{trace vN4} shows that $[\hat{O}]$ is independent of the choice of the bases $\{ | \varphi_\mu \rangle \}_{\mu =1}^\infty$ and $\{ | \psi_\nu \rangle \}_{\nu =1}^\infty$. Von Neumann (1927a, p.\ 37) initially introduced the quantity $[\hat{O}; \varphi_\mu; \psi_\nu] \equiv \sum_{\mu, \nu} | \langle \varphi_\mu | \hat{O} | \psi_\nu \rangle |^2$. He then showed that this quantity does not actually depend on $\varphi_\mu$ and $ \psi_\nu$, renamed it $[\hat{O}]$ (see Eq.\ \eqref{trace vN3} and note \ref{Aphipsi}), and called it the ``absolute value of the operator" $\hat{O}$ (ibid., p.\ 38).} Using the definitions of $[\hat{A}, \hat{B}]$ and $[\hat{O}]$ in Eqs.\ \eqref{trace vN2} and \eqref{trace vN4}, with $\hat{A} = \hat{F}(K)$, $\hat{B} = \hat{E}(I)$, and $\hat{O} = \hat{F}(K)^\dagger \hat{E}(I)$, we can rewrite Eq.\ \eqref{trace vN1} as
\begin{equation}
[\hat{F}, \hat{E}] = [\hat{F}^\dagger \hat{E}] = {\rm Tr}((\hat{F}^\dagger \hat{E}) (\hat{F}^\dagger \hat{E})^\dagger) = {\rm Tr}( \hat{F}^\dagger \hat{E}  \hat{E}^\dagger \hat{F}),
\label{trace vN5}
\end{equation}
where, to make the equation easier to read, we temporarily suppressed the value ranges $K$ and $I$ of $\hat{F}(K)$ and $\hat{E}(I)$. Using the cyclic property of the trace, we can rewrite the final expression in Eq.\ \eqref{trace vN5} as ${\rm Tr}(\hat{F} \hat{F}^\dagger \hat{E}  \hat{E}^\dagger)$. Since projection operators $\hat{P}$ are both Hermitian 
and idempotent,
we have $\hat{F} \hat{F}^\dagger = \hat{F}^2 = \hat{F}$ and $\hat{E} \hat{E}^\dagger = \hat{E}^2 = \hat{E}$. Combining these observations and restoring the arguments of $\hat{F}$ and $\hat{E}$, we can rewrite Eq.\ \eqref{trace vN5} as:
\begin{equation}
[\hat{F}(K), \hat{E}(I)] = {\rm Tr}(\hat{F}(K)\hat{E}(I)),
\label{trace formula vN6}
\end{equation}
which is just Eq.\ \eqref{trace formula} for ${\rm Pr}(x \; {\rm in} \; K | E_n \; {\rm in} \; I)$ found above.

From ${\rm Tr}(\hat{F}\hat{E}) =  {\rm Tr}(\hat{E}\hat{F})$,  it follows that
\begin{equation}
{\rm Pr}(x \; {\rm in} \; K | E_n \; {\rm in} \; I) = {\rm Pr}(E_n \; {\rm in} \; I | x \; {\rm in} \; K),
\label{symm vN MB}
\end{equation}
which is just the symmetry property imposed on Jordan's probability amplitudes in postulate B of {\it Neue Begr\"undung} I \citep[p.\ 813; see Section 2.1]{Jordan 1927b} and postulate II in {\it Neue Begr\"undung} II \citep[p.\ 6; see Section 4]{Jordan 1927f}.

Von Neumann generalized Eq.\ \eqref{trace formula} for a pair of quantities to a similar formula for a pair {\it of sets} of quantities such that the operators for all quantities in each set commute with those for all other quantities in that same set but not necessarily with those for quantities in the other set \citep[p.\ 45]{von Neumann 1927a}.\footnote{Von Neumann distinguished between the commuting of $\hat{R}_i$ and $\hat{R}_j$ and the commuting of the corresponding projection operators $\hat{E}_i(I_i)$ and $\hat{E}_j(I_j)$. For {\it bounded} operators, these two properties are equivalent. If both $\hat{R}_i$ and $\hat{R}_j$ are {\it unbounded}, \citet[p.\ 45]{von Neumann 1927a} cautioned, ``certain difficulties of a formal nature occur, which we do not want to go into here" (cf.\ note \ref{bounded/unbounded}).} Let $\{ \hat{R}_i \}_{i=1}^n$ and $\{ \hat{S}_j \}_{j=1}^m$ be two such sets of commuting operators: $[\hat{R}_{i_1}, \hat{R}_{i_2}]=0$ for all $1 \leq i_1, i_2 \leq n$; $[\hat{S}_{j_1}, \hat{S}_{j_2}]=0$ for all $1 \leq j_1, j_2 \leq m$.\footnote{Unlike von Neumann (see Eqs.\ \eqref{trace vN1}--\eqref{trace vN2}), we continue to use the notation $[.,.]$ for commutators.} 
Let $\hat{E}_i(I_i)$ ($i = 1, \ldots, n$) be the projection operators onto the space spanned by eigenstates of $\hat{R}_i$ with eigenvalues in the interval $I_i$ and let $\hat{F}_j(J_j)$ ($j = 1, \ldots, m$) likewise be the projection operators onto the space spanned by eigenstates of $\hat{S}_j$ with eigenvalues in the interval $J_j$ (cf.\ Eq.\ \eqref{E(I) and F(K)}). A straightforward generalization of von Neumann's trace formula \eqref{trace formula} gives the probability that the $\hat{S}_j$'s have values in the intervals $J_j$ given that  the $\hat{R}_i$'s have values in the intervals $I_i$:
\begin{equation}
{\rm Pr}(\hat{S}_j{\rm 's} \; {\rm in} \; J_j{\rm 's} | \hat{R}_i{\rm 's} \; {\rm in} \; I_i{\rm 's} )  = {\rm Tr}(\hat{E}_1(I_1) \ldots \hat{E}_n(I_n) \hat{F}_1(J_1) \ldots \hat{F}_m(J_m)).
\label{trace formula gen}
\end{equation}
The outcomes `$\hat{R}_i \; {\rm in} \; I_i$' are called the ``assertions" ({\it Behauptungen}) and the outcomes `$\hat{S}_j \; {\rm in} \; J_j$' are called  the ``conditions"
({\it Voraussetzungen}) \citep[p.\ 45]{von Neumann 1927a}. Because of the cyclic property of the trace, which we already invoked in Eq.\ \eqref{symm vN MB}, Eq.\ \eqref{trace formula gen} is invariant under swapping all assertions with all conditions. Since all $\hat{E}_i(I_i)$'s commute with each other and all $\hat{F}_j(J_j)$'s commute with each other, Eq.\ \eqref{trace formula gen} is also invariant under changing the order of the assertions and changing the order of the conditions. These two properties are given in the first two entries of a list of five properties, labeled $\alpha$ through $\theta$ (there are no points $\zeta$ and $\eta$), of the basic rule \eqref{trace formula gen} for probabilities in quantum mechanics \citep[pp.\ 45--47]{von Neumann 1927a}. 

Under point $\gamma$, von Neumann noted that projection operators $\hat{F}(J)$ and $\hat{E}(I)$  for ``empty" ({\it nichtssagende}) assertions and conditions, i.e., those for which the intervals $J$ and $I$ are $(- \infty, + \infty)$, can be added to or removed from Eq.\ \eqref{trace formula gen} without affecting the result.

Under point $\delta$, von Neumann, following Jordan, noted that the standard multiplication law of probabilities does not hold in quantum mechanics. Parenthetically, he added, 
``(what does hold is a weaker law corresponding to the ``superposition [{\it Zusammensetzung}] of probability amplitudes" in [the formalism of] Jordan, which we will not go into here)" \citep[p.\ 46]{von Neumann 1927a}. Note that von Neumann did not use Jordan's (1927b, p.\ 812) phrase ``interference of probabilities."\footnote{After reading Heisenberg's criticism of this aspect of Jordan's theory in the uncertainty paper \citep[pp.\ 183--184, p.\ 196; cf.\ note \ref{HvJ0}]{Heisenberg 1927b}, von Neumann changed his mind \citep[p.\ 246; cf.\ note \ref{HvJ3}]{von Neumann 1927b}.\label{HvJ2}}

Under point $\varepsilon$, \citet{von Neumann 1927a} wrote: ``The addition rule of probabilities is valid" (p.\ 46). In general, as \citet[p.\ 18]{Jordan 1927b} made clear in {\it Neue Begr\"undung} I (see Eq.\ \eqref{completeness} in Section 1), the addition rule does {\it not} hold in quantum mechanics. In general, in other words, ${\rm Pr}(A\;{\rm or}\;B) \neq {\rm Pr}(A) + {\rm Pr}(B)$, even if the outcomes $A$ and $B$ are mutually exclusive. Instead, Jordan pointed out, the addition rule, like the multiplication rule, holds for the corresponding probability {\it amplitudes}. Von Neumann, however, considered only a rather special case, for which the addition rule {\it does} hold for the probability themselves. Consider Eq.\ \eqref{trace formula} for the conditional probability that we find a particle in some region $K$ given that its energy $E$ has a value in some interval $I$. Let the region $K$ consist of two disjoint subregions, $K'$ and $K''$, such that $K = K' \cup K''$ and $K' \cap K'' = \emptyset$. Given that the energy $E$ lies in the interval $I$, the probability that the particle is {\it either} in $K'$ {\it or} in $K''$, is obviously equal to the probability that it is in $K$. Von Neumann now noted that 
\begin{equation}
{\rm Pr}(x \; {\rm in} \; K | E \; {\rm in} \; I) = {\rm Pr}(x \; {\rm in} \; K' | E \; {\rm in} \; I) + {\rm Pr}(x \; {\rm in} \; K'' | E \; {\rm in}
\; I).
\label{vN addition rule 1}
\end{equation}
In terms of the trace formula \eqref{trace formula}, Eq.\ \eqref{vN addition rule 1} becomes:
\begin{equation}
 {\rm Tr}(\hat{F}(K)\hat{E}(I)) =  {\rm Tr}(\hat{F}(K')\hat{E}(I)) +  {\rm Tr}(\hat{F}(K'')\hat{E}(I)).
 \label{vN addition rule 2}
\end{equation}
Similar instances of the addition rule obtain for the more general version of the trace formula in Eq.\ \eqref{trace formula gen}.

Under point $\vartheta$, finally, we find the one and only reference  to ``canonical transformations" in {\it Mathematische Begr\"undung}.
Von Neumann (1927a, pp.\ 46--47) defined a canonical transformation as the process of subjecting {\it all} operators $\hat{A}$ to the transformation $\hat{U}\hat{A}\hat{U}^\dagger$, where $\hat{U}$ is some unitary operator. The absolute value squared $[\hat{A}]$ is invariant under such transformations. Recall $[\hat{A}] = {\rm Tr}(\hat{A}\hat{A}^\dagger)$ (see Eq.\ \eqref{trace vN4}). Now consider $[\hat{U}\hat{A}\hat{U}^\dagger]$:
\begin{equation}
[\hat{U}\hat{A}\hat{U}^\dagger] 
= {\rm Tr}(\hat{U}\hat{A}\hat{U}^\dagger (\hat{U} \hat{A} \hat{U}^\dagger)^\dagger) 
= {\rm Tr}(\hat{U}\hat{A}\hat{U}^\dagger \hat{U} \hat{A}^\dagger \hat{U}^\dagger) 
= {\rm Tr}(\hat{A} \hat{A}^\dagger) = [\hat{A}].
\end{equation}
Traces of products of operators are similarly invariant. This definition of canonical transformations makes no reference in any way to sorting quantities into sets of conjugate variables.

\section{Von Neumann's {\it Wahr\-schein\-lich\-keits\-theoretischer Aufbau} (November 1927)}

On November 11, 1927, about half a year after the first installment, {\it Ma\-the\-ma\-ti\-sche Begr\"undung} \citep{von Neumann 1927a}, the second and the third installments of von Neumann's 1927 trilogy were presented to the G\"ottingen Academy \citep{von Neumann 1927b, von Neumann 1927c}.\footnote{These three papers take up 57, 28, and 19 pages. The first installment is thus longer than the other two combined. Note that in between the first and the two later installments,  {\it Neue Begr\"undung} II appeared (see Section 4), in which \citet{Jordan 1927f} responded to von Neumann's criticism in {\it Mathematische Begr\"undung}. Von Neumann made no comment on this response in these two later papers.} The second, {\it Wahr\-schein\-lich\-keits\-theo\-re\-ti\-scher Aufbau}, is important for our purposes; the third, {\it Thermodynamik quantenmechanischer Gesamtheiten}, is not.\footnote{For discussion of this third paper, see \citet[pp.\ 439--445]{Mehra Rechenberg}. See pp.\ 431--436 for their discussion of the second paper.} In {\it Ma\-the\-ma\-ti\-sche Begr\"undung}, as we saw in Section 5, von Neumann had simply taken over the basic rule for probabilities in quantum mechanics as stated by Jordan, namely that probabilities are given by the absolute square of the corresponding probability amplitudes, the prescription now known as  the Born rule. In {\it Wahr\-schein\-lich\-keits\-theo\-re\-ti\-scher Aufbau}, he sought to derive this rule from more basic considerations.

In the introduction of the paper, \citet[p.\ 245]{von Neumann 1927b}  replaced the old opposition between ``wave mechanics" and ``matrix mechanics" by a new distinction between ``wave mechanics"  on the one hand and what he called ``transformation theory" or ``statistical theory," on the other. 
By this time, matrix mechanics and Dirac's $q$-number theory had morphed into the Dirac-Jordan statistical transformation theory. The two names von Neumann used for this theory reflect the difference in emphasis between Dirac (transformation theory) and Jordan (statistical theory).\footnote{In Section 1.3, we already quoted his observation about the Schr\"odinger wave function, $\langle q|E\rangle$ in our notation: ``Dirac interprets it as a row of a certain transformation matrix, Jordan calls it a probability amplitude" \citep[p.\ 246, note 3]{von Neumann 1927b}.} Von Neumann mentioned  Born, Pauli, and London 
as the ones who had paved the way for the statistical theory and Dirac and Jordan as the ones responsible for bringing this development to a conclusion (ibid., p. 245; cf.\ note \ref{Dirac snub}).\footnote{He cited the relevant work by \citet{Dirac 1927} and \citet{Jordan 1927b, Jordan 1927f}. He did not give references for the other three authors but presumably was thinking of \citet{Born 1926a, Born 1926b, Born 1926c}, \citet{Pauli 1927a}, and \citet{London 1926b}. The reference to London is somewhat puzzling. While it is true that London anticipated important aspects of the Dirac-Jordan transformation theory (see Lacki, 2004; Duncan and Janssen, 2009), the statistical interpretation of the formalism is not among those. Our best guess is that von Neumann took note of Jordan's repeated acknowledgment of London's paper (most prominently perhaps in footnote 1 of {\it Neue Begr\"undung} I). In his book, \citet[p.\ 2, note 2; the note itself is on p.\ 238]{von Neumann 1932} cited papers by \citet{Dirac 1927}, \citet{Jordan 1927b}, and \citet{London 1926b} in addition to the book by \citet{Dirac 1930} for the development of transformation theory. In that context, the reference to London is entirely appropriate.}

Von Neumann was dissatisfied with the way in which probabilities were introduced in the Dirac-Jordan theory. He listed two objections. First, he felt that the relation between quantum probability concepts and ordinary probability theory needed to be clarified. Second, he felt that the Born rule was not well-motivated:
\begin{quotation}
The method hitherto used in statistical quantum mechanics was essentially {\it deductive}: the square of the norm of certain expansion coefficients of the wave function or of the wave function itself was  fairly {\it dogmatically} set equal to a probability, and agreement with experience was verified afterwards. A systematic derivation of quantum mechanics from empirical facts or fundamental probability-theoretic assumptions, i.e., an {\it inductive} justification, was not given. Moreover, the relation to the ordinary probability calculus was not sufficiently clarified: the validity of its basic rules (addition and multiplication law of the probability calculus) was not sufficiently stressed \citep[p.\ 246; our emphasis]{von Neumann 1927b}.\footnote{The last sentence is in response to Heisenberg's criticism in his uncertainty paper \citep[pp.\ 183--184; cf.\ note \ref{HvJ0}]{Heisenberg 1927b} of Jordan's concept of ``interference of probabilities" defined in {\it Neue Begr\"undung} I as ``the circumstance that not the probabilities themselves but their amplitudes obey the usual composition law of the probability calculus" \citep[p.\ 812]{Jordan 1927b}. Earlier von Neumann had endorsed this concept: \citet[p.\ 5; cf.\ note \ref{HvJ1}]{Hilbert-von Neumann-Nordheim} and \citet[p.\ 46; cf.\ note \ref{HvJ2}]{von Neumann 1927a}.\label{HvJ3}}
\end{quotation}
To address these concerns, von Neumann started by introducing probabilities in terms of selecting members from a large ensemble of systems.\footnote{In his book, \citet[p.\ 158, p.\ 255, note 156]{von Neumann 1932} referred to a book by Richard \citet{von Mises 1928} for this notion of an ensemble ({\it Gesamtheit} or {\it Kollektiv}) \citep[p.\ 308]{Lacki 2000}. Von Neumann may have picked up this notion from von Mises in the period leading up to his {\it Habilitation} in mathematics in Berlin in December 1927. Von Mises was one of his examiners \citep[p.\ 402]{Mehra Rechenberg}. As \citet{von Mises 1928} explained in its preface, his book elaborates on ideas he had been presenting for ``about fifteen years" in various talks, courses, and articles. Von Mises defined a collective ({\it Kollektiv}) as an ensemble ({\it Gesamtheit}) whose members are distinguished by some observable marker ({\it beobachtbares merkmal}). One of his examples is a group of plants of the same species grown from a given collection of seeds, where the individual plants differ from one another in the color of their flowers (ibid., pp. 12--13). In their discussion of von Neumann's {\it Wahr\-schein\-lich\-keits\-theoretischer Aufbau}, \citet[p.\ 306]{Born and Jordan 1930} also cited \citet{von Mises 1928}\label{von mises}}  He then presented his ``inductive" derivation of his trace formula for probabilities (see Eq.\ \eqref{trace formula}), which contains the Born rule as a special case, from two very general, deceptively innocuous, but certainly non-trivial assumptions about expectation values of properties of the systems in such ensembles. From those two assumptions, some key elements of the Hilbert space formalism introduced in {\it Ma\-the\-ma\-ti\-sche Begr\"undung}, and two assumptions about the repeatability of measurements not explicitly identified until the summary at the end of the paper (ibid., p.\ 271), Von Neumann indeed managed to recover Eq.\  \eqref{trace formula} for probabilities. He downplayed his reliance on the formalism of {\it Ma\-the\-ma\-ti\-sche Begr\"undung} by characterizing the assumptions taken from it as ``not very far going formal and material assumptions" (ibid., p.\ 246). He referred to sec.\ IX, the summary of the paper, for these assumptions at this point, but most of them are already stated, more explicitly in fact, in sec.\ II, ``basic assumptions" (ibid., pp.\ 249--252). 

Consider an ensemble $\{ \textfrak{S}_1, \textfrak{S}_2, \textfrak{S}_3, \ldots \}$ of copies of a system \textfrak{S}.
Von Neumann wanted to find an expression for the expectation value $\mathcal{E}(\textfrak{a})$ in that ensemble of some property \textfrak{a} of the system (we use $\mathcal{E}$ to distinguish the expectation value from the projection operator $\hat{E}$). He made the following basic assumptions about $\mathcal{E}$ \citep[pp.\ 249--250]{von Neumann 1927b}:
\begin{description}
\item[A.] {\it Linearity}: $\mathcal{E}(\alpha \, \textfrak{a} + \beta \, \textfrak{b} + \gamma \, \textfrak{c} + \ldots)
= \alpha \, \mathcal{E}(\textfrak{a}) + \beta \, \mathcal{E}(\textfrak{b}) + \gamma \, \mathcal{E}(\textfrak{c}) + \ldots$ (where $\alpha$, $\beta$, and $\gamma$ are real numbers).\footnote{Here von Neumann appended a footnote in which he looked at the example of a  harmonic oscillator in three dimensions. The same point can be made with a one-dimensional harmonic oscillator with position and momentum operators $\hat{q}$ and $\hat{p}$, Hamiltonian $\hat{H}$, mass $m$, and characteristic angular frequency $\omega$: ``The three quantities [$\hat{p}/2m, m \omega^2 \hat{q}^2/2, \hat{H} = \hat{p}/2m + m \omega^2 \hat{q}^2/2$] have very different spectra: the first two both have a continuous spectrum, the third
has a discrete spectrum. Moreover, no two of them can be measured simultaneously. Nevertheless, the sum of the expectation values of the first two equals the expectation value of the third" (ibid., p.\ 249). While it may be reasonable to impose condition (A) on directly measurable quantities, it is questionable whether this is also reasonable for hidden variables (see note \ref{bell 2}).\label{bell 1}}
\item[B.] {\it Positive-definiteness}. If the quantity \textfrak{a} never takes on negative values, then $\mathcal{E}(\textfrak{a}) \geq 0$.
\end{description}
To this he added two formal assumptions (ibid., p.\ 252):
\begin{description}
\item[C.] {\it Linearity of the assignment of operators to quantities}.  If the operators $\hat{S}$, $\hat{T}$, \ldots represent the quantities \textfrak{a}, \textfrak{b}, \ldots, then $\alpha \hat{S} + \beta \hat{T} + \ldots$ represents the quantity $\alpha \, \textfrak{a} + \beta \, \textfrak{b} + \ldots$\footnote{In von Neumann's own notation, the operator $\hat{S}$ and the matrix $S$ representing that operator are both written simply as $S$.}
\item[D.] If the operator $\hat{S}$ represents the quantity $\textfrak{a}$, then $f(\hat{S})$ represents the quantity $f(\textfrak{a})$.
\end{description}

In sec.\ IX, the summary of his paper, von Neumann once again listed the assumptions that go into his derivation of the expression for  $\mathcal{E}(\textfrak{a})$. He wrote:
\begin{quotation}
The goal of the present paper was to show that quantum mechanics is not only compatible with the usual probability calculus, but that, if it [i.e., ordinary probability theory]---along with a few plausible factual [{\it sachlich}] assumptions---is taken as given, it [i.e., quantum mechanics]  is actually the only possible solution. The assumptions made were the following:
\begin{description}
\item[{\rm 1.}] Every measurement changes the measured object, and two measurements therefore always disturb each other---except when they can be replaced by a single measurement.
\item[{\rm 2.}] However, the change caused by a measurement is such that the measurement itself retains its validity, i.e., if one repeats it immediately afterwards, one finds the same
result.
\end{description}
In addition, a formal assumption:
\begin{description}
\item[{\rm 3.}] Physical quantities are to be described by functional operators in a manner subject to a few simple formal rules.
\end{description}
These principles already inevitably entail quantum mechanics and its statistics \citep[p.\ 271]{von Neumann 1927b}.
\end{quotation}
Assumptions A and B of sec.\ II are not on this new  list in sec.\ IX. Presumably, this is because they are part of ordinary probability theory. Conversely, assumptions 1 and 2 of sec.\ IX are not among the assumptions A--D of sec.\ II. These two properties of measurements, as we will see below, are guaranteed in von Neumann's formalism by the idempotency of the projection operators associated with those measurements.\footnote{Assumption 2 is first introduced in footnote 30 on p.\ 262 of von Neumann's (1927b) paper: ``Although a measurement is fundamentally an intervention ({\it Eingriff}), i.e., it changes the system under investigation (this is what the ``acausal" character of quantum mechanics is based on, cf. [Heisenberg, 1927b, on the uncertainty principle]), it can be assumed that the change occurs for the sake of the experiment, i.e., that as soon as the experiment has been carried out the system is in a state in which the {\it same} measurement can be carried out without further change to the system. Or: that if the same measurement is performed twice (and nothing happens in between), the result is the same."  Von Neumann (1927c) reiterated assumptions 1 and 2 in the introduction of  {\it Thermodynamik quantenmechanischer Gesamtheiten} and commented (once again citing Heisenberg's uncertainty paper): ``1. corresponds to the explanation given by Heisenberg for the a-causal behavior of quantum physics; 2. expresses that the theory nonetheless gives the appearance of a kind of causality" (p.\  273, note 2).\label{neumann heisenberg}} 
Finally, the ``simple formal rules" referred to in assumption 3 are spelled out in assumptions C--D.

We go over the main steps of von Neumann's derivation of his trace formula from these assumptions. 
Instead of the general Hilbert space $\bar{\textfrak{H}}$, von Neumann considered $\textfrak{H}_0$, i.e., $l^2$ (von Neumann, 1927b, p.\ 253; cf.\ von Neumann, 1927a, pp.\ 14--15 [see Section 5]).
Consider some infinite-dimensional Hermitian matrix $S$, with matrix elements $s_{\mu\nu} = s^*_{\nu\mu}$, representing an Hermitian operator $\hat{S}$. 
This operator, in turn, represents some measurable quantity  \textfrak{a}. The matrix $S$ can be written as a linear combination of three types of infinite-dimensional matrices labeled $A$, $B$, and $C$. To show what these matrices look like, we write down their finite-dimensional counterparts:
$$
A_\mu \equiv \left(
\begin{array}{ccccc}
0 & \cdots & \cdots & \cdots & 0 \\ 
\vdots & 1 & & &  \vdots \\ 
\vdots& & \ddots & & \vdots \\ 
 \vdots & & & 0 &  \vdots \\ 
0 & \cdots & \cdots & \cdots & 0 \\ 
\end{array} \right),
B_{\mu\nu} \equiv \left(
\begin{array}{ccccc}
0 & \cdots & \cdots & \cdots & 0 \\ 
\vdots & 0 & & 1 &  \vdots \\ 
\vdots& & \ddots & & \vdots \\ 
 \vdots & 1 & & 0 &  \vdots \\ 
0 & \cdots & \cdots & \cdots & 0 \\ 
\end{array} \right),
C_{\mu\nu} \equiv \left(
\begin{array}{ccccc}
0 & \cdots & \cdots & \cdots & 0 \\ 
\vdots & 0 & & i &  \vdots \\ 
\vdots& & \ddots & & \vdots \\ 
 \vdots & -i & & 0 &  \vdots \\ 
0 & \cdots & \cdots & \cdots & 0 \\ 
\end{array} \right).
$$
The $A_\mu$'s have 1 in the $\mu^{\rm th}$ row and the $\mu^{\rm th}$ column and 0's everywhere else. The $B_{\mu\nu}$'s ($\mu < \nu$) have 1 in the $\mu^{\rm th}$ row and the $\nu^{\rm th}$ column and in the $\nu^{\rm th}$ row and the $\mu^{\rm th}$ column and 0's everywhere else. The $C_{\mu\nu}$'s ($\mu < \nu$) have $i$ in the $\mu^{\rm th}$ row and the $\nu^{\rm th}$ column and $-i$ in the $\nu^{\rm th}$ row and the $\mu^{\rm th}$ column and 0's everywhere else.  


The $A_\mu$'s have eigenvectors (eigenvalue 1) with 1 in the $\mu^{\rm th}$ row and 0's everywhere else; and infinitely many eigenvectors with eigenvalue 0.  The $B_{\mu\nu}$'s have eigenvectors (eigenvalue 1) with 1's in the $\mu^{\rm th}$ and the $\nu^{\rm th}$ row and 0's everywhere else; eigenvectors (eigenvalue $-1$) with 1 in the $\mu^{\rm th}$ row, $-1$ in the $\nu^{\rm th}$ row, and 0's everywhere else; and infinitely many eigenvectors with eigenvalue 0. The $C_{\mu\nu}$'s have eigenvectors (eigenvalue 1) with $i-1$ in the $\mu^{\rm th}$ row, $i+1$ in the $\nu^{\rm th}$ row, and 0's everywhere else; eigenvectors (eigenvalue $-1$) with $-1-i$ in the $\mu^{\rm th}$ row, $1-i$ in the $\nu^{\rm th}$ row, and 0's everywhere else; and infinitely many eigenvectors with eigenvalue 0.

For the counterpart of $B_{\mu\nu}$ in a simple finite case (with $\mu, \nu = 1,2,3$), we have: 
$$ 
\left( \begin{array}{ccc}
0 & 0 &1 \\  0 & 0 & 0 \\  1 & 0 & 0 \\ 
\end{array} \right)
\,
\left( \begin{array}{c}
1 \\  0 \\  1 \\ 
\end{array} \right)
=
\left( \begin{array}{c}
1 \\  0 \\  1 \\ 
\end{array} \right),
\quad \quad
\left( \begin{array}{ccc}
0 & 0 &1 \\  0 & 0 & 0 \\  1 & 0 & 0 \\ 
\end{array} \right)
\,
\left( \begin{array}{c}
1 \\  0 \\  -1 \\ 
\end{array} \right)
=
-\left( \begin{array}{c}
1 \\  0 \\  -1 \\ 
\end{array} \right).
$$
For the counterpart of $C_{\mu\nu}$ we similarly have:
$$ 
\left( \begin{array}{ccc}
0 & 0 & i \\  0 & 0 & 0 \\  -i & 0 & 0 \\ 
\end{array} \right)
\,
\left( \begin{array}{c}
i-1 \\  0 \\  i+1 \\ 
\end{array} \right)
=
\left( \begin{array}{c}
i-1 \\  0 \\  i+1 \\ 
\end{array} \right),
\quad \quad
\left( \begin{array}{ccc}
0 & 0 & i \\  0 & 0 & 0 \\  -i & 0 & 0 \\ 
\end{array} \right)
\,
\left( \begin{array}{c}
-1-i \\  0 \\  1-i \\ 
\end{array} \right)
=
- \left( \begin{array}{c}
-1-i \\  0 \\  1-i \\ 
\end{array} \right).
$$
The matrix $S$ can be written as a linear combination of $A$, $B$, and $C$:
\begin{equation}
S = \sum_\mu s_{\mu\mu} \cdot A_\mu + \sum_{\mu < \nu} {\rm Re} \, s_{\mu\nu} \cdot B_{\mu\nu}
+  \sum_{\mu < \nu} {\rm Im} \, s_{\mu\nu} \cdot C_{\mu\nu},
\end{equation}
where ${\rm Re} \, s_{\mu\nu}$ and ${\rm Im} \, s_{\mu\nu}$ and the real and imaginary parts of $s_{\mu\nu}$, respectively.

Using von Neumann's linearity assumption (A), we can write the expectation value of $S$ in the ensemble $\{ \textfrak{S}_1, \textfrak{S}_2, \textfrak{S}_3, \ldots \}$ as:
\begin{equation}
\mathcal{E}(S) =
\sum_\mu s_{\mu\mu} \cdot \mathcal{E}(A_\mu) + \sum_{\mu < \nu} {\rm Re} \, s_{\mu\nu} \cdot 
\mathcal{E}(B_{\mu\nu})
+  \sum_{\mu < \nu} {\rm Im} \, s_{\mu\nu} \cdot \mathcal{E}(C_{\mu\nu}).
\label{E(S) 0}
\end{equation}
Since the eigenvalues of $A_\mu$, $B_{\mu\nu}$, and $C_{\mu\nu}$ are all real, the expectation values $\mathcal{E}(A_\mu)$, $\mathcal{E}(B_{\mu\nu})$, and $\mathcal{E}(C_{\mu\nu})$ are also real. 
Now define the matrix $U$ (associated with some operator $\hat{U}$) with diagonal components $u_{\mu\mu} \equiv \mathcal{E}(A_\mu)$ and off-diagonal components ($\mu < \nu$):
\begin{equation}
 u_{\mu\nu} \equiv \frac{1}{2} \left( \mathcal{E}(B_{\mu\nu}) + i \, \mathcal{E}(C_{\mu\nu}) \right), \;\;\;
u_{\nu\mu} \equiv  \frac{1}{2} \left( \mathcal{E}(B_{\mu\nu}) - i  \, \mathcal{E}(C_{\mu\nu}) \right).
\label{u mu neq nu}
\end{equation}
Note that this matrix is Hermitian: $u^*_{\mu\nu}=u_{\nu\mu}$. With the help of this matrix $U$, the expectation value of $S$ can be written as
\citep[p.\ 253]{von Neumann 1927b}:
\begin{equation}
\mathcal{E}(S) = \sum_{\mu\nu} s_{\mu\nu} \, u_{\nu\mu}.
\label{E(S) = sum s u}
\end{equation}
To verify this, we consider the sums over $\mu = \nu$ and $\mu \neq \nu$ separately. For the former we find
\begin{equation}
\sum_{\mu} \, s_{\mu\mu} \, u_{\mu\mu} = \sum_{\mu} \,s_{\mu\mu} \cdot \mathcal{E}(A_\mu).
\label{trace mu = nu}
\end{equation}
For the latter, we have
\begin{equation}
\sum_{\mu \neq \nu} s_{\mu\nu} \, u_{\nu\mu} = \sum_{\mu < \nu} s_{\mu\nu} \, u_{\nu\mu} + \sum_{\mu >  \nu} s_{\mu\nu} \, u_{\nu\mu}.
\end{equation}
The second term can be written as  $\sum_{\nu > \mu} s_{\nu\mu} \, u_{\mu\nu} = \sum_{\mu < \nu} s^*_{\mu\nu} \, u^*_{\nu\mu}$, which means that
\begin{equation}
\sum_{\mu \neq \nu} s_{\mu\nu} \, u_{\nu\mu} =  \sum_{\mu < \nu} 2 \, {\rm Re} \, ( s_{\mu\nu} \, u_{\nu\mu} ).
\end{equation}
Now write $s_{\mu\nu}$ as the sum of its real and imaginary parts and use Eq.\ \eqref{u mu neq nu} for $u_{\nu\mu}$:
\begin{eqnarray}
\sum_{\mu \neq \nu} s_{\mu\nu} \, u_{\nu\mu} & = & 
\sum_{\mu < \nu} {\rm Re} \, \{ ( {\rm Re} \, s_{\mu\nu} + i \, {\rm Im} \, s_{\mu\nu}) \cdot ( \mathcal{E}(B_{\mu\nu}) - i   \mathcal{E}(C_{\mu\nu})  ) \}  \nonumber \\
& = & \sum_{\mu < \nu} {\rm Re} \, s_{\mu\nu} \cdot  \mathcal{E}(B_{\mu\nu})
+ \sum_{\mu < \nu}  {\rm Im} \, s_{\mu\nu} \cdot \mathcal{E}(C_{\mu\nu}).
\label{trace mu < nu}
\end{eqnarray}
Adding Eq.\ \eqref{trace mu = nu} and Eq.\ \eqref{trace mu < nu}, we arrive at
\begin{equation}
\sum_{\mu \nu} \, s_{\mu\nu} \, u_{\nu\mu}  = \sum_{\mu} \,s_{\mu\mu} \cdot \mathcal{E}(A_\mu)
+ \sum_{\mu < \nu} {\rm Re} \, s_{\mu\nu} \cdot  \mathcal{E}(B_{\mu\nu})
+ \sum_{\mu < \nu}  {\rm Im} \, s_{\mu\nu} \cdot \mathcal{E}(C_{\mu\nu}).
\end{equation}
Eq.\ \eqref{E(S) 0} tells us that the right-hand side of this equation is just $\mathcal{E}(S)$. This concludes the proof of Eq.\ \eqref{E(S) = sum s u}, in which one readily recognizes the trace of the product of $S$ and $U$:\footnote{Von Neumann (1927b, p.\ 255) only wrote down the first step of Eq.\ \eqref{E(S) = Tr SU}. It was only in  the third installment of his trilogy, that  \citet[p.\ 274]{von Neumann 1927c} finally introduced the notation trace ({\it Spur}), which we used here and in Eq.\ \eqref{trace formula} in Section 5. For another discussion of the derivation of the key formula \eqref{E(S) = Tr SU}, see \citet[pp.\ 433--434]{Mehra Rechenberg}.} 
\begin{equation}
\mathcal{E}(S) = \sum_{\mu\nu} s_{\mu\nu} \, u_{\nu\mu} =  \sum_{\mu} (SU)_{\mu\mu} = {\rm Tr}(US).
\label{E(S) = Tr SU}
\end{equation}
In other words, $U$ is what is now called a {\it density matrix}, usually denoted by the Greek letter $\rho$. It corresponds to a density operator $\hat{U}$ or $\hat{\rho}$.

The matrix $U$ characterizes the ensemble $\{ \textfrak{S}_1, \textfrak{S}_2, \textfrak{S}_3, \ldots \}$. Von Neumann (1927b, sec.\ IV, p.\ 255) now focused on ``pure" ({\it rein}) or ``uniform" ({\it einheitlich}) ensembles, in which every copy $\textfrak{S}_i$ of the system is in the exact same state. Von Neumann characterized such ensembles as follows: one cannot obtain a uniform ensemble ``by mixing ({\it vermischen}) two ensembles unless it is the case that both of these correspond to that same ensemble" (ibid., p.\ 256). 
Let the density operators $\hat{U}$, $\hat{U}^*$, and $\hat{U}^{**}$ correspond to the ensembles $\{ \textfrak{S}_i \}$, $\{ \textfrak{S}^*_j \}$, $\{ \textfrak{S}^{**}_k \}$, respectively. Suppose $\{ \textfrak{S}_i \}$ consists of $\eta \times  100\%$  $\{ \textfrak{S}^*_j \}$ and $\vartheta \times 100\%$ $\{ \textfrak{S}^{**}_k \}$. The expectation value of an arbitrary property represented by the operator $\hat{S}$ in $\{ \textfrak{S}_i \}$ is then given by (ibid.):
\begin{equation}
\mathcal{E}(\hat{S}) = \eta \, \mathcal{E}^*(\hat{S}) + \vartheta \, \mathcal{E}^{**}(\hat{S}),
\end{equation}
where $\mathcal{E}^*$ and $\mathcal{E}^{**}$ refer to ensemble averages over  $\{ \textfrak{S}^*_j \}$ and $\{ \textfrak{S}^{**}_k \}$, respectively. Using Eq.\ \eqref{E(S) = Tr SU}, we can write this as: 
\begin{equation}
{\rm Tr}(\hat{U}\hat{S}) =  \eta \, {\rm Tr}(\hat{U}^*\hat{S}) + \vartheta \, {\rm Tr}(\hat{U}^{**}\hat{S}).  
\end{equation}
Since $\hat{S}$ is arbitrary, it follows that $\hat{U}$, $\hat{U}^*$, and $\hat{U}^{**}$ satisfy (ibid.):
\begin{equation}
\hat{U} = \eta \, \hat{U}^* + \vartheta \, \hat{U}^{**}.
\end{equation}
Von Neumann now proved a theorem pertaining to uniform ensembles (ibid., pp.\ 257--258). That $\hat{U}$ is the density operator for a uniform ensemble can be expressed by the following conditional statement: If ($\hat{U} = \hat{U}^* + \hat{U}^{**}$) then ($\hat{U}^* \propto \hat{U}^{**} \propto \hat{U}$). Von Neumann showed that this is equivalent to the statement that there is a unit vector $|\varphi \rangle$ such that $\hat{U}$ is the projection operator onto that vector, i.e., $\hat{U} = \hat{P}_\varphi = | \varphi \rangle \langle \varphi |$.\footnote{The notation $\hat{P}_\varphi$ (except for the hat) is von Neumann's own (ibid., p.\ 257).} Written more compactly, the theorem says:
\begin{equation}
\{ \, (\hat{U} = \hat{U}^* + \hat{U}^{**}) \Rightarrow (\hat{U}^* \propto \hat{U}^{**} \propto \hat{U}) \, \} \Leftrightarrow \{ \, \exists \, | \varphi \rangle, \, \hat{U} = \hat{P}_{\varphi} = | \varphi \rangle \langle \varphi | \, \}.
\end{equation}
The crucial input for the proof of the theorem is the inner-product structure of Hilbert space. The theorem implies two important results, which, given the generality of the assumptions going into its proof, have the unmistakable flavor of a free lunch. First, pure dispersion-free states (or ensembles) correspond to unit vectors in Hilbert space.\footnote{This is the essence of von Neumann's later no-hidden variables proof \citep[Ch.\ 4, p.\ 171]{von Neumann 1932}, which was criticized by John \citet[pp.\ 1--5]{Bell 1966}, who questioned the linearity assumption (A), $\mathcal{E}(\alpha \, \textfrak{a} + \beta \, \textfrak{b}) = \alpha \, \mathcal{E}(\textfrak{a}) + \beta \, \mathcal{E}(\textfrak{b})$ (see note \ref{bell 1}). Bell argued, with the aid of explicit examples, that the linearity of expectation values was too strong a
requirement to impose on hypothetical dispersion-free states (dispersion-free via specification of additional ``hidden" variables). In particular, the dependence of spin expectation values on the (single) hidden variable in the explicit example provided by Bell is manifestly {\em nonlinear}, although the model reproduces exactly
the standard quantum-mechanical results when one averages (uniformly) over the hidden variable. For recent discussion, see \citet{Bacciagaluppi and Crull 2009} and \citet{Bub 2010}.\label{bell 2}} Second, the expectation value of a quantity $\textfrak{a}$ represented by the operator $\hat{S}$ in a uniform ensemble $\{ \textfrak{S}_i \}$ characterized by the density operator $\hat{U} = | \varphi \rangle \langle \varphi |$ is given by the trace of the product of the corresponding matrices:
\begin{equation}
\mathcal{E}(\hat{S})  = {\rm Tr}(\hat{U}\hat{S}) =   {\rm Tr}(| \varphi \rangle \langle \varphi | \hat{S}) = \langle \varphi | \hat{S} | \varphi \rangle,
\end{equation}
which is equivalent to the Born rule. 

Von Neumann was still not satisfied. In sec.\ V, ``Measurements and states," he noted that 
\begin{quotation}
our knowledge  of a system $\textfrak{S}'$, i.e., of the structure of a statistical ensemble $\{ \textfrak{S}'_1, \textfrak{S}'_2,$ $\ldots \}$, is never described by the specification of a state---or even by the corresponding $\varphi$ [i.e., the vector $| \varphi \rangle$]; but usually by the result of measurements performed on the system \citep[p.\ 260]{von Neumann 1927b}.
\end{quotation}
He considered the simultaneous measurement of a complete set of commuting operators and constructed a density operator for (the ensemble representing) the system on the basis of outcomes of these measurements showing the measured quantities to have values in certain intervals. He showed that these measurements can fully determine the state and that the density operator in that case is the projection operator onto that state.

Let $\{ \hat{S}_\mu \}$ ($\mu = 1, \ldots, m$) be a complete set of commuting operators with common eigenvectors, $\{ | \sigma_n \rangle \}$, with eigenvalues $\lambda_\mu(n)$:
\begin{equation}
\hat{S}_\mu \,  | \sigma_n \rangle =   \lambda_\mu(n) \, | \sigma_n \rangle.
\label{S mu}
\end{equation}
Now construct an operator $\hat{S}$ with those same eigenvectors and completely non-degenerate eigenvalues $\lambda_n$:
\begin{equation}
\hat{S} | \sigma_n \rangle = \lambda_n | \sigma_n \rangle,
 \label{S}
\end{equation}
with $ \lambda_n \neq  \lambda_{n'}$ if $n \neq n'$. Define the functions $f_\mu(\lambda_n) = \lambda_\mu(n)$.
Consider the action of $ f_\mu(\hat{S})$ on $| \sigma_n \rangle$:
\begin{equation}
f_\mu(\hat{S}) \,  | \sigma_n \rangle = f_\mu(\lambda_n) \,  | \sigma_n \rangle =  \lambda_\mu(n) \, | \sigma_n \rangle =  \hat{S}_\mu \, | \sigma_n \rangle.
\label{S_mu = S}
\end{equation}
Hence, $\hat{S}_\mu = f_\mu(\hat{S})$. It follows from Eq.\ \eqref{S_mu = S} that a measurement of $\hat{S}$ uniquely determines the state of the system. As \citet{von Neumann 1927b} put it: ``In this way measurements have been identified that uniquely determine the state of [the system represented by the ensemble] $\textfrak{S}'$" (p.\ 264).

As a concrete example, consider the bound states of a hydrogen atom.
These states are uniquely determined by the values of four quantum
numbers:  the principal quantum number $n$, the orbital quantum number
$l$, the magnetic quantum number $m_l$, and the spin quantum number
$m_s$. These four quantum numbers specify the eigenvalues of four
operators, which we may make dimensionless by suitable choices of units:
the Hamiltonian in Rydberg units ($\hat{H}$/Ry), the angular momentum squared
($\hat{L}^2/\hbar^{2}$), the $z$-component of the angular momentum
($\hat{L}_z/\hbar$), and the $z$-component of the spin  ($\hat{\sigma}_z/\hbar$).
In this case, in other words,
\begin{equation}
\{ \hat{S}_\mu \}_{\mu=1}^4 = (\hat{H}/{\rm Ry}, \hat{L}^2/\hbar^{2}, \hat{L}_z/\hbar, \hat{\sigma}_z/\hbar).
\end{equation}
The task now is to construct an operator $\hat{S}$  that is a function
of  the $\hat{S}_\mu$'s (which have rational numbers as eigenvalues) and that has a
completely non-degenerate spectrum. One measurement of $\hat{S}$ then
uniquely determines the (bound) state of the hydrogen atom. For example, choose
$\alpha,\beta,\gamma,$ and $\delta$ to be four real numbers, incommensurable over
the rationals (i.e., no linear combination of $\alpha,\beta,\gamma,\delta$ with rational
coefficients vanishes), and define
\begin{equation}
 \hat{S} = \alpha\hat{S}_{1}+\beta\hat{S}_{2}+\gamma\hat{S}_{3}+\delta\hat{S}_{4}.
\end{equation}
One sees immediately that the specification of the eigenvalue of $\hat{S}$ suffices to uniquely
identify the eigenvalues of $\hat{H},\hat{L}^{2},\hat{L}_{z}$ and $\hat{\sigma}_{z}$.

Von Neumann thus arrived at the typical statement of a problem in modern quantum mechanics. There is no need anymore for $\hat{q}$'s and $\hat{p}$'s, where the $\hat{p}$'s do not commute with the $\hat{q}$'s. Instead one identifies a complete set of commuting operators. Since all members of the set commute with one another, they can all be viewed as $\hat{q}$'s. The canonically conjugate $\hat{p}$'s do not make an appearance.

To conclude this section, we want to draw attention to one more passage in  {\it Wahr\-schein\-lich\-keits\-theoretischer Aufbau}. Both Jordan and von Neumann considered conditional probabilities of the form 
$$
{\rm Pr}(\hat{a} {\rm \; has \; the \; value} \; a \, | \, \hat{b} {\rm \; has \; the \; value} \; b),
$$ or, more generally, 
$$
{\rm Pr}(\hat{a} {\rm \; has \; a \; value \; in \; interval} \; I \, | \, \hat{b} {\rm \; has \; a \;value \; in \; interval} \; J).
$$
To test the quantum-mechanical predictions for these probabilities one needs to prepare a system in a pure state in which $\hat{b}$ has the value $b$ or in a mixed state in which $\hat{b}$ has a  value in the interval $I$, and then measure $\hat{a}$. 
What happens {\it after} that measurement? Elaborating on Heisenberg's (1927b) ideas in the uncertainty paper (cf. note \ref{neumann heisenberg}), von Neumann addressed this question in the concluding section of {\it Wahr\-schein\-lich\-keits\-theoretischer Aufbau}:
\begin{quotation}
A system left to itself (not disturbed by any measurements) has a completely causal time evolution [governed by the Schr\"odinger equation]. In the confrontation with experiments, however, the statistical character is unavoidable: for every experiment there is a state adapted [{\it angepa\ss t}] to it in which the result is uniquely determined (the experiment in fact produces such states if they were not there before); however, for every state there are ``non-adapted" measurements, the execution
of which demolishes [{\it zertr\"ummert}] that state and produces adapted states according to stochastic laws \citep[pp.\ 271--272]{von Neumann 1927b}.
\end{quotation}
As far as we know, this is the first time the infamous collapse of the state vector in quantum mechanics was mentioned in print. 

\section{Conclusion: Never mind your $p$'s and $q$'s}

The postulates of Jordan's {\it Neue Begr\"undung} papers
amount to a concise formulation of the fundamental tenets of the probabilistic interpretation of quantum mechanics. Building on insights of Born and Pauli, \citet{Jordan 1927b} was the first to state {\it in full generality} that probabilities in quantum mechanics are given by the absolute square of what he called probability amplitudes. He was also the first to recognize in full generality the peculiar rules for combining probability amplitudes.\footnote{In his uncertainty paper, however, \citet[pp.\ 183--184, p.\ 196]{Heisenberg 1927b} criticized Jordan's notion of the ``interference of probabilities" (see note \ref{HvJ0}).} However, after laying down these rules in a set of postulates, he did not fully succeed in constructing a satisfactory formalism realizing those postulates.

As we argued in this paper, Jordan was lacking the requisite mathematical tools to do so, namely abstract Hilbert space and the spectral theorem for operators acting in Hilbert space. Instead, Jordan drew on the canonical formalism of classical mechanics.
Jordan was steeped in this formalism, which had played a central role in the transition from the old quantum theory to matrix mechanics \citep{Duncan and Janssen 2007} as well as in the further development of the new theory, to which Jordan had made a number of significant contributions \citep{Duncan and Janssen 2008, Duncan and Janssen 2009}. Most importantly in view of the project Jordan pursued in {\it Neue Begr\"undung}, he had published two papers the year before \citep{Jordan 1926a, Jordan 1926b}, in which he had investigated the implementation of canonical transformations in matrix mechanics \citep{Lacki 2004, Duncan and Janssen 2009}. As he put it in his AHQP interview (see Section 2.2 for the full quotation), canonically conjugate variables and canonical transformations had thus been his ``daily bread" in the years leading up to {\it Neue Begr\"undung}. 

Unfortunately, as we saw in Sections 2 and 4, this formalism---the $p$'s and $q$'s of the title of our paper---proved ill-suited to the task at hand. As a result, Jordan ran into a number of serious problems. First, it turns out to be crucially important for the probability interpretation of the formalism that only Hermitian operators be allowed. Unfortunately, canonical transformations can turn $\hat{p}$'s and $\hat{q}$'s corresponding to Hermitian operators into new $\hat{P}$'s and $\hat{Q}$'s corresponding to non-Hermitian ones. Initially, Jordan hoped to make room in his formalism for quantities corresponding to non-Hermitian operators by introducing the so-called {\it Erg\"anzungsamplitude} (see Section 2.4). Eventually, following the lead of \citet{Hilbert-von Neumann-Nordheim}, he dropped the {\it Erg\"anzungsamplitude}, which forced him to restrict the class of allowed canonical transformations (rather arbitrarily from the point of view of classical mechanics) to those preserving Hermiticity. The difficulties facing Jordan's approach became particularly severe when, in {\it Neue Begr\"undung} II, \citet{Jordan 1927f} tried to extend his formalism, originally formulated only for quantities with purely continuous spectra, to quantities with wholly or partly discrete spectra. One problem with this extension was that, whereas canonical transformations do not necessarily preserve Hermiticity, they {\it do} preserve the spectra of the $\hat{p}$'s and $\hat{q}$'s to which they are applied. Hence, there is no canonical transformation that connects the generalized coordinate $\hat{q}$, which has a continuous spectrum, to, for instance, the Hamiltonian $\hat{H}$, which, in general, will  have at least a partly discrete spectrum. As Jordan's construction of a realization of his postulates hinged on the existence of a canonical transformation connecting $\hat{q}$ and $\hat{H}$, this presented an insurmountable obstacle. The newly introduced spin variables further exposed the limitations of Jordan's canonical formalism. To subsume these variables under his general approach, Jordan had to weaken his definition of canonically conjugate quantities to such an extent that the concept lost much of its meaning. Under Jordan's definition in {\it Neue Begr\"undung} II, any two of the three components $\hat{\sigma}_x$, $\hat{\sigma}_y$, and $\hat{\sigma}_z$ of spin angular momentum are canonically conjugate to each other.

All these problems can be avoided if the canonical formalism of classical mechanics is replaced by the Hilbert space formalism, even though other mathematical challenges remain. When Jordan's probability amplitudes $\varphi(a,b)$ for the quantities $\hat{a}$ and $\hat{b}$ are equated with `inner products' $\langle a | b \rangle$ of normalized `eigenvectors' of the corresponding operators $\hat{a}$ and $\hat{b}$, the rules for such amplitudes, as laid down in the postulates of Jordan's {\it Neue Begr\"undung}, are automatically satisfied. Probabilities are given by the absolute square of these inner products, and Jordan's addition and multiplication rules for probability amplitudes essentially reduce to the familiar completeness and orthogonality relations in Hilbert space (see Section 2.1, Eq.\ \eqref{completeness/orthogonality}). Once the Hilbert space formalism is adopted, the need to sort quantities into $\hat{p}$'s and $\hat{q}$'s disappears. Canonical transformations, at least in the classical sense as understood by Jordan, similarly cease to be important. Instead of canonical transformations $(\hat{p}, \hat{q}) \rightarrow (T\hat{p}T^{-1}, T\hat{q}T^{-1})$ of pairs of canonically conjugate quantities, one now considers unitary transformations $\hat{A} \rightarrow U\hat{A}U^{-1}$ of individual Hermitian operators. Such transformations get us from one orthonormal basis of Hilbert space to another, preserving inner products as required by the probability interpretation of quantum theory.

The Hilbert space formalism was introduced by \citet{von Neumann 1927a} in {\it Mathematische Begr\"undung}. However, von Neumann did not use this formalism to provide a realization of Jordan's postulates along the lines sketched in the preceding paragraph. As we saw in Section 5, Von Neumann had some fundamental objections to the approach of Jordan (and Dirac). The basic probability amplitude for $\hat{p}$ and $\hat{q}$ in Jordan's formalism, $\langle p | q \rangle = e^{-ipq/\hbar}$ (see Eq.\ \eqref{NB1-18}), is not a square-integrable function and is thus not an element of the space $L^2$ instantiating abstract Hilbert space. The delta function, unavoidable in the Dirac-Jordan formalism, is no function at all. Moreover, von Neumann objected to the phase-ambiguity of the probability amplitudes.

Jordan's response to this last objection illustrates the extent to which he was still trapped in thinking solely in terms of $p$'s and $q$'s. In {\it Neue Begr\"undung} II, he eliminated the phase-ambiguity of the probability amplitude for any two quantities by adding two indices indicating a specific choice of the quantities canonically conjugate to those two quantities (see Section 4, Eqs.\ \eqref{non-uniqueness 1}--\eqref{non-uniqueness 2}). Von Neumann's response to this same problem was very different and underscores that he was not wedded at all to the canonical formalism of classical mechanics. Von Neumann decided to avoid probability amplitudes altogether. Instead he turned to projection operators in Hilbert space, which he used both to formulate the spectral theorem and to construct a new formula for conditional probabilities in quantum mechanics (see Eq.\ \eqref{trace formula} and Eq.\ \eqref{trace formula gen}). 

Although von Neumann took Jordan's formula for conditional probabilities as his starting point and rewrote it in terms of projection operators, his final formula is more general than Jordan's in that it pertains both to pure and to mixed states. However, it was not until the next installment of his 1927 trilogy, {\it Wahr\-schein\-lich\-keits\-theoretischer Aufbau}, that \citet{von Neumann 1927b} carefully defined the difference between pure and mixed states. In this paper, von Neumann freed his approach from reliance on Jordan's even further  (see Section 6). He now derived his formula for conditional probabilities  in terms of the trace of products of projection operators from the Hilbert space formalism, using a few seemingly innocuous assumptions about expectation values of observables of  systems in an ensemble of copies of those systems characterized by a density operator. He then showed that the density operator for a uniform ensemble is just the projection operator onto the corresponding pure dispersion-free state. Such pure states can be characterized completely by the eigenvalues of a complete set of commuting operators. This led von Neumann to a new way of formulating a typical problem in quantum mechanics. Rather than identifying $\hat{p}$'s and $\hat{q}$'s for the system under consideration, he realized that it suffices to specify the values of a maximal set of commuting operators for the system. All operators in such sets can be thought of as $\hat{q}$'s. There is no need to find the $\hat{p}$'s canonically conjugate to these $\hat{q}$'s.

\section*{Coda: Return of the $p$'s and $q$'s in Quantum Field Theory}

In this paper, we emphasized the difficulties engendered by Jordan's insistence on the primacy of canonical $(\hat{p}, \hat{q})$ variables in expressing the dynamics of general  quantum-mechanical systems (see especially Section 4 on {\it Neue Begr\"undung} II). These difficulties became particularly acute in the case of systems with observables with purely or partially discrete spectra, of which the most extreme case is perhaps the treatment of electron spin (see Eqs.\ \eqref{example-e-1}--\eqref{example-e-3}). Here, the arbitrary choice of two out of the three spin components to serve as a non-commuting canonically conjugate pair clearly reveals the artificiality of this program. 

In a sense, however, Jordan was perfectly right to insist on the importance of a canonical approach, even for particles with spin. In hindsight, his error was simply to attempt to impose this structure at the level of non-relativistic quantum theory, where electron spin appears as an essentially mysterious internal attribute that must be grafted
onto the nonrelativistic kinematics (which does have a perfectly sensible canonical interpretation). Once electron spin was shown to emerge naturally at the relativistic level, and all aspects of the electron's dynamics were subsumed in the behavior of a local relativistic field, canonical `$p$ \& $q$' thinking could be reinstated in a perfectly natural way.  This was first done explicitly by \citet{Heisenberg and Pauli 1929} in their seminal paper on Lagrangian field theory. 
In modern notation, they introduced a relativistically invariant action for a spin-$\frac{1}{2}$ field, as a spacetime integral of a Lagrangian:
\begin{equation}
\label{SDirac}
   S = \int {\mathcal L} \, d^{4}x = \int \overline{\psi}(x)(i\gamma^{\mu}\partial_{\mu}-m)\psi(x)d^{4}x.
   \end{equation}
   Here the field $\psi(x)$ is a four-component field, with the $\gamma^{\mu}$ $(\mu=0,1,2,3)$ the $4 \times 4$ Dirac matrices.
   A conjugate momentum {\em field} $\pi^{\psi}(x) \equiv \partial{\mathcal L}/\partial\dot{\psi}$
   is then defined in the standard fashion, with canonical equal-time {\em anti}-commutation  relations imposed---as indicated by earlier work by \citet{Jordan 1927h} and \citet{Jordan and Wigner 1928}---in order to insert the desired fermionic
   statistics of the particles described by the field,
   \begin{equation}
   \label{anticomm}
     \{ \pi^{\psi}_{m}(\vec{x},t), \psi_{n}(\vec{y},t)  \} = i\hbar\delta_{mn}\delta^{3}(\vec{x}-\vec{y}).
      \end{equation}
   The transition from a Lagrangian to a Hamiltonian (density) is then carried out in the usual way
   \begin{equation}
     {\mathcal H} = \pi^{\psi}\dot{\psi} - {\mathcal L} = \bar{\psi}(i\vec{\gamma}\cdot\vec{\nabla}+m)\psi(x).
\end{equation}
  The spatial integral of this Hamiltonian energy density would within a few years be shown to describe exactly the free Hamiltonian
  for arbitrary multi-particle states of non-interacting electrons and positrons, degenerate in mass and
  each displaying the usual panoply of spin-$\frac{1}{2}$ behavior which had  finally been deciphered, in
  the non-relativistic context, by the atomic spectroscopy of the mid to late 1920s. The essential point is that
  the relevant $p$'s and $q$'s
  appearing in the theory are not associated with the first-quantized wave-functions or
  state vectors appropriate for a non-relativistic treatment, but rather with the {\em fields} that must replace them
  once a fully relativistic theory takes center stage.

\section*{Acknowledgments}
  
This paper originated in talks we gave at New Directions in the Foundations of Physics, Washington, D.C., April 30--May 2, 1910, and at the Third International Conference on the History of Quantum Physics (HQ3) at the {\it Max-Planck-Institut f\"{u}r Wissenschaftsgeschichte} (MPIWG), Berlin, June 28--July 2, 2010. We thank Wolf Beiglb\"ock, Olivier Darrigol, Gonzalo Gimeno, Christoph Lehner, Charles Midwinter, Serge Rudaz, Rob ``Ryno" Rynasiewicz, Jos Uffink, Mercedes Xipell, and an anonymous referee for helpful comments, discussion, and references. We thank Michael Jordan and Jochen Heisenberg for permission to quote from correspondence of and interviews with their fathers in the Archive for History of Quantum Physics (AHQP). We consulted the copy of the AHQP in Walter Library at the University of Minnesota and the electronic copy at the MPIWG. Michel Janssen gratefully acknowledges support of the MPIWG. The research of Anthony Duncan is supported in part by the National Science Foundation under grant PHY-0854782.

\end{document}